\documentclass{article}

\usepackage{arxiv}

\usepackage[utf8]{inputenc} % allow utf-8 input
\usepackage[T1]{fontenc}    % use 8-bit T1 fonts
\usepackage{hyperref}       % hyperlinks

\usepackage{booktabs}       % professional-quality tables
\usepackage{amsfonts}       % blackboard math symbols
\usepackage{nicefrac}       % compact symbols for 1/2, etc.
\usepackage{microtype}      % microtypography
\usepackage{lipsum}		% Can be removed after putting your text content
\usepackage{graphicx}
\usepackage{subcaption}
\usepackage{doi}

\usepackage{latexsym}
\usepackage{graphicx}
\usepackage{multicol,multirow}
\usepackage{mathtools}
\usepackage{amsmath,amssymb,amsfonts}
\usepackage{mathrsfs}
\usepackage{amsthm}
\usepackage{rotating}
\usepackage{appendix}
\usepackage{ifpdf}
\usepackage[T1]{fontenc}
\usepackage{times}
\usepackage{newtxmath}
\usepackage{textcomp}%
\usepackage{xcolor}%
\usepackage{lipsum}
\usepackage{needspace}
\usepackage{float}
\usepackage{chngcntr}
\usepackage[backend=biber]{biblatex}
\usepackage{stackrel}
\usepackage{todonotes}
\presetkeys{todonotes}{color=orange!25}{}
\newcommand{\ind}{\mathrel{\perp\!\!\!\perp}}

\newcommand{\tZ}{\tilde{Z}}

\newcommand{\var}{\operatorname{Var}}
\usepackage{booktabs}

\usepackage{xspace}

\newcommand{\ourmethod}{\textit{EnScale}\xspace}
\newcommand{\condenscoreloss}{EnScale-loss\xspace}
\newcommand{\ourmethodmarg}{\textit{EnScale}\xspace}
\newcommand{\ourmethodtemp}{\textit{EnScale-t}\xspace}

\newcommand{\temporalcoarsemodel}{temporally-consistent coarse model\xspace}

\usepackage{enumerate}
\usepackage{siunitx}

\providecommand{\citeA}[1]{\cite{#1}}

\title{EnScale: Temporally-consistent multivariate generative downscaling via proper scoring rules}

\author{
Maybritt Schillinger \\ ETH Zurich \\ maybritt.schillinger@stat.math.ethz.ch
\and
Maxim Samarin \\ Swiss Data Science Center, \\ EPFL and ETH Zurich \\
\and
Xinwei Shen \\ University of Washington
\and 
Reto Knutti \\ ETH Zurich
\and
Nicolai Meinshausen \\ ETH Zurich 
}

\date{\today}

\begin{document}
\maketitle

\begin{abstract}
The practical use of future climate projections from global circulation models (GCMs) is often limited by their coarse spatial resolution, requiring downscaling to generate high-resolution data. Regional climate models (RCMs) provide this refinement, but are computationally expensive. To address this issue, machine learning (ML) models can learn the downscaling function, mapping coarse GCM outputs to high-resolution fields. Among these, generative approaches aim to capture the full conditional distribution of RCM data given coarse-scale GCM data, which is characterized by large variability and thus challenging to model accurately. We introduce \textit{EnScale}, a generative ML framework emulating the full GCM-to-RCM map by training on multiple pairs of GCM and corresponding RCM data. It first adjusts large-scale mismatches between GCM and coarsened RCM data, followed by a super-resolution step to generate high-resolution fields. To efficiently model the high-dimensional output, the super-resolution step employs a novel class of sparse local stochastic layers. Both steps employ generative models optimized with the energy score, a proper scoring rule. Compared to state-of-the-art ML downscaling approaches, our setup reduces computational cost by about one order of magnitude. \textit{EnScale} jointly emulates multiple variables -- temperature, precipitation, solar radiation, and wind -- spatially consistent over Central Europe. In addition, we propose a variant \textit{EnScale-t} that enables temporally consistent downscaling. We establish a comprehensive evaluation framework across various categories including calibration, spatial and temporal structure, extremes, and multivariate dependencies. Comparison with diverse benchmarks demonstrates \textit{EnScale(-t)}'s competitive performance and computational efficiency, offering a promising approach for accurate and temporally consistent RCM emulation.
\end{abstract}

%in capturing spatial variability, multivariate dependencies, and temporal structure. 
%Our method 
%compares favorably with a state-of-the-art diffusion model, and the best-performing model typically depends on the specific variable and evaluation metric. 

%\tableofcontents
%\section{Title ideas}
%\quest{So far, ``EnScale: Temporally-consistent multivariate generative downscaling with proper scoring rules'' won the discussion. The only thing that I don't like about it is the looong chain of adjectives at the start. Hence, I have a small preference for ``EnScale: Temporally-consistent multivariate downscaling with generative modeling via proper scoring rules'', but the downside is that it's a little bit longer. What do you think? You can still convince me otherwise :D}
%\mytodo{Nicolai: short is better; Anya: longer is better}

%in capturing spatial variability, multivariate dependencies, and temporal structure. 
%Our method 
%compares favorably with a state-of-the-art diffusion model, and the best-performing model typically depends on the specific variable and evaluation metric. 

\section[Introduction]{Introduction}

General circulation models (GCMs) provide us with future climate projections that enable assessing potential changes in climate under different greenhouse gas emission scenarios. Multiple research groups around the world develop and maintain GCMs that incorporate different modeling assumptions and parameterizations to simulate the climate system of the Earth. By virtue of international efforts such as the Coupled Model Intercomparison Project (CMIP) \cite{Eyring2016OverviewOrganization, Taylor2012AnDesign}, large ensembles of GCM data are available for various Representative Concentration Pathway (RCP) / Shared Socioeconomic Pathway (SSP) scenarios. 

For computational reasons, these physics-based models are typically operated at a coarse spatial resolution. GCMs are able to simulate large-scale climate patterns, but often lack the spatial resolution needed to capture climate variations at impact-relevant regional and local scales \cite{Benestad2010DownscalingExtremes, Maraun2016BiasReview, Rampal2024EnhancingLearning,  Wilby1997DownscalingLimitations}. \textit{Downscaling} describes the process of increasing the spatial and/or temporal resolution of GCM outputs. The common approach of \textit{dynamical downscaling} relies on process-based regional climate models (RCMs) that use GCM data as boundary conditions, i.e.,~RCMs are ``driven'' by the corresponding GCM. They refine the GCM's large-scale climate information and yield high-resolution projections \cite{Giorgi2019ThirtyNext}. Due to necessary simplifications and parameterizations, regional climate models differ in their representation of processes, resulting in different RCMs, and initiatives such as the Coordinated Regional Climate Downscaling Experiment (CORDEX) \cite{Giorgi2015RegionalInitiative} provide data from multiple models for several regions around the globe.

However, RCMs remain computationally costly. To address this challenge, \textit{statistical downscaling} establishes a statistical relationship between low-resolution (LR) and high-resolution (HR) data. In addition to more traditional approaches (e.g.~\cite{Maraun2010PrecipitationUser, Maraun2018StatisticalResearch}), machine learning (ML) for resolution enhancement has been applied to predict high-resolution outputs from coarse data in various contexts. This includes downscaling GCMs to RCM resolutions \cite{Rampal2024EnhancingLearning}, targeting convection-permitting models \cite{Addison2022} or high-resolution observational data \cite{Harris2022AForecasts, Leinonen2020, Price2022IncreasingModels, Vosper2023DeepScales}, often using varying sets of predictors.

Many different setups for ML-based downscaling have been considered. Among these, some attempts directly approximate the full path from GCM to RCM data: the LR predictors stem from GCM data and the predictands from the RCM \cite{Bano-Medina2021}. We refer to this as \textit{RCM emulation}. Since there are modeling biases between global and regional outputs, downscaling GCMs to RCMs requires learning a transformation map in addition to the pure resolution refinement \cite{
vanderMeer2023DeepFrameworks}. Similar mismatches can arise in other setups, for example with predictors from a biased numerical weather model and targets from radar measurements \cite{Harris2022AForecasts, Price2022IncreasingModels}. Alternative setups include \textit{super-resolution} and \textit{perfect prognosis downscaling}. In the super-resolution setting, the inputs for the ML model are artificially coarsened versions of the HR targets, so that the predictor is exactly a degraded form of the target field \cite{Harder2023Hard-ConstrainedDownscaling, Leinonen2020, Stengel2020, Vosper2023DeepScales}. Other works instead use different climate variables as predictors, e.g.~by learning the relationship between large-scale atmospheric predictors and a local HR target. In perfect prognosis downscaling, this relationship is learned from observational data \cite{Bano-Medina2020ConfigurationDownscaling}, whereas \citeA{Doury2022} and \citeA{Doury2024OnPrecipitation} learn it from RCM data. Note that they also refer to this task as ``RCM emulation'', which is different from our RCM emulation: they consider both predictors and targets from the RCM, whereas we use predictors from the GCM and targets from the RCM. The literature shows mixed results: Some studies conclude that emulators can successfully learn the GCM-RCM map directly \cite{vanderMeer2023DeepFrameworks}, whereas others find better performance for the pure super-resolution task compared to a setup that additionally needs to learn large-scale shifts \cite{Harris2022AForecasts}, and some highlight the potential of learning the transformation map and the downscaling map separately \cite{Price2022IncreasingModels}.

Many downscaling approaches focus on producing a deterministic map at the fine resolution, given LR predictors \cite{Bano-Medina2020ConfigurationDownscaling, Bano-Medina2021, Doury2022}. However, to capture the full variability, including extreme events, a stochastic approach is needed. In this case, the goal is to model the conditional distribution of the HR target given the LR input. Generative ML methods provide a way to achieve this by ``learning'' a complex, high-dimensional distribution from training data. After training, generative models can generate samples that reflect the variability present in the training distribution. Examples of generative models are Generative Adversarial Networks (GANs) \cite{Goodfellow2020GenerativeNetworks} and diffusion models \cite{Ho2020DenoisingModels, KarrasElucidatingModels, Song2019GenerativeDistribution, Song2020Score-BasedEquations}. In past studies, both approaches have been used successfully, e.g.~to downscale precipitation \cite{Addison2022, Harris2022AForecasts, Leinonen2020, Vosper2023DeepScales} or wind \cite{Miralles2022, Stengel2020, White2019}. While GANs can often generate realistic samples, their optimization can be unstable \cite{Arjovsky2017TowardsNetworks}, and they are prone to mode collapse, where the model fails to represent the full diversity of the target distribution. Diffusion models, in contrast, are effective in capturing complex distributions, but are computationally costly, especially during inference. Reducing computational latency is an active area of research, e.g., via efficient sampling algorithms \cite{Song2022DenoisingModels, Lu2022DPM-Solver:Steps}, latent diffusion models \cite{Rombach2022High-ResolutionModels}, and distillation methods \cite{Salimans2022ProgressiveModels, Song2023ConsistencyModels}. While these approaches can accelerate inference, they may introduce trade-offs in sample quality depending on compression levels or the number of denoising steps.
% Efficient sampling algorithms reduce the number of required denoising steps without retraining \cite{Song2020ddim, Lu2020dpm}. Other promising directions include latent diffusion models \cite{Rombach2022stablediff}, which operate in a compressed latent space and reduce per-step computational costs, and distillation methods \cite{Salimans2022progressive, Song2023consistency}, which learn to efficiently approximate the full multi-step diffusion process to perform few-step or even one-step generation. While these approaches can accelerate inference, they may introduce trade-offs in sample quality depending on compression levels or the number of denoising steps.

Besides GANs and diffusion models, it is possible to train a generative model by minimizing a proper scoring rule \cite{Shen2024Engression:Regression, Alet2025SkillfulMarginals, Chen2022GenerativePost-processing, Kochkov2024NeuralClimate, Lang2024AIFS-CRPS:Score, Pacchiardi2024ProbabilisticMinimization}. Such a scoring rule compares the generated distribution by the machine learning model with the target RCM distribution. It is minimal if and only if the generated and target distribution agree \cite{Gneiting2007StrictlyEstimation}. This loss function allows for easy optimization and fast training. In addition, it is flexible and can be combined with different model architectures. While the Continuous Ranked Probability Score (CRPS) recently gained attention in AI-based weather forecasting, its multivariate generalization, the energy score, has been adopted as a training objective only in a few recent studies \cite{Shen2023Engression:Regression, Shen2025ReverseDistributions, Pacchiardi2024ProbabilisticMinimization} and remains relatively underexplored. To our knowledge, training via proper scoring rules has not yet been rigorously investigated the case of climate downscaling.

Temporal consistency of downscaled data is important for accurate representation of multi-day statistics, for example in the assessment of heatwaves or accumulated precipitation. Yet, most existing methods do not consider temporal consistency. A few studies perform refinement in a temporal dimension, generating multiple time steps on a high-resolution scale for one or more input fields \cite{Glawion2023SpateGAN:Approach, Serifi2021, Bassetti2024DiffESM:Models}. However, they still process each input patch separately of previous ones and, thus, do not ensure temporal consistency of longer time series on the high-resolution scale. In contrast, AI-based weather forecasting commonly employs an autoregressive approach: previous time steps are included as predictors, and an iterative application of the ML model can provide temporally consistent predictions. 

In this study, we emulate the conditional distribution of the RCM data giving the corresponding GCM input for a domain covering Central Europe. We train our model on a joint dataset combining multiple pairs of RCMs and GCMs. To make best use of this pooled dataset, we introduce a novel approach that splits the problem into several physically meaningful steps, similar to \citeA{Price2022IncreasingModels} and \citeA{Lopez-Gomez2025Dynamical-generativeEnsembles}. We first correct for large-scale mismatches between the driving GCM and a coarsened version of the target RCM. Second, we perform super-resolution from the coarsened RCM to the high-resolution RCM. For this, we progressively refine the resolution in multiple stages \cite{Shen2025ReverseDistributions}. In addition, to handle the high-dimensional output in a computationally efficient manner, we introduce a new architecture with sparse local stochastic layers. The generated values at a specific location only depend on the values at the lower-resolution in a local neighborhood. This architecture enables learning location-specific features while at the same time limiting the number of model parameters. For each step, we train a generative model minimizing the \textit{energy score}. Accordingly, we call our method \textit{EnScale} (energy score + downscaling).

Our contributions are as follows: 
\begin{enumerate}
    \item We demonstrate that fully data-driven emulation of RCMs from GCM data in a multivariate and generative setup is feasible, both capturing the variability of RCM targets as well as multivariate dependencies by jointly emulating temperature (\textit{tas}), precipitation (\textit{pr}), solar radiation (\textit{rsds}), and surface wind (\textit{sfcWind}).
    \item We propose new methods for RCM emulation: \textit{EnScale} and \textit{EnScale-t}, which combine the idea of downscaling via coarse correction with a novel generative model architecture, optimized for using the energy score as a loss function. This loss allows accurate representation of the target distribution, while at the same time yielding computationally lightweight models.
    \item \textit{EnScale} is our version that downscales each day independently of previous days, whereas \textit{EnScale-t} conditions on the previous day and allows for generating temporally consistent time series.
    \item We establish a comprehensive evaluation framework with metrics for overall performance, spatial and temporal structure, behavior in the extremes, variable dependencies and climate change signal. For comparison, we propose a set of benchmarks, including physically motivated approaches and state-of-the-art generative models.
\end{enumerate}

\section{Setup and methods}

\subsection{Downscaling as a generative modeling problem}
\label{sec:setup:downscaling-generative}

We view downscaling as a generative modeling problem as follows. Let $X$ denote the daily GCM data, consisting of multiple climate variables across all grid cells in Europe, with distribution $p_X$. Likewise, we call $Y$ the corresponding downscaled RCM data on our target area in Central Europe, with distribution $p_Y$. We denote all probability distributions by their corresponding density $p$. We aim to model the conditional distribution $p_{Y|X}$. Intuitively, this conditional $p_{Y|X}$ represents the distribution that we would obtain if we could re-run the same RCM with slightly perturbed GCM boundary conditions. The RCM output has much variability that is not present in the GCM input, as we discuss in Sec.~\ref{sec:results:variability}. 
Physically, this can be interpreted as internal variability: different plausible realizations of local weather that are consistent with the same boundary forcing. Thus, multiple draws from the conditional distribution $p_{Y|X}$ correspond to such alternative realizations of internal variability. When these draws are produced by our ML model (or a baseline), we will refer to them as ``samples'', or alternatively as ``predictions''. Note that, in our case, ``predictions'' refer to the output of our downscaling models, not to forecasts of future climate (downscaled fields always correspond to the same time point as the GCM input). The predictions from our model can capture differences between RCMs, and we investigate various sources of variability later in Sec.~\ref{sec:results:variability}.

\begin{figure}
	\centering
	\includegraphics[width=0.7\linewidth]{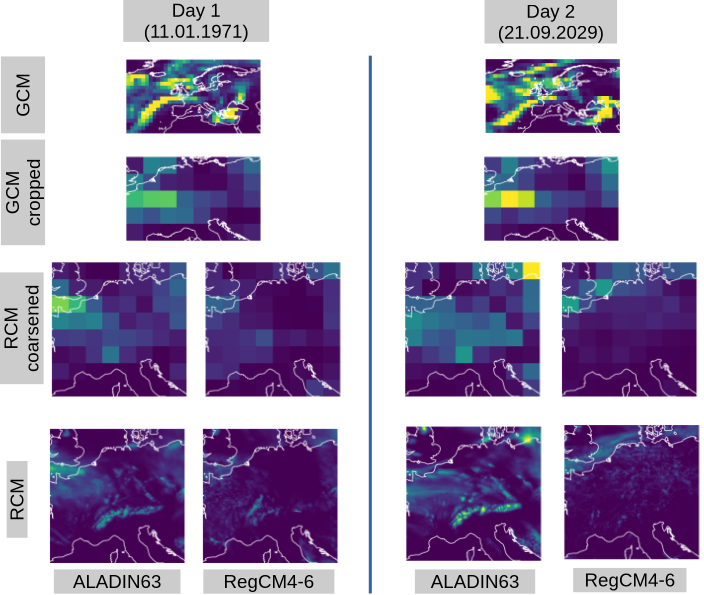}
	\caption{Illustration of the dataset. This highlights the difference between RCMs, the mismatches of GCM and RCM even at a coarse scale, and the large variability of RCM fields given GCM conditions. The first row shows the GCM data from CNRM-CM5 on two example days with similar precipitation fields, displayed on the full spatial extent that is used as the model input. The second row presents the same GCM data cropped to approximately match the target area used in this study. The fourth row shows the corresponding target RCM data (ALADIN63 and RegCM4-6) that downscaled this GCM. The third row presents these RCM fields manually coarsened to a resolution comparable to the GCM data. Small differences between shape and domain of cropped GCM (row two) and coarsened RCM (row three) are due to small resolution differences and misaligned grids.}
	\label{fig:setup:data-example}
\end{figure}

\subsection{Downscaling with coarse correction}
\label{sec:setup:2step-marginal}
Our model is trained jointly on the entire dataset comprising multiple pairs of GCM data $X$ and corresponding RCM data $Y$. They stem from three different GCMs, and each driving GCM was downscaled by several RCMs (between two and three RCMs for one GCM, see Sec.~\ref{sec:methods:data_processing}). Despite training a single unified model, our goal is to capture RCM-specific characteristics. We aim to learn the conditional $p^\iota_{Y|X}$. Here, $\iota$ is an index for the RCM, emphasizing that we target the distribution for each RCM individually. Fig.~\ref{fig:setup:data-example} shows example GCM data on two selected days as well as corresponding downscaled data from two RCMs. These examples help to illustrate several challenges present in our heterogeneous dataset, as explained in the following.

Firstly, the GCM data differs from the RCM not only on local scales. Even when the RCM data is coarsened to a similar resolution as the GCM data, vast disagreements remain for some variables and days. For example, comparing the cropped GCM data with the pooled RCM data in Fig.~\ref{fig:setup:data-example} reveals distinct mismatches. Fig.~\ref{fig:setup:data-example} also illustrates the large variability of the conditional distribution of the coarsened RCM data given the GCM data. Even though the GCM input $X$ is similar for both days, the coarsened RCM data show considerable differences (see Sec.~\ref{sec:results:variability}). These differences are due to internal variability of the climate system, as the GCM's boundary conditions do not uniquely determine the weather conditions simulated by the RCM.

Secondly, each GCM-RCM pair follows a different distribution. While we train the ML model jointly with all pairs to maximize the available dataset size, we still aim for the ML model to learn the correct downscaling function for each pair individually by conditioning on the model on the RCMs considered. We also tested conditioning on the GCM information. However, we found that this does not improve the result. Hence, to allow our model to be applicable to unseen GCMs, we decided to only add a label for the RCM. Fig.~\ref{fig:setup:data-example} compares the downscaled fields by two different RCMs, given the same driving GCM. We find that responses from multiple RCMs to the same GCM input differ considerably. Partly, this can be due to internal variability, as discussed above. In addition, there are disagreements between RCMs. This is because each RCM has parameterizations describing the effect of small-scale processes (e.g., clouds) that differ from those of other RCMs and the driving GCM \cite{Giorgi2019ThirtyNext}. The land surface may be fixed in one model and respond to the atmosphere in another.

\begin{figure}
	\centering
	\includegraphics[width=0.7\linewidth]{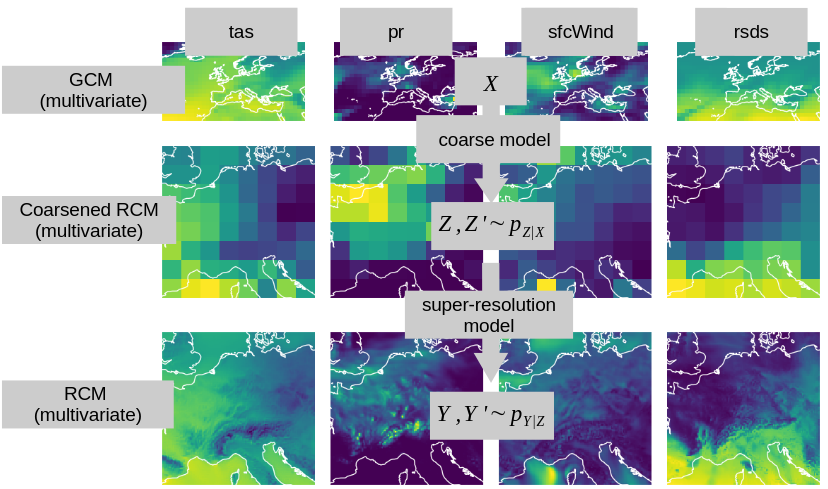}
	\caption{Downscaling via coarse correction for \ourmethod. We approximate the conditional distribution of RCM data $Y$ given GCM data $X$ with a two-step approach. In the second row, $Z$ represents RCM data manually coarsened through average pooling. The map learning the conditional $p_{Z|X}$ is called the \textit{coarse model}, and the map for the conditional $p_{Y|Z}$ the \textit{super-resolution model}. All $X, Z, Y$ include multiple climate variables and grid points, and represent daily data.}
	\label{fig:methods:2step}
\end{figure}

To solve these challenges, we split the problem into two parts, separating the correction on coarse scales and the super-resolution task \cite{Price2022IncreasingModels, Wan2024StatisticalModels}, as visualized in Fig.~\ref{fig:methods:2step}. In the first step, we train a generative model to learn the conditional distribution of coarsened RCM data $Z$ given the GCM inputs $X$, $p^\iota_{Z|X}$. For this, the RCM data is artificially coarsened through average pooling over patches of size $16\times16$, producing additional data $Z$ for each day. As the resolutions of $X$ ($\approx \SI{250}{km}$) and $Z$ ($\approx \SI{170}{km}$) are similar, predicting $Z$ from $X$ does not include any ``downscaling''. Instead, this step captures large-scale mismatches between GCM and RCM. The output of this step is relatively low-dimensional (dimension $8\times8$ per variable) and we call this the \textit{coarse model}.

For the second step, we learn the conditional distribution of high-resolution RCM data $Y$ given coarsened RCM data $Z$, $p^\iota_{Y|Z}$. This is a pure super-resolution problem, as $Z$ is the average-pooled version of $Y$, excluding the large-scale mismatches between $X$ and $Y$. Hence, we refer to the corresponding generative model as the \textit{super-resolution model}. The transformation from $Z$ to $Y$ is high-dimensional, but the training data is more homogeneous across GCM-RCM pairs: To assess this, we tested training a single joint model for all pairs without conditioning on the model identifier. We found that this primarily affects the loss for the \textit{coarse model}, while we only observe marginal performance decrease for the \textit{super-resolution model}. This suggests that different GCM-RCM pairs behave similarly in the super-resolution step.

We split the \textit{super-resolution model} again into several resolution steps, each one learning a simpler conditional distribution \cite{Shen2025ReverseDistributions}. In particular, each step increases resolution by a factor of two in both spatial dimensions, i.e.,~we model $p^\iota_{Z^{(k+1)}|Z^{(k)}}$, where $Z^{(k)}, Z^{(k+1)}$ are average-pooled versions of the target $Y$ with dimensions $2^k \times 2^k$ and $2^{k+1} \times 2^{k+1}$, respectively. The first step uses $k=3$, i.e. $Z = Z^{(3)}$, corresponding to the dimension of $Z$ ($8 \times 8$). Then, $k$ increases by 1 for each step, so that the last step uses $k=6$, as $Y$ has dimension $128 \times 128$.

In each step of our modeling chain, we train jointly on the entire dataset, but add predictor variables indicating the RCM considered (see App.~\ref{sec:appendix:model-architecture} for how these predictors are integrated in the model architectures).

All steps are trained independently. We tested more complex training procedures but found no improvement in performance. In addition, each step uses the variables defined for that stage, for example, the \textit{super-resolution model} only uses the coarsened RCM data $Z$ as input and not also the GCM data $X$. This simplification approximates the full distribution $p^\iota_{Y|X}$ accurately if $Z$ contains all the relevant information about the target $Y$ or, in other words, the map from $Z$ to $Y$ does not depend on $X$ anymore. This approximation is formalized through conditional independence and checked heuristically in App.~\ref{sec:appendix:marginal-cond-ind}. Empirically, we found no (e.g.,~precipitation) or at most 2 \% (e.g.,~temperature) decrease in performance due to this approximation.

\subsection{Modeling temporal consistency}
\label{sec:time-series}

\label{sec:setup:time-series}
In the previous paragraph, we considered predicting each RCM day independently, ignoring the temporal context. However, our architecture can naturally be extended to generating temporally consistent samples. For time series generation, we aim to learn the distribution of the RCM data $Y_t$ for a given day $t$, given the GCM input $X_t$ on the same day and the RCM data $Y_{t-1}$ from the previous day. We restrict temporal dependence to one-day lags. This follows the general approach in ML-based weather forecasting which typically models only low-order temporal dependence, e.g. with a two-step lag \cite{Lam2023LearningForecasting, Price2025ProbabilisticLearning}. More long-term memory effects are potentially still present in the GCM input.

We implement this autoregressive structure at each spatial resolution, following our stepwise modelling approach as above (Sec.~\ref{sec:setup:2step-marginal}). At the coarsest scale, the \textit{temporal coarse model} predicts $Z_t$ given $Z_{t-1}$ and $X_t$. This models autocorrelation on large spatial scales, e.g.~in pressure systems. During training, we obtain $Z_{t-1}$ via average pooling the RCM data $Y_{t-1}$. 
This extends naturally to further resolution steps: For each intermediate resolution $Z^{(k)}$, we predict $Z_t^{(k)}$ from the previous timestep at the same resolution, $Z_{t-1}^{(k)}$, and the same timestep at a coarser resolution, $Z^{(k-1)}_t$ (see App.~\ref{sec:appendix:model-architecture} for details on how the conditioning is integrated in the model architecture).

\begin{figure}
	\centering
	\includegraphics[width=0.7\linewidth]{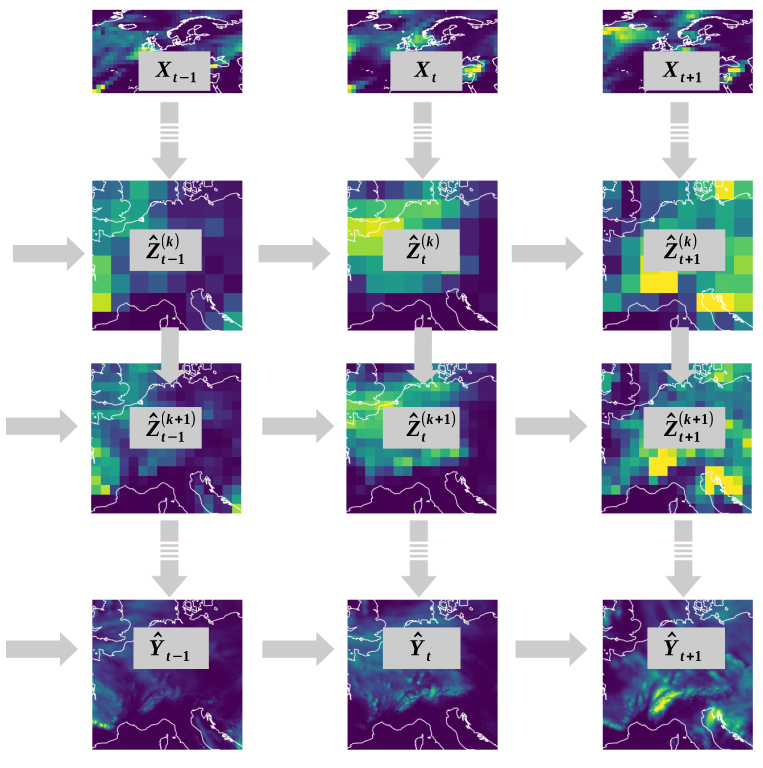}
	\caption{Time series generation with temporal consistency in \ourmethodtemp using an autoregressive roll-out at each spatial resolution. At the coarsest level, the model predicts $\hat{Z}_t$ using the sample from the previous day, $\hat{Z}_{t-1}$, and the GCM data from the same day, $X_t$. This time series is then progressively refined to higher resolutions, where each step again conditions on the previous time step at the same scale and the coarser prediction from the current day.
    }
	\label{fig:temporal}
\end{figure}

At inference time, we generate time series with an ``autoregressive rollout'' iteratively for each step: At the coarsest level, starting from a sample $\hat{Z}_{t-1}$ for time step $t-1$, we predict $\hat{Z}_t$ with the \textit{temporal coarse model} using $\hat{Z}_{t-1}$ and GCM input $X_t$. This time series is then iteratively refined through the super-resolution steps, each predicting a higher spatial resolution, until a time series for $Y_t$ is generated, as shown in Fig.~\ref{fig:temporal}.

We also tested approximating temporal consistency only on the coarse scale, i.e. using the \textit{temporal coarse model} and then applying the \textit{super-resolution model} independently on each day. We find that the results are very similar and this coarse approximation already models the temporal structure rather accurately in our setting.

\subsection{Generative model architecture}
\label{sec:methods:architecture}

We extend model architectures to our stochastic and high-dimensional setting by combining dense and sparse architectures. The \textit{coarse model} uses a dense multilayer perceptron (MLP), following \citeA{Shen2024Engression:Regression} (for details see App.~\ref{sec:appendix:model-architecture}).

Secondly, we introduce a sparse architecture which models each conditional distribution $p^\iota_{Z^{(k+1)}|Z^{(k)}}$. Our architecture is a middle-ground between flexibility of dense connections and efficiency of convolutional architectures. 
\begin{itemize}
    \item \textbf{Breaking translation invariance:} Unlike fully convolutional models, which are translation-invariant and apply the same weights everywhere, part of our model weights are specific to each location. This allows the model to capture unique local characteristics that vary by coordinate, such as topography or land-sea contrast. These local effects enter only through linear maps, which limits the risk of overfitting.
    \item \textbf{Structural efficiency:} We keep memory size manageable even for high resolutions. First, the value of $Z^{(k+1)}$ in one location is approximated to only depend on the values in a local neighborhood grid cells in $Z^{(k)}$, with learnable, linear, and location-specific weights. This restriction reduces connections compared to dense networks. Second, we share non-linear transformations across all locations. 
    \item \textbf{Training performance:} Finally, our design reduces training times compared to deep fully convolutional networks substantially (as discussed in Sec.~\ref{sec:results:computational-cost}). 
\end{itemize}
 
\begin{figure}
	\centering
	\includegraphics[width=0.9\linewidth]{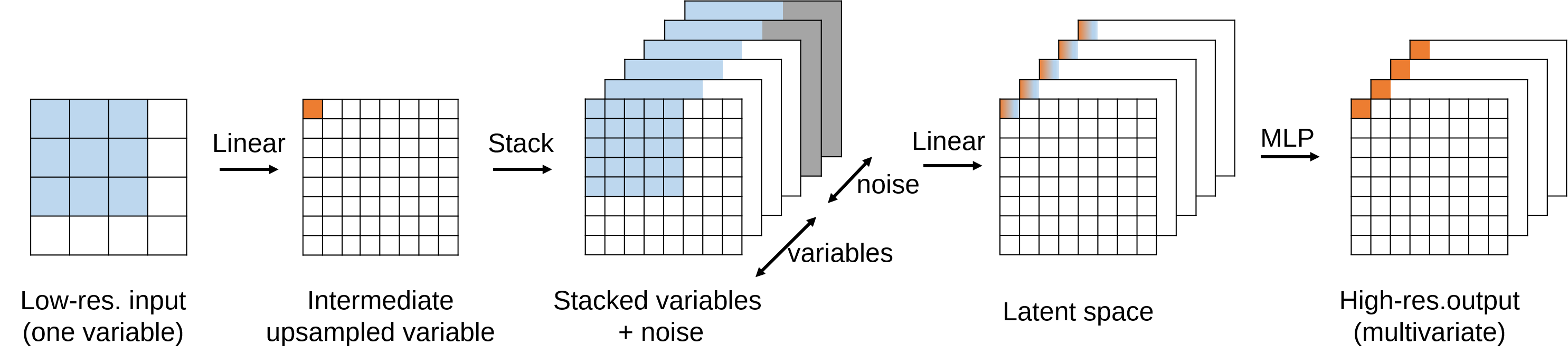}
	\caption{Sparse local stochastic layers from \ourmethod's \textit{super-resolution model}, refining the low-resolution input by a factor of 2 in each dimension. As an example, we demonstrate modeling the distribution in an example pixel of interest (top left corner, orange); the same procedure is applied to all pixels. For each intermediate map (small arrows), light-blue-shaded pixels serve as inputs and orange pixels as the targets. First, a deterministic upsampling step processes each target variable (tas, pr, sfcWind, rsds) separately, linearly interpolating nearest pixels in the low-resolution input with learnable weights. Second, the intermediate upsampled variables are stacked and concatenated with noise channels. Next, each pixel again interpolates linearly from its nearest neighbors with learnable location-specific weights (this time using all variables and the noise as inputs). Finally, an MLP is applied to each pixel independently, using the same weights at each location, yielding the high-resolution output.}
	\label{fig:methods:sparselayers}
\end{figure}

For our generative modeling task, the ML model must generate multiple outputs for one single GCM input $X$. To achieve this, we integrate Gaussian noise into our model as a source of stochasticity. We do this separately for each subpart of our model (\textit{coarse model} and each component of the \textit{super-resolution model}). For the dense MLP architectures, the noise is concatenated to hidden units in each hidden layer. This facilitates learning complex representations of the random inputs and yields spatially correlated variability in the samples, in line with independently developed approaches such as \citeA{Alet2025SkillfulMarginals}, \citeA{Lang2024AIFS-CRPS:Score}. For the sparse layers, the stochastic super-resolution proceeds in two stages (see Fig.~\ref{fig:methods:sparselayers}). A first deterministic upsampling step is done separately for each target variable, and interpolates each high-resolution pixel in $Z^{(k+1)}$ linearly from its nearest low-resolution neighbors in $Z^{(k)}$ with learnable location-specific weights. The second step performs a stochastic refinement. The intermediate upsampled variables are stacked and concatenated with several channels of Gaussian noise in each pixel; this concatenation is used as input to another local aggregation. Each pixel again has its distinct set of linear weights, applied to inputs from a local neighborhood, using both the upsampled variables as well as the noise channels. This step enables generating spatially correlated variability and capturing multivariate dependencies. Finally, a shared MLP is used for all pixels (weight-sharing across space), where each pixel is processed independently of others, but all dimensions are considered jointly. More details on all model architectures are given in App.~\ref{sec:appendix:model-architecture}.

\subsection{Generative Models with Proper Scoring Rules}

\subsubsection{Connection to forecasts and forecast evaluation}
\label{sec:methods:generative-vs-forecasts}

Our method, along with our benchmarks and evaluation techniques, is inspired by the correspondence between generative modeling and forecast evaluation. Consider the example of weather forecasts. Given the state of the atmosphere at a certain time, weather forecasts aim to predict the weather a few hours ahead. However, rather than issuing a single best estimate, they also aim to quantify uncertainties by predicting the full probability distribution of possible future weather for the next time step \cite{Gneiting2005WeatherMethods}. Typically, this is done by running an ensemble of simulations from a numerical weather model. Afterwards, a post-processing step corrects biases and variance discrepancies of the ensemble \cite{Vannitsem2021StatisticalWorld}. This post-processing is trained based on previous forecasts and corresponding observations. The goal is to produce a probability distribution that should be consistent with the target, i.e.,~the observed data.

A key challenge in evaluating such weather forecasts is that, at each time step, only a single ground truth from observations is available. Thus, tools are needed to compare a single observed value with the full predicted distribution. Averaging over multiple forecast-observation pairs then allows evaluating the quality of the distributional forecasts \cite{Gneiting2007ProbabilisticSharpness}.

This setup is analogous to generative modeling: for each GCM input, the generative ML model outputs an ensemble of predictions (multiple samples). The model's samples serve as an approximation to the underlying (continuous) probability distribution of possible outcomes. We wish to evaluate these predictions relative to the target data from the RCM, where for each RCM we only have a single realization available. This comparison between forecast evaluation and generative modeling is crucial both during training of the generative model, as well as in its evaluation.

%\mytodo{I should at least mention the model architecture and reference \ref{sec:appendix:model-architecture} somewhere.}

\subsubsection{Energy score loss}
\label{sec:methods:energy-score-loss}

For simplicity, we first consider the unconditional generation of a target distribution $p_Y$ and neglect GCM predictors $X$. We denote the distribution from the ML model by $q_Y$. Recall that we do not have explicit access to the ML's output distributions, but instead have samples for $p_Y$ and can generate samples for $q_Y$ via the generative ML model. We denote true data by $Y$ and two random predictions from the model by $\hat{Y}, \hat{Y}'$.

Our loss minimizes the error in the predictions, $\hat{Y}$, while ensuring that the estimated uncertainty in the samples is consistent with the prediction error (e.g., is sufficiently wide and not overconfident). To achieve this, the loss balances predictive accuracy and variability. The first term, $E_{Y \sim p_Y, \hat{Y} \sim q_Y}(\lVert Y - \hat{Y} \rVert)$, describes the prediction error. Minimizing only this term would result in deterministic predictions with $\hat{Y} = \hat{Y}'$. For example, in the one-dimensional case, $\hat{Y} = \hat{Y}' = \text{median}(p_Y)$. A generative model, however, should sample from the full distribution. Hence, a second term is introduced, $E_{\hat{Y}, \hat{Y}' \sim q_Y}(\lVert \hat{Y} - \hat{Y}' \rVert)$, to account for diversity in samples.

The total loss then presents a trade-off between prediction error (first term) and the scaled variability of the samples (second term)
\begin{align}
	\text{Loss}_{\text{uncond}} = E_{Y \sim p_Y, \hat{Y} \sim q_Y}(\lVert \hat{Y} - Y \rVert) - \frac{1}{2} E_{\hat{Y}, \hat{Y}' \sim q_Y}(\lVert \hat{Y} - \hat{Y}' \rVert).
	\label{eq:exp-loss-uncond}
\end{align}
Here, the subscript ``uncond'' refers to the fact that the loss does not take into account any conditioning yet. The norms in this score are Euclidean norms that can be applied directly to high-dimensional targets $Y$. Both $\hat{Y}$ and $Y$ typically are full spatial fields for multiple climate variables at once. When combining multiple variables with different physical units, a data transformation for standardization is crucial, as we introduce in Sec.~\ref{sec:methods:data-transformation}.

Our loss can uniquely identify the correct distribution: $\text{Loss}_{\text{uncond}}$ is minimal if and only if the modeled distribution $q_Y$ follow the target distribution $p_Y$ ($q_Y = p_Y$). More precisely, it holds that, for all $p_Y, q_Y$
\begin{align*}
	&E_{Y \sim p_Y, \hat{Y} \sim q_Y}(\lVert Y - \hat{Y} \rVert)- \frac{1}{2} E_{\hat{Y}, \hat{Y}' \sim q_Y}(\lVert \hat{Y} - \hat{Y}' \rVert) \\
	&\geq E_{Y \sim p_Y, \hat{Y} \sim p_Y}(\lVert Y - \hat{Y} \rVert) - \frac{1}{2} E_{\hat{Y}, \hat{Y}' \sim p_Y}(\lVert \hat{Y} - \hat{Y}' \rVert),
\end{align*}
and equality holds if and only if $p_Y = q_Y$ \cite{Rizzo2016EnergyDistance, SzekelyE-Statistics:Distances}. This implies that minimizing this loss ensures that the generated samples follow the target distribution.

If $p_Y = q_Y$, both terms of the expected energy score are indeed equal, i.e.,~
\begin{align*}
	E_{Y \sim p_Y, \hat{Y} \sim q_Y}(\lVert \hat{Y} - Y \rVert) = E_{\hat{Y}, \hat{Y}' \sim q_Y}(\lVert \hat{Y} - \hat{Y}' \rVert).
\end{align*}
However, the contrary is not true, i.e.,~equal scores do not imply that the score is minimal. In other words, equality is a necessary but not sufficient criterion for reaching the minimum.

This loss is related to the energy score in forecast evaluation. For a single target data point $y$ and a predicted distribution $q_Y$, the energy score is defined as
\begin{align*}
	S(q_Y, y) \coloneqq \underbrace{E_{\hat{Y} \sim q_Y}(\lVert y - \hat{Y} \rVert)}_{\text{ES}_\text{pred}} - \frac{1}{2} \underbrace{E_{\hat{Y}, \hat{Y}' \sim q_Y}(\lVert \hat{Y} - \hat{Y}' \rVert)}_{\text{ES}_\text{var}}.
\end{align*}
This is a multivariate generalisation of the Continuous Ranked Probability Score (CRPS) \cite{Gneiting2007StrictlyEstimation, Szekely2013EnergyDistances} frequently used in numerical weather prediction. For one-dimensional $Y$, the energy score and CRPS are equivalent. The CRPS is typically applied per variable and per grid point, whereas we use the energy score jointly over all locations (and, unless stated otherwise, also across all target variables), capturing dependencies between them. We call the prediction error (first term) $\text{ES}_\text{pred}$ and the variability (second term) $\text{ES}_\text{var}$.

The above loss function $\text{Loss}_{\text{uncond}}$ corresponds to the expected energy score over multiple target data points, for example several days in the RCM: $\text{Loss}_{\text{uncond}} = E_{Y \sim p_Y}(S(q_Y, Y))$.

To account for the GCM inputs, our loss is adjusted for the conditional case by averaging over multiple GCM inputs $X$ and drawing samples from the generative model, which is conditioned on $x$ and the index for the GCM-RCM pair (see Sec.~\ref{sec:methods:gen-model-energy-score:loss-cond} for more details).
$\text{ES}_\text{pred}$ and $\text{ES}_\text{var}$ are defined as averages over GCM inputs $X$ accordingly.

\section{Data}
\label{sec:methods:data_processing}

\subsection{Data preprocessing}

\begin{table}[h!]
\caption{Regional climate model and global circulation model pairs (EURO-CORDEX) used in our study.}
\centering
\begin{tabular}{lcccc}
\toprule
\textbf{GCM $\backslash$ RCM} & ALADIN63 & CCLM4-8-17 & REMO2015 & RegCM4-6 \\
\midrule
CNRM-CM5   & $\checkmark$ & $\checkmark$ &  & $\checkmark$ \\
MIROC5     &  & $\checkmark$ & $\checkmark$ &  \\
MPI-ESM-LR & $\checkmark$ & $\checkmark$ &  & $\checkmark$ \\
\bottomrule
\end{tabular}
\label{tab:gcm_rcm_matrix}
\end{table}

For training and evaluation, we use data from eight GCM-RCM pairs from the EURO-CORDEX framework \cite{Jacob2014EURO-CORDEX:Research}, see Table~\ref{tab:gcm_rcm_matrix}. We consider daily 2m temperature (\textit{tas}), total precipitation (\textit{pr}), surface wind (\textit{sfcWind}) and surface downwelling shortwave (solar) radiation (\textit{rsds}) from both driving GCMs and corresponding downscaled RCMs. In addition, we add sea level pressure (\textit{psl}) from the GCMs as a predictor because it provides valuable information about the large scale weather pattern. The selection of climate models was based on the availability of daily surface wind data, which not all models publish. We chose a set of driving GCMs for which multiple corresponding downscaled datasets from different RCMs were available. We use daily data for the years 1971-2099, from both the historical and the RCP8.5 experiment.

We also tested extrapolation to GCMs unseen during training: For this, we use ALADIN63 -- HadGEM2-ES and CCLM4-8-17 -- CanESM2 in addition to the eight training pairs (see Sec.~\ref{sec:discussion}).

\subsubsection{Targets}

The targets are the RCM data in Central Europe for all four variables mentioned above. The domain spans $0^\circ E$ to $18.9^\circ E$ and $42^\circ N$ to $54.8^\circ N$ and includes land as well as sea areas and the Alps. For most of the RCMs output, the data are provided on the EUR-11 grid. Otherwise, we re-gridded them with first-order conservative remapping.

Different RCMs write data in different units. We rescaled temperature to $^{\circ}$Celsius, precipitation to mm/day, solar radiation to $\mathrm{W} / \mathrm{m}^2$ and wind to m/s.

When both the driving GCM and the corresponding RCM dataset include the 29th of February in leap years, we left the data as is. However, when the GCM uses a no-leap calendar with 365 days in each year and the RCM has a calendar including leap days, we removed all 29th of February in the RCM data. After preprocessing, driving GCM and corresponding RCM dataset always have the same length.

\subsubsection{Predictors}

% note on preprocessing - didn't find it for CMIP5, but found in CMIP6-ng gitlab that it should be remapcon2
The inputs for the ML model are the GCM data from the driving GCM. The GCMs were regridded via second-order conservative remapping to a regular 2.5 degree grid. We select a domain which is larger than the target domain and covers Europe, spanning $40^\circ W$ to $50^\circ E$ in longitude and $25^\circ N$ to $75^\circ N$ in latitude.

We tested the addition of further variables as predictors, e.g.~humidity, wind and geopotential height at 500 and 850 hPa, but found no improvement in performance on the test loss.

We add additional predictors for seasonality information. The day of the year is passed as an integer together with its sine and cosine to facilitate learning seasonal information. We do not input the year explicitly in order to allow for extrapolation to years that have not been seen in the training data.

\subsubsection{Train-test split}

The total runs span the years 1971-2099, and we use 1971-2029 and 2040-2089 as training data. We pool the training data across all GCM-RCM pairs and then select a random subset of 10 \% of this data as the validation set. The two test data sets are 2030-2039 and 2090-2099. The first test data set represents the ``interpolation'' test case, whereas the second is the ``extrapolation'' case with test data outside the time period covered in the training data. This yields approximately 320k data points for training and around 30k points for each test set.

\subsection{Data transformation for training}
\label{sec:methods:data-transformation}

To combine multiple variables in a loss function for model training, the data must be transformed to avoid issues arising from different scales or units. For the target data from the RCM, we transform each variable and location to a normal distribution. 

First, we approximate the empirical distribution function (ECDF), $\hat{F}$, on a subsample of 10 \% of the training data. The RCM's target data from all GCM-RCM pairs are pooled, but we fit the ECDF separately for each variable and location. Applying the ECDF to the training data, again in a univariate manner, yields probabilities $\hat{F}(Y)$ that are uniformly distributed on $(0,1)$. In the case of precipitation, however, the ECDF introduces a point mass at zero rainfall and the uniformity only holds above this threshold. Second, we map to a normal distribution by computing $\Phi^{-1}(\hat{F}(Y))$, where $\Phi$ is the CDF of the standard normal distribution. This maps the probabilities $\hat{F}(Y)$ from $(0,1)$ to $\mathbb{R}$. This transformation ensures that all locations and variables follow the same distribution, allowing joint application of the loss to all variables. 

%The motivation for the transformation to normal scores instead of probabilities is that directly applying the \condenscoreloss on the scale of probabilities would penalize mistakes in high quantiles too little. For example, assume the model predicts $\hat{p} = 0.5$, where the truth is $p=0.55$ and $\hat{p} = 0.9$ where $p=0.95$. The errors are the same on the probability-scale, but the difference between the 55 \% and 50 \% quantiles is usually much smaller than the difference between the 95 \% and 90 \% quantiles in the original data. Without the normal transformation, errors in high quantiles would have disproportionally influence on the loss.

The ML model's outputs, $\hat{Y}$, lie in $\mathbb{R}$. We first apply the inverse transformation $\hat{F}_Y^{-1}(\Phi(\hat{Y}))$, where $\Phi(\hat{Y})$ maps the real-valued output $\hat{Y}$ to probabilities in $(0,1)$. Then, the inverse ECDF $\hat{F}_Y^{-1}$ converts the probabilities back to the original data scale. These transformations are also considered separately for each grid point and variable.

For the input data, the precise transformation is less crucial, so we apply a simpler and computationally more efficient normalization: each variable is standardized by subtracting the mean and standard deviation. Here, mean and SD were computed using the entire spatial field and training data, again from all GCMs.

\section{Methods for evaluation}
\subsection{Benchmarks}
\label{sec:benchmarks}

\subsubsection{Deterministic neural network}
\label{sec:det-nn}

The most simple deterministic benchmark is a deterministic neural network (\textit{NN-det} in the following) with an MSE loss. As input, we use five GCM variables, the four target variables plus sea level pressure. For simplicity, we have separate ML models for each target variable. The model architecture is the same as \ourmethod's \textit{coarse model}, using a dense MLP but without additional noise input. More details are given in App.~\ref{sec:appendix:details-benchmarks}.

\subsubsection{EasyUQ}

EasyUQ is a tool to convert predictions from a deterministic statistical model into probabilistic ones \cite{Walz2024EasyOutput}. We use it to obtain predicted probability distributions from the above simple NN-det (Sec.~\ref{sec:det-nn}). We introduce EasyUQ because, based on statistical theory, it is expected to issue good probabilistic predictions for each grid point and variable separately, as a baseline comparison for the performance of \ourmethod on at the level of individual locations. However, EasyUQ breaks spatial and inter-variable dependencies in the output. 

EasyUQ it is applied as a post-processing step after training the NN-det, independently for each GCM-RCM pair. The NN-det's predictions, denoted by $\hat{\mu}$, approximate the conditional mean $\mu \coloneqq E[Y|X]$. Given training data consisting of pairs of NN-det's predictions $\hat{\mu}$ and the RCM target data $Y$, EasyUQ quantifies the uncertainty in the NN-det's predictions. It returns full probability distributions.
More details on theoretical properties and the implementation are given in App.~\ref{sec:appendix:details-benchmarks}.

\subsubsection{Analogues}

Analogue-based downscaling \cite{Boe2023AValue} is a statistical-dynamical downscaling method which resamples the existing RCM data. It is based on a nearest neighbor search in the GCM data, and stochastic downscaling is approximated by selecting five similar GCM days and using their corresponding RCM days as probabilistic predictions. The main strength of the analogue approach is that, by construction, it yields samples with the correct spatial pattern, as they are all drawn from the existing RCM data. However, it only reproduces the training data and can never create truly new random fields, and the predicted conditional distributions are only discrete. Finally, due to a lack of perfect ``near neighbors'', the analogue benchmark is expected to have a larger prediction error between target RCM data and predicted samples than the true RCM's conditional distribution. Details on our algorithm and potential limitations are given in Sec.~\ref{sec:appendix:details-analogues}.

\subsubsection{Generative Adversarial Network}

Generative Adversarial Networks (GANs) have successfully been used to downscale precipitation fields \cite{Harris2022AForecasts, Leinonen2020, Price2022IncreasingModels}. As an additional benchmark, we consider the approach presented in \citeA{Harris2022AForecasts}, which is a Wasserstein GAN with gradient norm penalty (\textit{WGAN-GP}). In the generator objective function, it includes an additional content loss term, more specifically, the mean-squared error of the mean prediction (over eight generated samples per input field) and the ground truth high-resolution output. We adjust \citeA{Harris2022AForecasts}\footnote{Code: \url{https://github.com/ljharris23/public-downscaling-cgan}} to our multivariate setting. As for \textit{EnScale}, we use five input fields (\textit{tas}, \textit{pr}, \textit{sfcWind}, \textit{rsds}, \textit{psl}) as the conditioning inputs to the WGAN-GP. Additionally, we include the one-hot-encoded index of the considered GCM-RCM pair and concatenate it to the five input fields. 
The outputs are the four target fields (\textit{tas}, \textit{pr}, \textit{sfcWind}, \textit{rsds}). Similar to \citeA{Harris2022AForecasts}, we scale the precipitation and surface wind values logarithmically with $\log(1+x)$. In addition, we use min-max normalization to ensure that all variables have values on similar scales. For implementation reasons, we confine the input domain of the GCM inputs to $-11.25^\circ E$ to $26.25^\circ E$ and $31.25^\circ N$ to $68.75^\circ N$, i.e.,~a smaller extent than covered in \textit{EnScale}.
We use the same hyperparameters as in \cite{Harris2022AForecasts} with a larger batch size of 16 and train until convergence, reached after 2.4 million training samples.

\subsubsection{Corrective Diffusion Model}
\label{sec:methods:corrdiff}

As another state-of-the-art ML downscaling approach, we consider \textit{CorrDiff}, which was shown to perform skillfully in multivariate downscaling \cite{Mardani2025ResidualDownscaling}. The model
consists of a UNet to predict the (deterministic) conditional mean $E[Y|X]$ and a diffusion model to predict the residual variation around the mean. We adopt the CorrDiff implementation provided in NVIDIA's \textit{PhysicsNeMo} library\footnote{Code: \url{https://github.com/NVIDIA/physicsnemo/tree/main/examples/weather/corrdiff}, commit 469d6aa09354ec91ebe1f388a8247c96e502e9c8} for the same setting as in the previous GAN experiment, i.e.,~five input fields concatenated with the one-hot-econded GCM-RCM pair indices, four output fields, and the confined GCM-input-domain. 

As in \citeA{Mardani2025ResidualDownscaling}, we standardize all variables to z-scores and use the same hyperparameter settings. In particular, we use a base embedding size of 128 for the UNet, channels are multiplied with the factors $[ 1,2,2,2,2]$, and the attention resolution is defined as 8 (as our input is of size $128\times 128$). We use a smaller batch size of 32, and train until convergence: first, the UNet for 4 million training samples and, afterwards, the diffusion model for 26 million training samples.

In it's original implementation, CorrDiff generates each ensemble member by drawing a single random noise realization that is shared across all time points. This construction implicitly imposes temporal dependence in the generated time series, but variability across realizations is limited. For this reason, we also tested an adapted version which we call \textit{CorrDiff-s}. In this variant, we randomize the sampling scheme: For each day, noise is drawn independently of previous days and of other ensemble members. This version represents the variance of samples more faithfully. However, it removes temporal autocorrelation, leading to negative errors in autocorrelations, similar to the other benchmarks.

%\newcorrdiff{In it's original implementation, CorrDiff generates each ensemble member by drawing a single random noise realization that is shared across all time points. This construction implicitly imposes temporal dependence in the generated time series, but variability across realizations is limited. For this reason, we also tested an adapted version which we call \textit{CorrDiff-s}. In this variant, we randomize the sampling scheme: For each day, noise is drawn independently of previous days and of other ensemble members. This version represents the variance of samples more faithfully. However, it removes temporal autocorrelation, leading to negative errors in autocorrelations, similar to the other benchmarks (not shown).}

\subsection{Validation}

To design our validation metrics, we again leverage the connection between generative models and forecast evaluation (Sec.~\ref{sec:methods:generative-vs-forecasts}). Two important properties of the generated samples by the ML model are \textit{calibration} and \textit{sharpness} \cite{Gneiting2007ProbabilisticSharpness}. 

Calibration is also called ``reliability'' as it checks whether the model's samples have the right ``spread'' and correctly capture all possible outcomes compared to the target RCM data. In other words, a well-calibrated model should not systematically over- or under-represent typical or extreme values. However, \textit{calibrated} samples are not necessarily informative: for example, if the ML model ignored the GCM input $X$ and generated samples from the climatology or from the marginal distribution across all days of the year, independently of $X$, the samples would be calibrated. But the samples are not very useful, as they carry little information about the particular day of interest. 

Samples with small variance are called \textit{sharp}. Sharpness does not imply that the samples are correct: a narrow range that is always far away from the truth would also be sharp. Calibrated samples balance prediction error and variability and often achieve $\text{ES}_\text{pred} = \text{ES}_\text{var}$, but only samples that are also sharp (small $\text{ES}_\text{var}$) are optimal (see our explanation of the loss, Sec.~\ref{sec:methods:energy-score-loss}).

We assess calibration and sharpness for the samples of our model in general and with a particular focus on extremes. In addition, we consider and temporal structure, as well as variable dependencies.

\subsubsection{Energy score and CRPS}
\label{sec:methods:metrics}

To evaluate the ML model, we use proper scoring rules, in particular the CRPS and the energy score. These assess calibration and sharpness jointly.

We consider the multivariate \condenscoreloss for the full spatial field for each variable separately (see Sec.~\ref{sec:methods:gen-model-energy-score:loss-cond}), i.e., treating all grid cells as a multivariate vector. In addition, we calculate CRPS values, which coincide with the energy score in the univariate case. We calculate the CRPS for each variable and in each location and then average it across the spatial field.

\subsubsection{Spatial structure}

To check the spatial structure, we combine the locations-wise CRPS calculation with a pre-pooling operation. We first apply max-pooling over patches of the data and then compute CRPS values for the pooled patches. Results are shown for a kernel size of 10. The conclusions are consistent across different kernel sizes.

\label{sec:methods:psd}
In addition, we compute power spectral densities (PSD). We follow \citeA{Harris2022AForecasts} and consider the radially-average PSD, calculated with \textit{pysteps}. Following \citeA{Harris2022AForecasts}, we compare the sampled PSDs with the RCM's PSDs via the log spectral distance, called RALSD:

\begin{equation}
\label{eq:methods:ralsd}
    \text{RALSD} = \sqrt{\frac{1}{N} \sum_{i=1}^N \Bigl( 10 \text{log}_{10} \frac{\overline{P}_{\text{RCM, i}}}{\overline{P}_{\text{gen, i}}} \Bigr) ^2}
\end{equation}
Here, $\overline{P}_{\text{RCM, i}}$ and $\overline{P}_{\text{gen, i}}$ are the power spectra for the RCM data and one generated field at one wavelength $i$, where the 2D spectra were averaged radially. $N$ is the number of wavelengths considered. The scaling by factor 10 is chosen such that the resulting units are in decibel. 

We compute PSDs separately for each test day and consider two options for averaging PSDs and spectral distances: (i) For the RALSD-metric, we compute the RALSD on daily PSDs as in Eq. \eqref{eq:methods:ralsd}. Finally, we average over all days in the test set, as in \citeA{Harris2022AForecasts}. (ii) Alternatively, we average radially-averaged PSDs across all days, plot these and compute the log spectral density of the averaged spectra. This distance emphasizes systematic errors in spatial structure, and we refer to this metric as RALSD-avg. 

PSDs $\overline{P}_{\text{gen, i}}$ for different random samples can vary substantially even on the same day, due to the large variability in the conditional $p_{Y|X}$. Thus, RALSD is also sensitive to deviations from individual samples from the target, whereas RALSD-avg focuses more directly on systematic differences in the average spatial structure.

\subsubsection{Calibration}
\label{sec:methods:validation:rh}

We use rank histograms (RHs) to evaluate the calibration of our model's samples. RHs consider ranks of the target RCM data $y$ within an ordered set of samples, e.g. $y_1 < ... < y_9$. These ranks are visualized in a histogram. If the samples are calibrated, all ranks from 1 to 10 are equally likely, implying a flat RH. This simultaneously checks if the samples are unbiased compared to the truth and if they have the right variability, i.e.,~are neither over- nor underconfident. Deviations from flatness indicate which type of violation occurs: for example, a U-shaped histogram suggests overconfident samples (too little variance), while a dome-shaped histogram indicates underconfident samples (too much variance).

RHs can only consider one-dimensional targets and samples. Thus, we reduce the high-dimensional spatial field to a one-dimensional quantity by taking the mean or maximum before plotting. Additionally, we compute rank histograms at each location and quantify miscalibration using the miscalibration measure (MCB). The MCB is calculated as the sum of the absolute deviations between the generated RH and a uniform (flat) rank histogram. Thus, an MCB of zero indicates a perfectly calibrated model, and larger values indicate increasing miscalibration. By averaging the MCB across all locations, we obtain a general measure of calibration quality. However, the MCB can not assess which type of violation occurs, e.g.~whether the samples are over- or underdispersed.

Alternatively, ranks could be pooled across spatial locations. However, opposite biases at different locations could cancel out in the pooled histogram, thus making the result difficult to interpret. Hence, we instead compute the MCB for each location individually, before averaging this across locations.

\subsubsection{Extremes}

\label{sec:methods:validation:extremes}
We consider extreme quantiles across all days in a season. For example, we compute the 99 \% - quantile for all samples on summer days in the test set. These quantiles are computed in each location separately. Afterwards, the absolute errors of the sampled vs. true RCM's quantiles are averaged across all locations.

Additionally, we analyze behaviour in the extremes of the predicted conditional distribution with a miscalibration measure, focussing on the calibration of the most extreme samples. We investigate calibration of high extremes in a single location as follows: for each day, we consider 9 samples of the generative models, $y_1,...,y_9$, select the highest value, say $y_9$, and compare it to the target RCM data $y$. If the samples capture extremes correctly, the target RCM data $y$ should lie above $y_9$ in 10~\% of the days. Hence, we calculate the ratio of occurrences of the case $y_9 < y$ (out of all days) and then consider the absolute deviation from this ratio to 0.1. This is a miscalibration measure similar to the general MCB metric from Sec.~\ref{sec:methods:validation:rh}, but now only considers the last bin of the RH. We average the extremal MCB scores across all locations, and proceed likewise for the low extremes. 

\subsubsection{Variable dependencies}
\label{sec:methods:validation:variable-dep}
We evaluate pairwise dependencies between several pairs of the four target variables. Firstly, we consider Pearson correlation coefficients in each location for both the RCM as well as the samples. We report absolute errors of the correlation coefficients, averaged across all locations.

In addition, we consider a novel metric of calibration: we compute RHs and MCB metrics (see Sec.~\ref{sec:methods:validation:rh}) for the difference of two variables in the samples. However, as directly comparing variables with different physical units is not meaningful, we first apply a data transformation to standardize them. Specifically, for each variable and each dataset (RCM or sampled), we transform the data to a uniform distribution via the ECDFs (similar to the above data transformation). This ensures that the marginal distributions for each variable are identical, eliminating differences to units. In addition, it removes mismatches between RCM and sampled data in individual variable distributions, such that the remaining discrepancies arise only from different dependence structures between the variables. After transformation, we calculate the differences between two variables. We compute RHs and MCBs for each grid point separately and then averaged across locations.

\subsubsection{Temporal structure}

We evaluate the temporal structure of the samples by computing auto-correlations. For each variable and each location, we calculate temporal autocorrelation functions (ACFs) for the generated time series as well as the RCM target for lags up to 20 days. We consider both raw as well as high-pass filtered time series. For the latter, we subtract a running mean over 30 days first and calculate ACFs on the filtered time series. We report both signed as well as absolute errors of the ACFs in the samples compared to the RCM target.

\subsection{Contributions to variability in \ourmethod}

We aim to better understand the variability \ourmethod's samples and present decompositions of its conditional variance $\var(Y|X)$, with $Y$ being the high-resolution RCM data and $X$ the GCM input, as above. As discussed (Sec.~\ref{sec:setup:downscaling-generative}), there is a wide range of possible realizations of high-resolution fields given a particular GCM input. \ourmethod allows us to quantify this variability. In this section, all variances and expected values refer to those of \ourmethod's samples, assuming already that \ourmethod emulates the true RCM's distributions accurately. In addition, $\var$ and $E$ denote quantities for the pooled distribution across multiple GCM-RCM pairs, $\iota$ denotes the index for an RCM, quantities with a superscript $\iota$, such as $\var^\iota$ and $E^\iota$, refer to distributions by an individual RCM.

\subsubsection{Coarse-scale vs. local-scale variance}
\label{sec:methods:variability:coarse-super}

We aim to decompose the internal variability of \ourmethod's full GCM-RCM path into the variability on coarse and local scales. Consider an input $X$ from one of the GCMs. Recall that $Z$ denotes the coarsened RCM data. For the input $X$, we split the variance of \ourmethod, $\var^\iota(Y|X)$, into contributions from (i) variance of the \textit{coarse model}, $\var^\iota(Z|X)$, which we call coarse-scale variability, and (ii) variance of the \textit{super-resolution model}, $\var^\iota(Y|Z)$, which we call local-scale variability (see Sec.~\ref{sec:setup:2step-marginal}). As $Z$ and $Y$ have different dimensions and physical meanings, $\var^\iota(Z|X)$ and $\var^\iota(Y|Z)$ are not directly comparable. Instead, we can formally decompose the conditional variance $\var^\iota(Y|X)$ as follows:
\begin{equation}
\label{eq:variability:coarse-local}
	\var^\iota(Y|X) = \underbrace{E^\iota[\var^\iota(Y|Z) | X]}_{\text{local-scale variance}} + \underbrace{\var^\iota(E^\iota[Y|Z] | X)}_{\text{coarse-scale variance}},
\end{equation}
where the first term is the variance of $Y|Z$, averaged over all possible $Z$ (conditional on $X$). The second term describes the variance of the conditional mean $E^\iota[Y|Z]$, again for different possible $Z$. This is conceptually very similar to $\var^\iota(Z|X)$, except it is ``evaluated in $Y$''. A proof is given in App.~\ref{sec:appendix:var-coarse-super}. This decomposition holds separately in each grid point and for each variable. We average each of the terms over all grid points as well as over all $X$ in the test set (i.e. over all days, but also over all GCM runs in our dataset).

We calculate the quantities above as follows: for a given GCM input $X$, we generate 9 realizations of $Z$ with the \textit{coarse model}. For each $Z$, we again generate 9 realizations of $Y$ with the \textit{super-resolution model}, and compute $\var^\iota(Y|Z)$ and $\var^\iota(E^\iota[Y|Z]|X)$ for each grid point. Afterwards, we average across $X$, grid points and GCM-RCM pairs. To restrict memory usage, the calculation is only done for the first half of the test set.

\subsubsection{Internal variability vs. model differences}
\label{sec:methods:variability:groups}

\ourmethod allows us to compare the magnitude of model differences between RCMs with the intrinsic variability in individual RCMs. As an example, consider a fixed GCM run that is dynamically downscaled by two RCMs. Assume we train \ourmethod to emulate both RCMs, such that it can generate samples from each of them, yielding a pooled dataset. Using these samples, we quantify contributions to the variance in this pooled data, $\var(Y|X)$, from (i) internal variability, present in each RCM, and (ii) structural differences between the two RCM. 

Formally, let the first RCM correspond to index $\iota=1$ and the second RCM to $\iota=2$. For example, $\var^{\iota =1}(Y | X)$ denotes the variance of the first RCM. We decompose $\var(Y|X)$ as follows:
\begin{align}
\label{eq:variability:internal-model-diff}
	\var(Y|X) = & \underbrace{\frac{1}{2} \big (\var^{\iota =1}(Y | X) + \var^{\iota =2}(Y | X)}_{{\text{Internal variability in individual RCMs}}} \big ) \\
    &+ \underbrace{\frac{1}{2} \big ( (E^{\iota =1}[Y | X] - E[Y|X])^2 + (E^{\iota =2}[Y | X] - E[Y|X])^2}_{\text{Model differences between RCMs}} \big ).
\end{align}
Thus, the total variance can be written as an average of the variances of the individual RCMs (internal variability) plus the squared deviations of the means of each RCM from the mean of the pooled data (model differences). This decomposition analogous to classical ANOVA. For a proof, see \citeA{Casella2024StatisticalInference}. As above, we consider this decomposition separately per grid point and per variable, and average over grid points and predictors $X$. In this case, we only report results for one driving GCM (and two corresponding RCM datasets).

\newpage

\section{Evaluating the performance of \ourmethod}

\subsection{Visual evaluation of samples}
\label{sec:results:example-samples}
\begin{figure}
    \centering
    \includegraphics[width=0.75\linewidth]{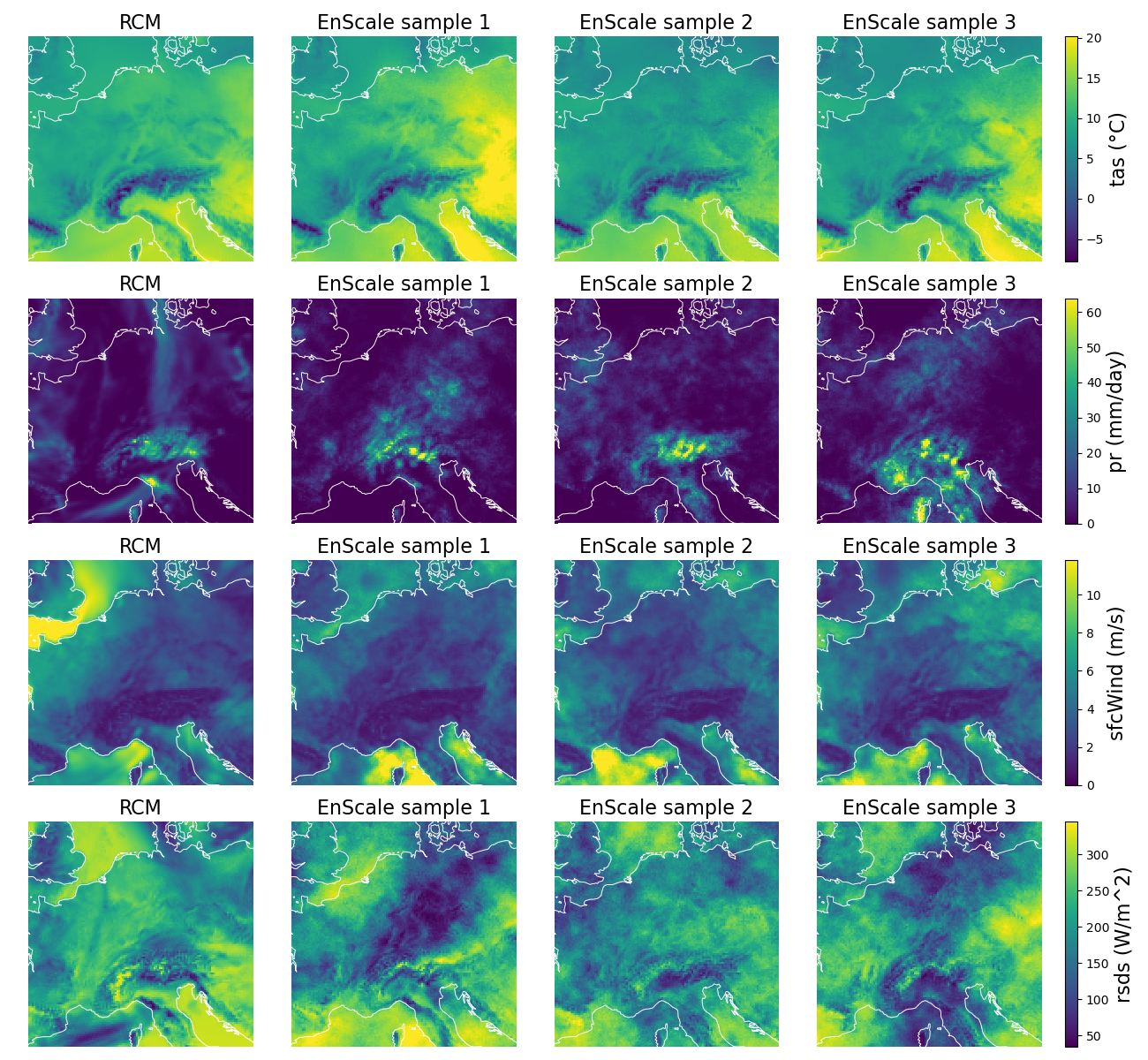}
    \caption{Samples from \ourmethod for all variables. The first column presents the unseen target RCM data by ALADIN63 driven by CNRM-CM5 on day 2035-05-06. Columns 2-4 show three corresponding random samples from \ourmethod. We chose a day with average performance, i.e.,~the \condenscoreloss is roughly equal to the median score of all days in the interpolation test set.}
    \label{fig:results:samples-all}
\end{figure}

Qualitatively, \ourmethod outputs realistic spatial fields that resemble the RCM data as illustrated in Fig.~\ref{fig:results:samples-all} for one GCM input. Examples for two other GCM inputs are shown in Fig.~\ref{fig:appendix:enscale-good} and \ref{fig:appendix:enscale-bad}. For example, Fig.~\ref{fig:appendix:enscale-good} demonstrates a case where samples by \ourmethod are particularly similar to the RCM data (i.e.,~small EnScale-loss).

The goal of \ourmethod is not to match the RCM data directly, but to generate plausible realizations given the GCM boundary conditions. We find that key features like topography and land-sea contrast are captured even though not provided explicitly as inputs. For temperature, samples align well with the RCM, with some differences in the Alps and in the eastern part of the domain, where variability is largest. For the other variables, the sampled fields broadly agree with the RCM, but exhibit substantial variability between samples for this particular GCM input. The samples differ both in location and intensity. Surface wind has a pronounced land-sea contrast, both in the RCM as well as in the samples, with higher magnitudes and variability over sea than over land. We further analyze and quantify sources for this variability in Sec.~\ref{sec:results:variability}. 

Per GCM-RCM pair, only one RCM field per GCM input is available. Thus, it is not possible to compare the variability in \ourmethod outputs with the raw data directly. Instead, one can follow the nearest-neighbor approach as in our analogue benchmark. Samples of this benchmark are shown in Fig.~\ref{fig:appendix:analogues}. In this case, the variability stems partly from the small training dataset and imperfect near-neighbors (App.~\ref{sec:appendix:details-analogues}). Nevertheless, the examples give a first impression of the variability that is present in the raw GCM-RCM data.

We present samples for the other machine learning methods (GAN and CorrDiff) in Fig.~\ref{fig:appendix:gan} and \ref{fig:appendix:corrdiff}. We find similar conclusions as for \ourmethod: while the benchmark's spatial fields resemble the RCM, they all exhibit considerable variability. GAN outputs tend to be blurry, whereas CorrDiff samples capture fine spatial details more realistically.

\subsection{Comparative overview of method performances}

%\begin{figure}
%    \centering
%     \includegraphics[width=0.99\linewidth]{figures/overview_heatmap_combined_vreview1_enscalet.png}
%     \caption{Summary of performance of \ourmethod compared to the benchmarks in several selected categories, shown for the interpolation test period (2030-39). Energy score (see Sec.~\ref{sec:results:overall_performance}), Calibration (see Sec.~\ref{sec:results:calibration}), Spatial structure (see Sec.~\ref{sec:results:spatial_structure}), Temporal structure (see Sec.~\ref{sec:results:temporal}), Extremes (see Sec.~\ref{sec:results:extremes}), Multivariate dependencies (Sec.~\ref{sec:results:dependencies}). The chosen metrics for the categories are outlined in more detail in the main text. First, all individual metrics are normalized by the value of \ourmethodtemp, such that \ourmethodtemp attains a score of 1 for each metric, and all other scores are expressed relative to it. The normalized metrics are then averaged in each category. In all categories, lower values indicate better performance. As calibration for deterministic models is not meaningful, we do not show it for NN-det.}
%     \label{fig:results:overview-summary}
% \end{figure}

\begin{figure}
    \centering
    \includegraphics[width=0.99\linewidth]{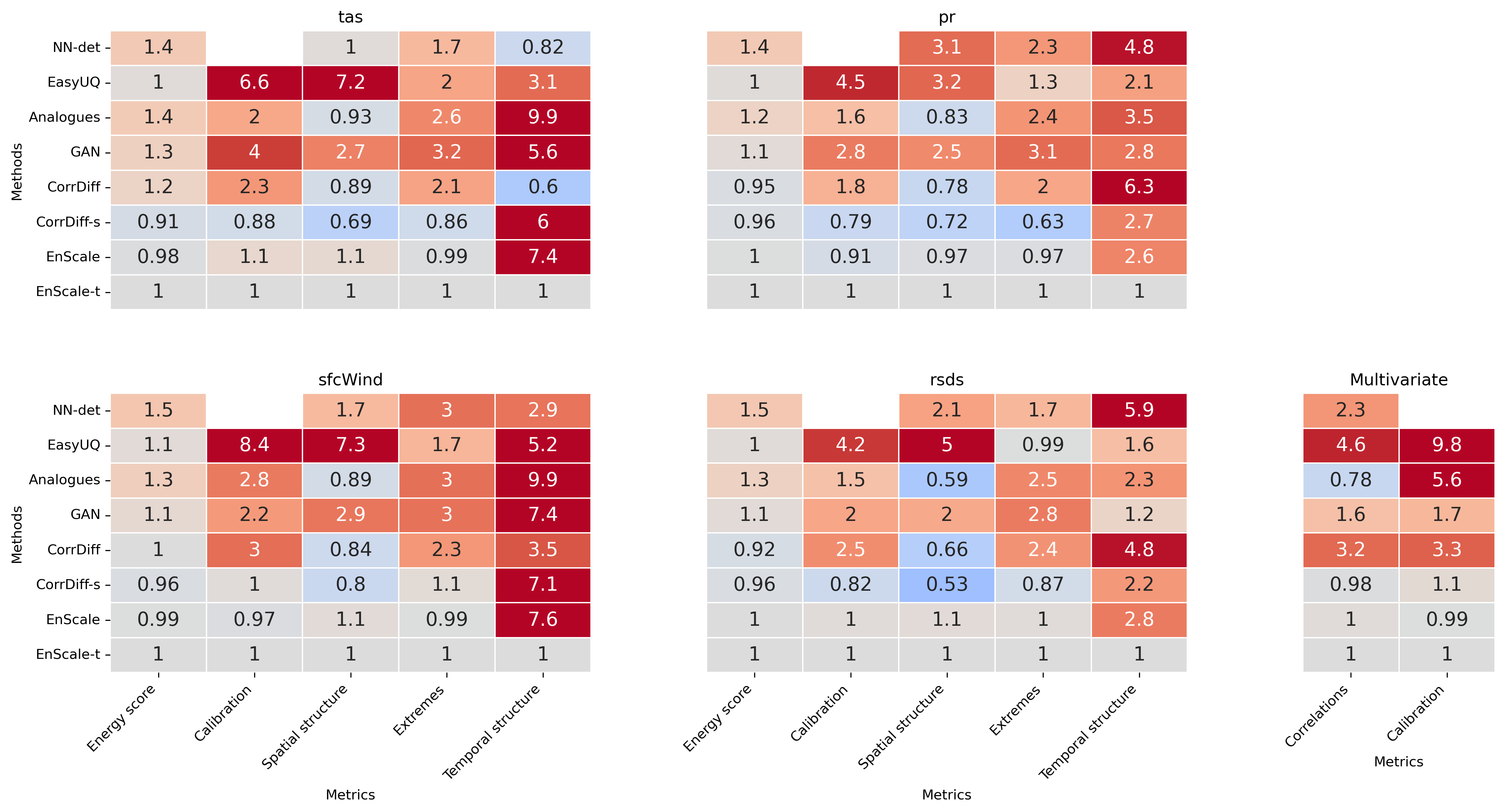}
    \caption{Summary of performance of \ourmethod compared to the benchmarks in several selected categories, shown for the interpolation test period (2030-39). Energy score (see Sec.~\ref{sec:results:overall_performance}), Calibration (see Sec.~\ref{sec:results:calibration}), Spatial structure (see Sec.~\ref{sec:results:spatial_structure}), Temporal structure (see Sec.~\ref{sec:results:temporal}), Extremes (see Sec.~\ref{sec:results:extremes}), Multivariate dependencies (Sec.~\ref{sec:results:dependencies}). The chosen metrics for the categories are outlined in more detail in the main text. First, all individual metrics are normalized by the value of \ourmethodtemp, such that \ourmethodtemp attains a score of 1 for each metric, and all other scores are expressed relative to it. The normalized metrics are then averaged in each category. In all categories, lower values indicate better performance. As calibration for deterministic models is not meaningful, we do not show it for NN-det.}
    \label{fig:results:overview-summary}
\end{figure}

Fig.~\ref{fig:results:overview-summary} provides an overview of the performance of \ourmethod compared to the benchmarks for several important categories. Scores are normalized such that \ourmethodtemp receives a 1 in each metric, and all other methods are expressed relative to it. The category ``Energy score'' refers to the energy score computed over the full spatial field (all grid cells jointly, separately for each variable, Tab.~\ref{tab:results:eval1-all}). The score for ``Spatial structure'' is calculated from the max-pooled CRPS (Tab.~\ref{tab:results:eval1-all}) and the radially-average log spectral density (RALSD and RALSD-avg, Tab.~\ref{tab:results:eval1-all} and Tab.~\ref{tab:appendix:lsd-avg}), ``Temporal structure'' refers to the error in autocorrelation (Tab.~\ref{tab:appendix:temporal-acf-both-errors}), and ``Extremes'' contains the errors in marginal quantiles and miscalibration metrics for quantiles (Tab.~\ref{tab:results:quantiles}).\footnote{We removed the error in the 1 \% quantile for precipitation, as small absolute errors were blown up to unrealistically large numbers with the relative scaling.} To evaluate variable dependencies, we compute correlations for multiple variable pairs and assess calibration of their differences (Tab.~\ref{tab:results:multivariable}). The results are then averaged and summarized as ``correlations'' and ``calibration''. These values are intended as a summary for comparison purposes and do not include formal uncertainty estimates. Standard errors are reported in the metric-specific tables below. All metrics are analyzed in more detail in the respective sections below. Unless otherwise stated, all metrics in this and the following sections are shown on the interpolation test set (2030-2039) and averaged across all GCM-RCM pairs.

% We find that \ourmethodmarg and \ourmethodtemp reach very good scores for many categories and variables in comparison to the benchmarks. It particularly stands out with respect to calibration, extremes, and multivariate dependencies. For temporal structure, \ourmethodtemp outperforms benchmarks by a large margin.

% CorrDiff yields mixed results: despite good energy score and realistic spatial structure, its performance for calibration, extremes and multivariate dependencies is poor. We find that these deficiencies are a consequence of the sampling scheme in CorrDiff, which is intended to provide ensemble members. The current implementation yields highly autocorrelated time series and restricts the diversity of samples. Thus, we adapted the sampling to allow a broader exploration of the space of possible high-resolution fields. This new version, which we call \textit{CorrDiff-s}, substantially improves on CorrDiff's performance in most metrics, as shown in App.~\ref{sec:appendix:corrdiff-stochastic}.

We find that \ourmethodmarg and \ourmethodtemp achieve consistently competitive scores across all categories and variables. In the majority of cases, they rank among the top three methods and beat most benchmarks. They particularly stands out with respect to calibration, extremes, multivariate dependencies. For temporal structure, \ourmethodtemp outperforms all benchmarks by a large margin.

CorrDiff-s often reaches best results except for temporal structure. It beats all other methods for calibration, spatial structure and extremes, and achieves results similar to \ourmethod for energy scores and multivariate dependencies. CorrDiff yields mixed results: despite good energy score and realistic spatial structure, its performance for calibration, extremes and multivariate dependencies is poor. We find that these deficiencies are a consequence of the sampling scheme in CorrDiff, which is intended to provide ensemble members. Due to the large conditional variance in the distribution of RCM fields given GCM inputs, the original CorrDiff sampling scheme is suboptimal, as it yields highly autocorrelated time series and restricts the diversity of samples. An independent sampling strategy, as in CorrDiff-s, better captures the spread of possible high-resolution fields.

The analogues benchmark performs well in many metrics as is to be expected. This benchmark resamples existing fields. For this reason, it perfectly reproduces the spatial characteristics and variable correlations of the RCM, thus reaching excellent scores in these categories. The GAN is inferior compared to the other ML-approaches. EasyUQ performs well on metrics that consider each location separately, but leads to the worst performance regarding calibration and spatial structure. As expected, NN-det is inferior to the benchmarks for the probabilistic metrics, in particular for the extremes and for the energy score. Also, its spatial structure for the variables with higher variance (all except temperature) is worse than for \ourmethod.

Fidelity regarding the temporal structure is of particular relevance for some statistics. The extension to \ourmethodtemp exhibits the best scores of all stochastic methods for temporal autocorrelations. This is because \ourmethodtemp is the only method that explicitly models temporal structure. All other methods downscale each day independently of previous days, with CorrDiff being one exception due to shared random initializations between different time steps. Generated time series also show that \ourmethodtemp and CorrDiff produce smoother, and thus more realistic, trajectories than \ourmethod (see Sec.~\ref{sec:results:temporal}) and CorrDiff-s.

\subsection{Evaluation of performance for individual variables}

\subsubsection{Overall performance for conditional distributions}
\label{sec:results:overall_performance}
% \begin{table*}
% \caption{Evaluation for all variables on the interpolation test regarding mean squared error (MSE), energy score loss (ES), prediction error (ES$_\text{pred}$), variability (ES$_\text{var}$), CRPS averaged over all grid points (CRPS) and after max-pooling across patches of size 10 (CRPS-mp), as well as Radially-averaged Log Spectral Distance (RALSD).$^a$ \red{new}}
% \input{tables/table_energy_scores_crps_lsd_vreview1}
% \label{tab:results:eval1-all}
% \end{table*}

\begin{table*}
\caption{Evaluation for all variables on the interpolation test regarding mean squared error (MSE), energy score loss (ES), prediction error (ES$_\text{pred}$), variability (ES$_\text{var}$), CRPS averaged over all grid points (CRPS) and after max-pooling across patches of size 10 (CRPS-mp), as well as Radially-averaged Log Spectral Distance (RALSD).$^a$}
\begin{tabular}{llllllll}
\midrule
\multicolumn{7}{c}{tas} \\
\midrule
method & MSE $\downarrow$ & ES $\downarrow$ & ES$_\text{pred}$ & ES$_\text{var}$ & CRPS $\downarrow$ & CRPS-mp $\downarrow$ & RALSD $\downarrow$ \\
\midrule
NN-det & \textbf{4.5} ± \footnotesize{0.08} & 245.5 ± \footnotesize{1.85} & 245.4 & 0.0 & 1.54 ± \footnotesize{0.01} & 1.51 ± \footnotesize{0.01} & \textbf{0.9} ± \footnotesize{0.01} \\
EasyUQ & 6.5 ± \footnotesize{0.08} & 182.2 ± \footnotesize{1.56} & 310.1 & 255.9 & 1.22 ± \footnotesize{0.01} & 2.25 ± \footnotesize{0.02} & 6.2 ± \footnotesize{0.02} \\
Analogues & 7.5 ± \footnotesize{0.13} & 241.5 ± \footnotesize{2.72} & 410.6 & 338.2 & 1.67 ± \footnotesize{0.01} & 1.63 ± \footnotesize{0.01} & 1.3 ± \footnotesize{0.01} \\
GAN & 9.2 ± \footnotesize{0.1} & 223.4 ± \footnotesize{1.46} & 376.0 & 305.3 & 1.48 ± \footnotesize{0.01} & 1.41 ± \footnotesize{0.01} & 2.4 ± \footnotesize{0.02} \\
CorrDiff & 6.5 ± \footnotesize{0.1} & 212.4 ± \footnotesize{2.23} & 396.9 & 368.7 & \textbf{1.09} ± \footnotesize{0.01} & \textbf{1.06} ± \footnotesize{0.01} & \underline{1.0} ± \footnotesize{0.01} \\
CorrDiff-s & 7.7 ± \footnotesize{0.14} & \textbf{159.4} ± \footnotesize{1.71} & 319.9 & 321.0 & \underline{1.09} ± \footnotesize{0.01} & \underline{1.06} ± \footnotesize{0.01} & \textit{1.0} ± \footnotesize{0.01} \\
EnScale & \underline{4.6} ± \footnotesize{0.08} & \underline{170.5} ± \footnotesize{1.7} & 351.3 & 361.4 & \textit{1.20} ± \footnotesize{0.01} & \textit{1.19} ± \footnotesize{0.01} & 1.3 ± \footnotesize{0.01} \\
EnScale-t & \textit{4.9} ± \footnotesize{0.09} & \textit{174.3} ± \footnotesize{1.76} & 350.6 & 352.8 & 1.23 ± \footnotesize{0.01} & 1.20 ± \footnotesize{0.01} & 1.2 ± \footnotesize{0.01} \\
\midrule
\multicolumn{7}{c}{pr} \\
\midrule
method & MSE $\downarrow$ & ES $\downarrow$ & ES$_\text{pred}$ & ES$_\text{var}$ & CRPS $\downarrow$ & CRPS-mp $\downarrow$ & RALSD $\downarrow$ \\
\midrule
NN-det & \textbf{30.9} ± \footnotesize{0.61} & 610.3 ± \footnotesize{5.83} & 610.1 & 0.0 & 2.34 ± \footnotesize{0.02} & 6.22 ± \footnotesize{0.06} & 10.1 ± \footnotesize{0.1} \\
EasyUQ & 49.5 ± \footnotesize{0.72} & 447.1 ± \footnotesize{4.32} & 816.8 & 740.9 & 1.94 ± \footnotesize{0.02} & 11.37 ± \footnotesize{0.07} & 9.6 ± \footnotesize{0.04} \\
Analogues & 39.5 ± \footnotesize{0.69} & 535.4 ± \footnotesize{6.24} & 892.8 & 717.6 & 2.35 ± \footnotesize{0.02} & 5.93 ± \footnotesize{0.05} & 5.2 ± \footnotesize{0.07} \\
GAN & 45.9 ± \footnotesize{0.94} & 460.4 ± \footnotesize{4.88} & 756.0 & 592.1 & 1.94 ± \footnotesize{0.02} & 5.16 ± \footnotesize{0.05} & 8.8 ± \footnotesize{0.07} \\
CorrDiff & 47.9 ± \footnotesize{0.77} & \textbf{414.7} ± \footnotesize{4.24} & 792.9 & 756.5 & \underline{1.87} ± \footnotesize{0.02} & \underline{4.68} ± \footnotesize{0.04} & \textbf{3.6} ± \footnotesize{0.05} \\
CorrDiff-s & 53.5 ± \footnotesize{0.96} & \underline{415.4} ± \footnotesize{3.99} & 812.4 & 794.2 & \textbf{1.82} ± \footnotesize{0.02} & \textbf{4.55} ± \footnotesize{0.05} & \underline{4.2} ± \footnotesize{0.06} \\
EnScale & \textit{31.9} ± \footnotesize{0.56} & \textit{433.0} ± \footnotesize{4.49} & 825.7 & 785.2 & 1.90 ± \footnotesize{0.02} & 4.88 ± \footnotesize{0.05} & 5.1 ± \footnotesize{0.06} \\
EnScale-t & \underline{31.3} ± \footnotesize{0.59} & 434.5 ± \footnotesize{4.45} & 806.0 & 743.5 & \textit{1.89} ± \footnotesize{0.02} & \textit{4.83} ± \footnotesize{0.05} & \textit{5.0} ± \footnotesize{0.06} \\
\midrule
\multicolumn{7}{c}{sfcWind} \\
\midrule
method & MSE $\downarrow$ & ES $\downarrow$ & ES$_\text{pred}$ & ES$_\text{var}$ & CRPS $\downarrow$ & CRPS-mp $\downarrow$ & RALSD $\downarrow$ \\
\midrule
NN-det & \underline{2.8} ± \footnotesize{0.03} & 204.7 ± \footnotesize{1.07} & 204.8 & 0.0 & 1.18 ± \footnotesize{0.01} & 1.50 ± \footnotesize{0.01} & 2.1 ± \footnotesize{0.02} \\
EasyUQ & 4.4 ± \footnotesize{0.03} & 147.3 ± \footnotesize{0.78} & 262.8 & 231.1 & 0.92 ± \footnotesize{0.0} & 2.25 ± \footnotesize{0.01} & 9.0 ± \footnotesize{0.02} \\
Analogues & 4.0 ± \footnotesize{0.04} & 183.0 ± \footnotesize{1.46} & 309.0 & 251.9 & 1.14 ± \footnotesize{0.01} & 1.41 ± \footnotesize{0.01} & 2.0 ± \footnotesize{0.02} \\
GAN & 4.9 ± \footnotesize{0.04} & 155.8 ± \footnotesize{0.86} & 275.4 & 239.1 & 0.97 ± \footnotesize{0.0} & 1.19 ± \footnotesize{0.01} & 3.7 ± \footnotesize{0.03} \\
CorrDiff & 4.0 ± \footnotesize{0.04} & 141.2 ± \footnotesize{0.83} & 276.5 & 270.7 & \underline{0.85} ± \footnotesize{0.0} & \textbf{1.04} ± \footnotesize{0.01} & \textbf{1.5} ± \footnotesize{0.01} \\
CorrDiff-s & 4.7 ± \footnotesize{0.05} & \textbf{134.2} ± \footnotesize{0.95} & 264.1 & 260.1 & \textbf{0.85} ± \footnotesize{0.0} & \underline{1.04} ± \footnotesize{0.01} & \underline{1.6} ± \footnotesize{0.02} \\
EnScale & \textbf{2.8} ± \footnotesize{0.03} & \underline{138.7} ± \footnotesize{0.91} & 272.9 & 268.4 & \textit{0.89} ± \footnotesize{0.0} & \textit{1.10} ± \footnotesize{0.01} & 1.8 ± \footnotesize{0.02} \\
EnScale-t & \textit{2.9} ± \footnotesize{0.03} & \textit{139.8} ± \footnotesize{0.91} & 272.5 & 265.2 & 0.89 ± \footnotesize{0.0} & 1.10 ± \footnotesize{0.01} & \textit{1.8} ± \footnotesize{0.02} \\
\midrule
\multicolumn{7}{c}{rsds} \\
\midrule
Method & MSE $\downarrow$ & ES $\downarrow$ & ES$_\text{pred}$ & ES$_\text{var}$ & CRPS $\downarrow$ & CRPS-mp $\downarrow$ & RALSD $\downarrow$ \\
\midrule
NN-det & \textbf{2033.3} ± \footnotesize{31.94} & 5036.9 ± \footnotesize{43.55} & 5041.7 & 0.0 & 31.02 ± \footnotesize{0.26} & 26.17 ± \footnotesize{0.21} & 3.8 ± \footnotesize{0.03} \\
EasyUQ & 3364.8 ± \footnotesize{39.66} & 3622.1 ± \footnotesize{33.25} & 6756.6 & 6292.3 & 24.32 ± \footnotesize{0.22} & 47.43 ± \footnotesize{0.33} & 10.3 ± \footnotesize{0.02} \\
Analogues & 2629.0 ± \footnotesize{37.71} & 4351.3 ± \footnotesize{48.99} & 7297.6 & 5903.9 & 28.70 ± \footnotesize{0.23} & 20.91 ± \footnotesize{0.19} & \textit{1.8} ± \footnotesize{0.02} \\
GAN & 3049.4 ± \footnotesize{44.51} & 3646.7 ± \footnotesize{31.35} & 6377.1 & 5468.2 & 24.02 ± \footnotesize{0.2} & \textit{17.76} ± \footnotesize{0.15} & 3.9 ± \footnotesize{0.03} \\
CorrDiff & 3644.4 ± \footnotesize{50.49} & \textbf{3179.7} ± \footnotesize{30.93} & 6463.3 & 6570.6 & \underline{22.79} ± \footnotesize{0.2} & \underline{16.00} ± \footnotesize{0.14} & \textbf{1.5} ± \footnotesize{0.01} \\
CorrDiff-s & 3414.7 ± \footnotesize{48.79} & \underline{3335.0} ± \footnotesize{30.28} & 6549.1 & 6430.1 & \textbf{21.92} ± \footnotesize{0.2} & \textbf{15.38} ± \footnotesize{0.16} & \underline{1.5} ± \footnotesize{0.01} \\
EnScale & \underline{2050.7} ± \footnotesize{30.38} & 3475.0 ± \footnotesize{32.31} & 7031.4 & 7104.6 & 24.02 ± \footnotesize{0.2} & 21.05 ± \footnotesize{0.14} & 2.9 ± \footnotesize{0.02} \\
EnScale-t & \textit{2064.4} ± \footnotesize{31.62} & \textit{3463.4} ± \footnotesize{31.28} & 7060.6 & 7179.1 & \textit{23.84} ± \footnotesize{0.2} & 21.39 ± \footnotesize{0.15} & 2.7 ± \footnotesize{0.02} \\ \bottomrule
\multicolumn{8}{@{}p{0.99\textwidth}}{$^{a}$ Metrics are first averaged across all days. Afterwards both metrics and standard errors are averaged across all GCM-RCM pairs to evaluate the overall performance on the entire dataset. Bold, underlined and italic values highlight the best, second / third-best results, respectively.}
\end{tabular}
\label{tab:results:eval1-all}
\end{table*}

Moving from visual evaluation to more quantitative results, Tab.~\ref{tab:results:eval1-all} demonstrates the performance of \ourmethod in comparison to the various benchmarks explained in Sec.~\ref{sec:benchmarks}. To assess skill overall and in each location, we show MSE, energy scores (ES), and CRPS (see Sec.~\ref{sec:methods:metrics}) separately for each variable. %In general, we find that \ourmethod reaches good results compared to the benchmarks, often achieving the best or second-best results. \ourmethodtemp shows similar results to \ourmethod, with the ordering depending on the variable. 
In general, we find that \ourmethod and \ourmethodtemp reaches good results compared to the benchmarks. One of the two methods often is among the top three methods, and the other only slightly worse. The ordering between \ourmethod and \ourmethodtemp depends on the variable. A dedicated evaluation of the temporal structure follows in Sec.~\ref{sec:results:temporal}. In the following, we analyze individual metrics in more detail.

The deterministic neural network excels regarding MSE for all variables. This is to be expected, as this approach explicitly minimizes the MSE during training. \ourmethodmarg and \ourmethodtemp perform only slightly worse. For all stochastic approaches (i.e.,~EasyUQ, analogues, GAN, CorrDiff/-s, and \ourmethod), we approximate the conditional expectation $E^\iota[Y|X]$ as the average over nine samples corresponding to the particular GCM input $X$. With more samples per day, the estimate of the conditional mean improves, and lower scores are likely attainable, but the qualitative picture should persist.

%When considering the energy score averaged over multiple GCM inputs $X$, CorrDiff or \ourmethod perform best, with the best method depending on the variable. 
When considering the energy score averaged over multiple GCM inputs $X$, CorrDiff or CorrDiff-s perform best, often followed by \ourmethod, with the best method depending on the variable.

The energy score allows us to investigate the trade-off between prediction error ($\text{ES}_\text{pred}$) and variability ($\text{ES}_\text{var}$) quantitatively. If a method learned the correct RCM conditional distribution $Y|X$, these two scores should be equal ($\text{ES}_\text{pred} = \text{ES}_\text{var}$, see Sec.~\ref{sec:methods:gen-model-energy-score:loss-cond}). For \ourmethod, we slightly underestimate variability for precipitation and surface wind ($\text{ES}_\text{pred} > \text{ES}_\text{var}$), but overestimate variability for temperature and solar radiation ($\text{ES}_\text{pred} < \text{ES}_\text{var}$). The latter could result from suboptimal convergence, whereas underestimation of variability can stem from mild overfitting. If $\text{ES}_\text{pred} = \text{ES}_\text{var}$ on the training data, $\text{ES}_\text{pred}$ usually increases on the test data, whereas $\text{ES}_\text{var}$ stays nearly constant, leading to $\text{ES}_\text{pred} > \text{ES}_\text{var}$ in test case.

We find that some other benchmarks (analogues, GAN) often severely underestimate variability relative to the prediction error ($\text{ES}_\text{pred} > \text{ES}_\text{var}$). For applications, this would imply overconfidence, i.e.,~a lack of spread in the generated samples, which from a risk perspective is an undesirable property of a prediction system. For GANs, for example, the lack of variance is a well-known problem \cite{Salimans2016ImprovedGANs}. For analogues, the relationship between $\text{ES}_\text{pred}$ and $\text{ES}_\text{var}$ depends on the number of near neighbors chosen, as alluded to in Sec.~\ref{sec:benchmarks}. In our setting, $k=5$ neighbors still underestimate $\text{ES}_\text{var}$. 

Crucially, a lower prediction error ($\text{ES}_\text{pred}$) is not always desirable if it comes at the cost of reduced variability. For instance, NN-det has the lowest $\text{ES}_\text{pred}$, but their lack of variability leads to the poorest scores. Hence, if one is interested in the full conditional distribution, the goal is not only to minimize $\text{ES}_\text{pred}$ but to strike the right balance between prediction error and variability.
We investigate this trade-off also visually, using the same example day that was presented above in Fig.~\ref{fig:results:samples-all}. Fig.~\ref{fig:appendix:enscale-pred-diff} shows both the prediction error for one sample as well as the variability across several samples of \ourmethod. Both terms show similar magnitude and spatial structure, highlighting that \ourmethod balances prediction error and variability terms. For comparison, Fig.~\ref{fig:appendix:analogues-pred-diff} shows the same for the analogues, with qualitatively similar conclusions.

Furthermore, we assess distributions in each location separately with the CRPS. CorrDiff or CorrDiff-s reaches best scores for all variables, closely followed by \ourmethod or \ourmethodtemp. Comparison with the EasyUQ benchmark serves as another sanity check: EasyUQ is expected to perform well as it explicitly aims to best capture the pointwise CRPS values. \ourmethod reaches slightly better scores than the EasyUQ benchmark, showing that we successfully learned the distributions in each location despite the high-dimensional target.

The values of energy score and CRPS can be interpreted physically. For example, for temperature and \ourmethod, the average deviation between one random sample and the RCM data is around \SI{2.74}{\degreeCelsius}. This conclusion can be reached with the following calculation: the square of $\text{ES}_\text{pred}$ yields the squared differences of one sample with the RCM data, summed over all locations and averaged across multiple $X$. Dividing by the number of locations gives the average difference per location. In our case, we calculate $\sqrt{(356.4^2) / (128^2)}$). This roughly agrees with the values for the CRPS: the average CRPS value in each location is \SI{1.19}{\degreeCelsius}, and twice this value should approximately give the average difference. Importantly, these differences are not a quantification of the errors of \ourmethod. Even if the model were perfect, the scores would be non-zero. Instead, this is again a quantification of the variability that is present in the data and the samples.

\subsubsection{Spatial structure}
\label{sec:results:spatial_structure}

Next, we assess the spatial structure of the samples. We pool pixels spatially and compute the CPRS on these pooled results. Tab.~\ref{tab:results:eval1-all} shows the results where we applied max pooling across patches of size $10\times 10$ before calculating CRPS. These findings remain consistent for other kernel sizes for pooling. The benefit of our spatially multivariate approach becomes apparent in these results. The EasyUQ benchmark, in contrast to good pointwise CRPS scores, fails to capture correct spatial structures and leads to large values for this metric. %CorrDiff, \ourmethod and \ourmethodtemp again show good performance and are superior to the other benchmarks. 
CorrDiff and CorrDiff-s shows best performance, and also \ourmethod and \ourmethodtemp reach low scores and are superior to the other benchmarks.

In addition, we compute power spectral densities (PSDs) for all days and generated samples in the test set and report radially-averaged log spectral distance (RALSD, see Sec.~\ref{sec:methods:psd}). 
%CorrDiff reaches best results, except for temperature, where NN-det has lowest RALSD. \ourmethod or analogues are second-best. 
CorrDiff and CorrDiff-s reach best results, except for temperature, where NN-det has lowest RALSD. \ourmethod or analogues are next strongest methods. Performance with RALSD depends on both the spatial quality as well as the similarity of samples and truth. Alternatively, we average power spectra across multiple days before calculating log spectral distances (RALSD-avg, see Sec.~\ref{sec:methods:psd}), results are shown in Tab.~\ref{tab:appendix:lsd-avg}. Here, the analogues reach lowest errors as expected because they reproduce correct spatial fields.

%Plots for the PSDs in Fig.~\ref{fig:appendix:psd} confirm that both CorrDiff and \ourmethod closely matches power spectra of the RCM. We find only a slight overestimation of the power for precipitation and solar radiation on small scales. Other results are as expected: EasyUQ's have too much small-scale variance, overestimating the power at small wavelengths and underestimating it at larger wavelengths. NN-det's outputs are too smooth, generally underestimating the PSDs across wavelengths, most strikingly for precipitation and wind.

Plots for the PSDs in Fig.~\ref{fig:appendix:psd} confirm that both \ourmethod and \ourmethodtemp closely matches power spectra of the RCM. We find only a slight overestimation of the power for precipitation and solar radiation on small scales. Other results are as consistent with previous findings: CorrDiff and CorrDiff-s model power spectra accurately, with CorrDiff-s sometimes even indistinguishable from the RCM's PSD. EasyUQ's have too much small-scale variance, overestimating the power at small wavelengths and underestimating it at larger wavelengths. NN-det's outputs are too smooth, generally underestimating the PSDs across wavelengths, most strikingly for precipitation and wind.

\subsubsection{Marginal distribution evaluation with histograms}

\begin{figure}
    \centering
    \includegraphics[width=0.99\linewidth]{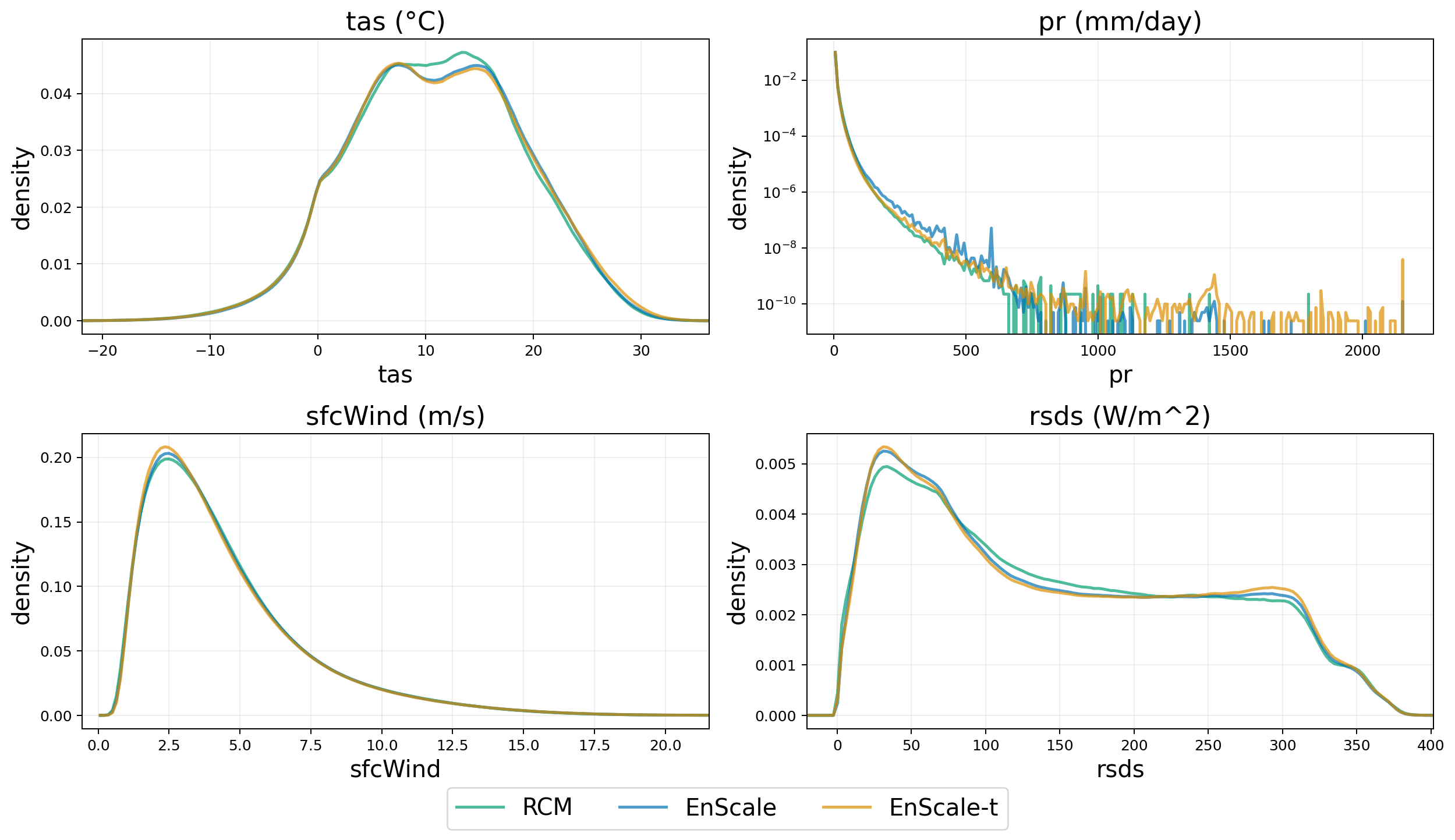}
    \caption{Distributions per variable for the target RCM data as well as \ourmethod and \ourmethodtemp. We pool data from all locations, all GCM-RCM pairs and all days in the test set. For our methods, data is additionally pooled across 9 samples per day.}
    %\caption{Histograms. We pool data from all locations, all GCM-RCM pairs and all days in the test set. We plot distributions for the target RCM data as well as \ourmethod and \ourmethodtemp, pooled across 9 samples per day.}
    \label{fig:results:histograms}
\end{figure}

We provide a qualitative assessment of the marginal distributions of \ourmethod's samples against the true RCM data with histograms (Fig.~\ref{fig:results:histograms}). We find that \ourmethod and \ourmethodtemp produce very similar marginal distributions and closely match the true RCM distribution across all variables. This highlights that our multivariate approach is able to flexibly learn distributions of different climate variables despite their vastly different characteristics and shapes. The plots also suggest that both the bulk as well as the tails of the RCM's distributions are reproduced by our ML models.

For temperature, both methods capture the tails well, with only a small underestimation of the density in the bulk of the distribution by \ourmethod and \ourmethodtemp. In case of precipitation, there is broad agreement: Both \ourmethod and \ourmethodtemp are able to generate very heavy rainfall events, which are also present in the true data. \ourmethodtemp even tends to overproduce very extreme precipitation events, as suggested by the heavy upper tail. The distribution of surface wind is reproduced accurately. For solar radiation, the general shape of the distribution is well captured by our methods, although there is a small overestimation of the density at low and high values, and a small underestimation in the middle of the range.

We present histograms for some selected benchmarks in Fig.~\ref{fig:results:histograms-benchmarks}. Analogues, CorrDiff and CorrDiff-s also capture the marginal distributions accurately, with small mismatches for CorrDiff, e.g., for large precipitation or high solar radiation values. The GAN, however, shows larger errors for several variables, e.g., for the mode of the surface wind or the general shape of the solar radiation distribution.

\subsubsection{Calibration}
\label{sec:results:calibration}

% \begin{figure}
% \includegraphics[width=0.99\linewidth]{figures/rh_joint_colorblind_mcb-per_variable_vvreview1_4rows.png}
% 	\caption{Rank histograms for evaluating calibration. For each day, we calculate the spatial mean and the spatial maximum for the RCM data and for 9 samples of each method, and construct rank histograms for these spatial statistics (first and second rows show the spatial mean, third and fourth row the spatial max.). The flat dashed line represents the uniform distribution, indicating perfect calibration. Miscalibration (MCB) scores in the legend quantify deviations from the uniform distribution at each individual grid point, averaged across all locations. \red{new}
%     }
% 	\label{fig:results:rh-mean-max}
% \end{figure}

\begin{figure}
\includegraphics[width=0.99\linewidth]{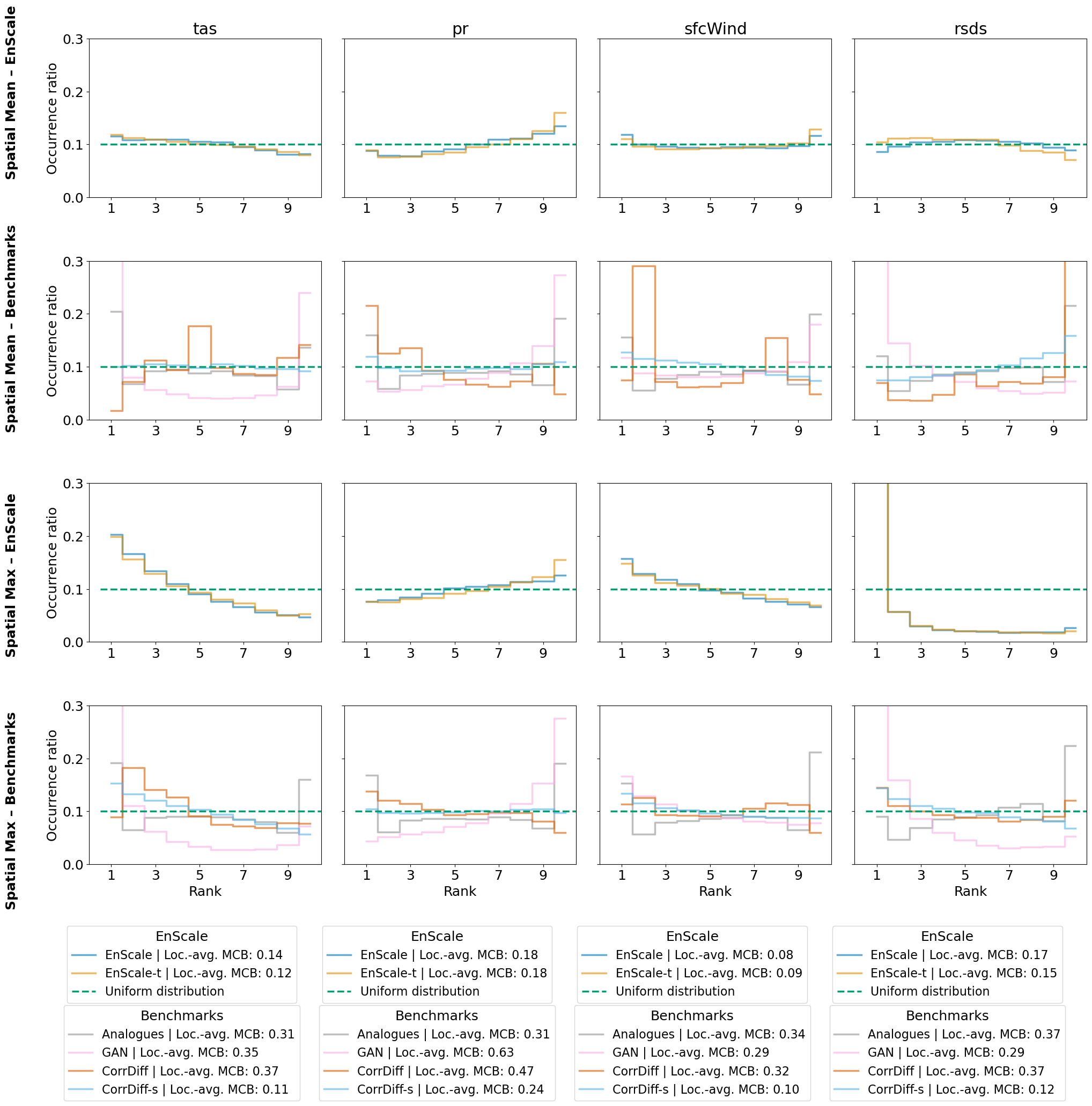}
	\caption{Rank histograms for evaluating calibration. For each day, we calculate the spatial mean and the spatial maximum for the RCM data and for 9 samples of each method, and construct rank histograms for these spatial statistics (first and second rows show the spatial mean, third and fourth row the spatial max.). The flat dashed line represents the uniform distribution, indicating perfect calibration. Miscalibration (MCB) scores in the legend quantify deviations from the uniform distribution at each individual grid point, averaged across all locations.
    }
	\label{fig:results:rh-mean-max}
\end{figure}

A related evaluation tool are rank histograms, which assess the calibration of the conditional distributions of the samples (see Sec.~\ref{sec:methods:validation:rh}). First, we analyze calibration of spatially-aggregated quantities, the spatial mean and maximum, in Fig.~\ref{fig:results:rh-mean-max}. We find that most approaches fail to achieve good calibration for most variables, with varying degrees of violation and biases. In general, calibration for the spatial mean is often better than that for the spatial maximum.

Considering the spatial mean, we find that \ourmethod only shows mild violations from calibration. The samples tend to underestimate the RCM's values for precipitation and surface wind (higher counts for larger ranks), and are slightly overdispersed for solar radiation (dome-shape of the rank histogram). However, the benchmarks do not perform better: 
CorrDiff-s preforms similarly to \ourmethod, but CorrDiff shows large deviations from calibration. We find that analogues are usually underdispersed, which is in line with the scores in Tab.~\ref{tab:results:eval1-all} ($\text{ES}_\text{var} < \text{ES}_\text{pred}$), and the GAN also shows errors in dispersion or biases.

The spatial maximum is more difficult to capture for some variables. Most approaches are not calibrated. For \ourmethod, we observe underestimation of the spatial maximum for precipitation and overestimation of that for temperature, wind and solar radiation.

%We compare the average miscalibration metric (MCB) across locations (see Sec.~\ref{sec:methods:validation:rh}), see scores in the legend of Fig.~\ref{fig:results:rh-mean-max}. \ourmethod and \ourmethodtemp show errors in calibration with MCB values between 0.08 and 0.18, outperforming CorrDiff. Also analogues and GAN reach higher MCB scores than \ourmethod. For reference, MCB scores for the spatial mean vary between 0.08 (temperature) and 0.2 (precipitation). Locationwise calibration is generally best for temperature and wind and worst for solar radiation and precipitation. \newcorrdiff{[replace two sentences above] \ourmethod and \ourmethodtemp show errors in calibration with MCB values between 0.08 and 0.18, similar to CorrDiff-s and outperforming CorrDiff, analogues and GAN.}
We compare the average miscalibration metric (MCB) across locations (see Sec.~\ref{sec:methods:validation:rh}), see scores in the legend of Fig.~\ref{fig:results:rh-mean-max}. \ourmethod and \ourmethodtemp show errors in calibration with MCB values between 0.08 and 0.18, similar to CorrDiff-s and outperforming CorrDiff, analogues and GAN. For reference, MCB scores for the spatial mean vary between 0.08 (temperature) and 0.2 (precipitation). Locationwise calibration is generally best for temperature and wind and worst for solar radiation and precipitation.

Tracing down the root of the calibration issues in \ourmethod is challenging due to the multi-step approach. We find that the \textit{coarse model} is close to calibration for the spatial mean, but the issues for the spatial maximum are already evident in this step. The several steps of the \textit{super-resolution model} show deviations from calibration for both the spatial mean and maximum.

These results show room for improvement for all ML approaches and point to limitations of the energy score as a loss function. In principle, a method that reaches the minimal energy score also has calibrated samples. However, in practice, the energy score is not equally sensitive to all aspects of misspecification \cite{Pinson2013DiscriminationScore}, and in some cases, a slightly sharper yet miscalibrated forecast may achieve a better energy score than a well-calibrated but less sharp one. This finding is in line with other works in weather post-processing \cite{Wilks2018EnforcingPostprocessing, Wessel2025EnforcingModels}. It suggests that additional measures may be needed to explicitly assess and improve calibration in generative approaches.

\subsubsection{Extremes}
\label{sec:results:extremes}

% \begin{figure}
%     \centering
%     \includegraphics[width=0.75\linewidth]{figures/quantiles_error_CNRM-CM5_ALADIN63_summer_q0.99_swapped_vreview1_coastlines.png}
%     \caption{Extreme quantiles for the summer season (June, July, August) compared to ALADIN63 driven by CNRM-CM5. The first column shows the 99 \% - quantile of the RCM calculated separately in each location. The remaining columns show the error for different methods, calculated as the signed difference of quantiles in the samples minus the RCM's quantiles. Additionally, the last column provides an assessment of ``model uncertainty'' where the difference of ALADIN63 to CCLM4-8-17 (both driven by CNRM-CM5) is shown. Similar results are obtained for pairs other than ALADIN63 driven by CNRM-CM5. \red{new}
%     }
%     \label{fig:results:extreme-quantiles}
% \end{figure}

\begin{figure}
    \centering
    \includegraphics[width=0.55\linewidth]{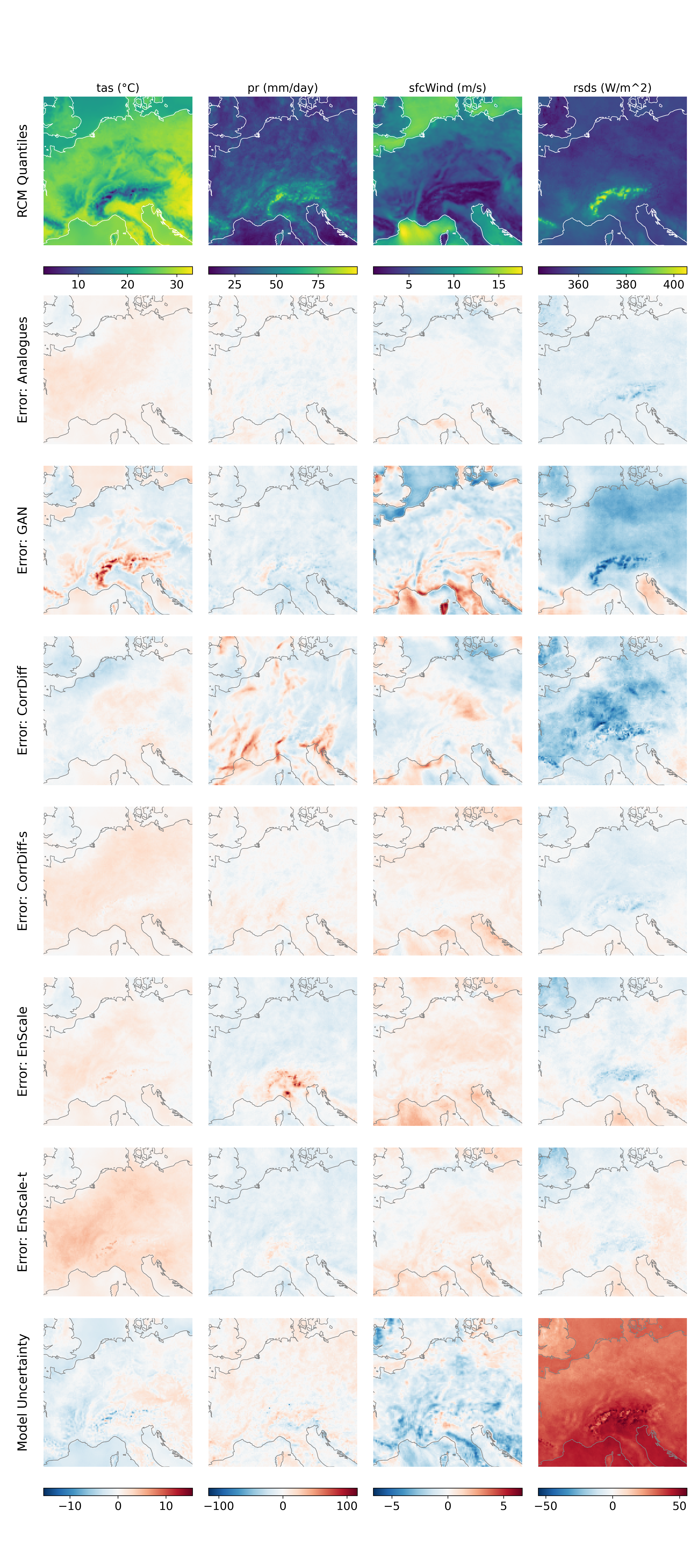}
    \caption{Extreme quantiles for the summer season (June, July, August) compared to ALADIN63 driven by CNRM-CM5. The first row shows the 99 \% - quantile of the RCM calculated separately in each location. The remaining rows show the error for different methods, calculated as the signed difference of quantiles in the samples minus the RCM's quantiles. Additionally, the last row provides an assessment of ``model uncertainty'' where the difference of ALADIN63 to CCLM4-8-17 (both driven by CNRM-CM5) is shown. Similar results are obtained for pairs other than ALADIN63 driven by CNRM-CM5.
    }
    \label{fig:results:extreme-quantiles}
\end{figure}

We focus on the capability of \ourmethod and the benchmark to capture extreme events. As an example, we show the errors for the 99 \% - quantile in summer months and for the first GCM-RCM pair in Fig.~\ref{fig:results:extreme-quantiles}. 
The figure shows the true quantile in the RCM data (first row), which varies spatially, e.g. with largest extremes for precipitation and solar radiation over the Alps. Also, the magnitude and direction of errors vary across locations and variables. For example, \ourmethod overestimates extreme temperatures but underestimates extreme solar radiation in the majority of locations. For precipitation, absolute errors are highest in the Alpine region, and for surface wind, errors are largest over the Sea, where also the variability of the data is highest. Again, \ourmethod performs well in comparison to the benchmarks, with errors of similar magnitude as for the analogues and CorrDiff-s, and often lower than for GAN and CorrDiff.

We note, however, that error estimates are conservative -- per season, there are approximately $1000$ days in the test set, leading to estimation errors for extreme quantiles. Thus, the error presented is a mixture of mistakes due to finite-sample approximation and actual limitation of the ML models and benchmarks.

In order to provide an additional perspective on these results, we compare to uncertainties across different RCMs. As an example, we show differences between the quantiles of two RCMs, both driven by the same GCM. We find that the mismatches between RCMs are of similar magnitude to the errors of the ML models, suggesting that the emulators produce results falling into the range spanned by multiple RCMs. 

In principle, if the ML models perfectly emulated each RCM individually, one would expect disagreements between RCMs to exceed the errors of the ML methods. %This is mostly the case for \ourmethod, \ourmethodtemp, CorrDiff \newcorrdiff{and CorrDiff-s}. 
This is mostly the case for \ourmethod, \ourmethodtemp, CorrDiff and CorrDiff-s.
However, the joint training across multiple GCM-RCM pairs likely also influences the learned distribution:  Especially for extreme events, where there is only little data per individual RCM, our method seems to partly emulate the pooled distribution across multiple pairs instead of exactly reproducing the RCM-specific characteristics. This effect could also be related to our relatively simple approach for RCM conditioning (Sec.~\ref{sec:appendix:model-architecture}) and could possibly be improved in future work.

Further evaluation is shown in Tab.~\ref{tab:results:quantiles}. We assess locationwise quantiles. \ourmethod performs well compared to the benchmarks, but best scores are often reached by the analogues and EasyUQ. Notably, the GAN and CorrDiff struggle to reproduce extreme quantiles accurately. In particular, they miss to predict the zero precipitation events correctly. The true 1 \% -quantile in winter is zero, but the ML approaches do not capture this. \ourmethod successfully predicts dry days with zero rain. This is an advantage of our data transformation (Sec.~\ref{sec:methods:data-transformation}).

In addition, we consider calibration of low / high quantiles of the predicted conditional distribution (see ``MCB Q'' columns in Tab.~\ref{tab:results:quantiles} and Sec.~\ref{sec:methods:validation:extremes} for details). Here, \ourmethod, \ourmethodtemp and CorrDiff-s usually reach the lowest errors, with the exact ordering of these three methods depending on the variable and quantile considered. This again exemplifies that \ourmethod is sometimes inferior to the analogues with respect to metrics evaluating the marginal distributions $p_Y$. For the conditional $p_{Y|X}$, however, \ourmethod is superior.

Evaluating \ourmethod for impact-relevant extremes leaves multiple possible directions of future work. Preliminary analysis suggests that \ourmethod can capture compound extremes \cite{Effenberger2025BridgingEurope}. We are aware that our choice of quantiles is only moderately extreme, and evaluating events with higher return periods would be relevant, but also require a longer test data set. Improving performance for extremes with, e.g., weighted proper scoring rules as loss functions is also a promising direction \cite{Wessel2025ImprovingRules, Wessel2025EnforcingModels}.

\subsection{Variable dependencies}

\label{sec:results:dependencies}
So far, our evaluation considered each target variable separately. Next, we evaluate the capability of our model and the benchmarks to capture the full distribution jointly for the four targets in Tab.~\ref{tab:results:multivariable}. We find that \ourmethod and \ourmethodtemp are often the best-performing ML methods, reaching similar scores as CorrDiff-s. More details can be found in App.~\ref{sec:appendix:dependencies}.

We note that the above metrics only evaluate pairwise distributions between variables. Assessing the full joint distribution is not considered. In addition, we do not explicitly evaluate physical consistency of variable dependencies. However, as the RCM's output is physically consistent, similarity of the ML's samples to the RCM implicitly evaluates this. Future work could focus more on physics-based evaluation.

\subsection{Temporal structure}
\label{sec:results:temporal}

% \begin{figure}
%     \centering
%     \includegraphics[width=0.9\linewidth]{figures/time_series_three_panels_vreview1.png}
%     \caption{Time series examples. Solid lines show the RCM time series (green) and a single randomly chosen realization of \ourmethodmarg (blue), \ourmethodtemp (yellow), and CorrDiff (orange). Shaded areas indicate the minimum and maximum from 9 samples of each model. The figure shows spatially averaged temperature for the first year in the interpolation test set (2030) and for the first GCM-RCM pair (CNRM-CM5 and ALADIN63). \red{new}}
%     \label{fig:results:overview-time-series}
% \end{figure}

\begin{figure}
    \centering
    \includegraphics[width=0.9\linewidth]{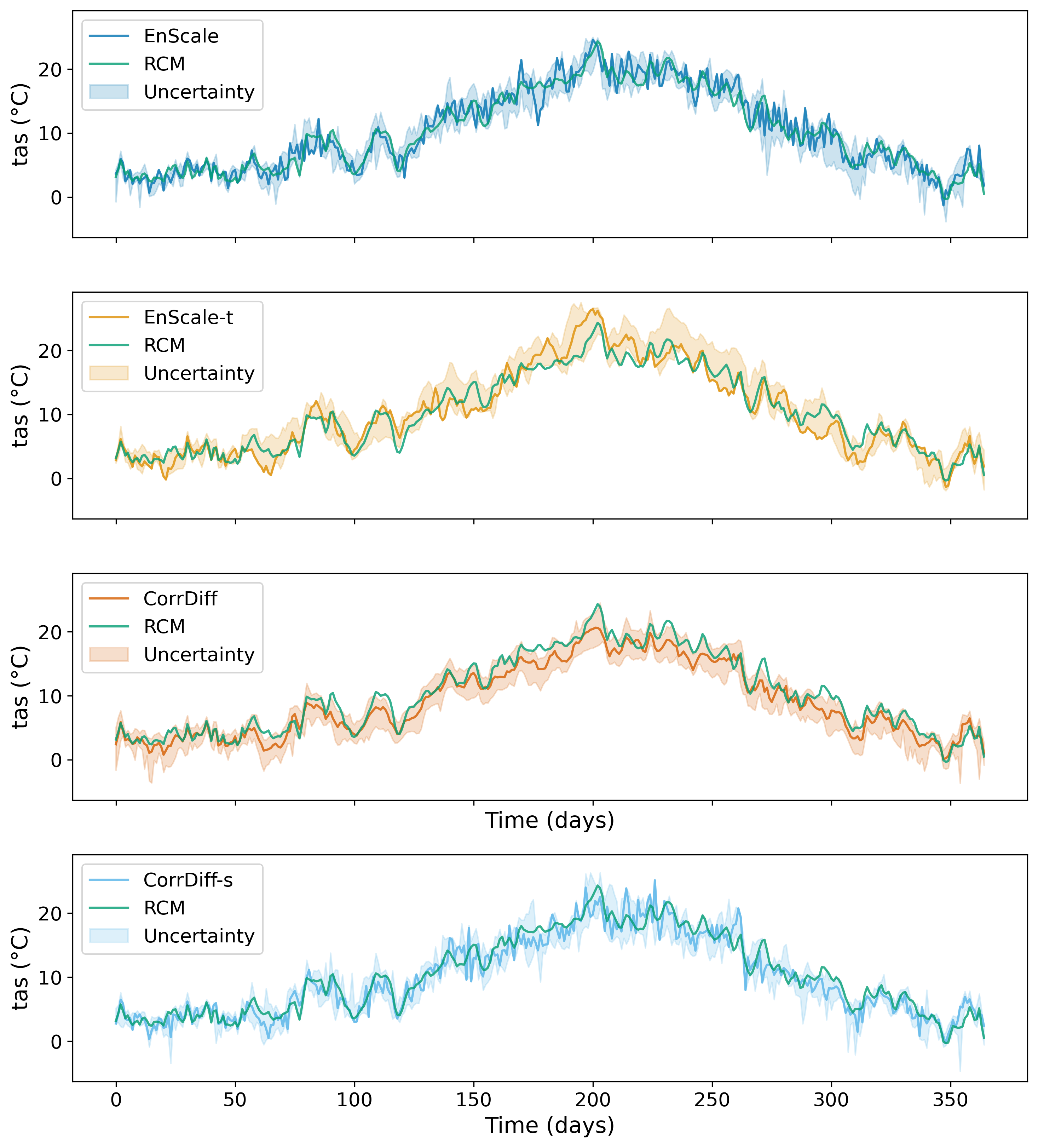}
    \caption{Time series examples. Solid lines show the RCM time series (green) and a single randomly chosen realization of \ourmethodmarg (blue), \ourmethodtemp (yellow), CorrDiff (orange) and CorrDiff-s (light blue). Shaded areas indicate the minimum and maximum from 9 samples of each model. The figure shows spatially averaged temperature for the first year in the interpolation test set (2030) and for the first GCM-RCM pair (CNRM-CM5 and ALADIN63).}
    \label{fig:results:overview-time-series}
\end{figure}

We investigate the temporal structure of the generated samples, that is, how accurately the methods capture correlation between consecutive time steps. Fig.~\ref{fig:results:overview-time-series} presents example time series for the spatial mean of temperature over one year. We compare \ourmethodmarg, \ourmethodtemp, CorrDiff, CorrDiff-s, and the underlying RCM.
\footnote{The remaining variables and time series for \ourmethodmarg, \ourmethodtemp, and the underlying RCM are shown in Fig.~\ref{fig:results:time-series}.}
Most importantly, we find that generated time series by both, \ourmethodmarg and \ourmethodtemp, closely match the ground truth over the entire year and capture the seasonal cycle. On most days, the truth lies within the range of the 9 samples.

The figure also highlights qualitative differences between generated time series by \ourmethodtemp and \ourmethodmarg. The time series for the \ourmethodtemp tend to be smoother because they take into account the temporal structure, especially for temperature. The time series by \ourmethodmarg is not necessarily always consistent over time as each day is generated independently of the previous. The uncertainty bands are similar for both approaches. In times where temporal autocorrelation is high (especially visible for temperature), the uncertainty range for the \ourmethodtemp is larger because including temporal structure allows us to sample a larger variety of possible trajectories. Comparing CorrDiff and CorrDiff-s highlights the effect of the sampling scheme: The original implementation for CorrDiff generates smooth time series with large temporal correlations, while independent sampling in CorrDiff-s leads to more ``noisy'' time series, qualitatively similar to \ourmethod.

These results also show the stability of the time series of the autoregressive rollout for our \temporalcoarsemodel. Both the individual time series as well as the uncertainty range remain stable over time. We also tested longer rollout of our generated time series and found that they are stable even for several years (not shown). Both the time series as well as the loss do not increase over time. The stability of rollout is an open discussion in the case of ML-based weather forecasting \cite{Price2025ProbabilisticLearning, Karlbauer2024AdvancingMesh}. In our case, the additional input from the GCM likely helps to stabilize the rollout and prevent potential drifts.

\begin{figure}
    \centering
    \includegraphics[width=0.99\linewidth]{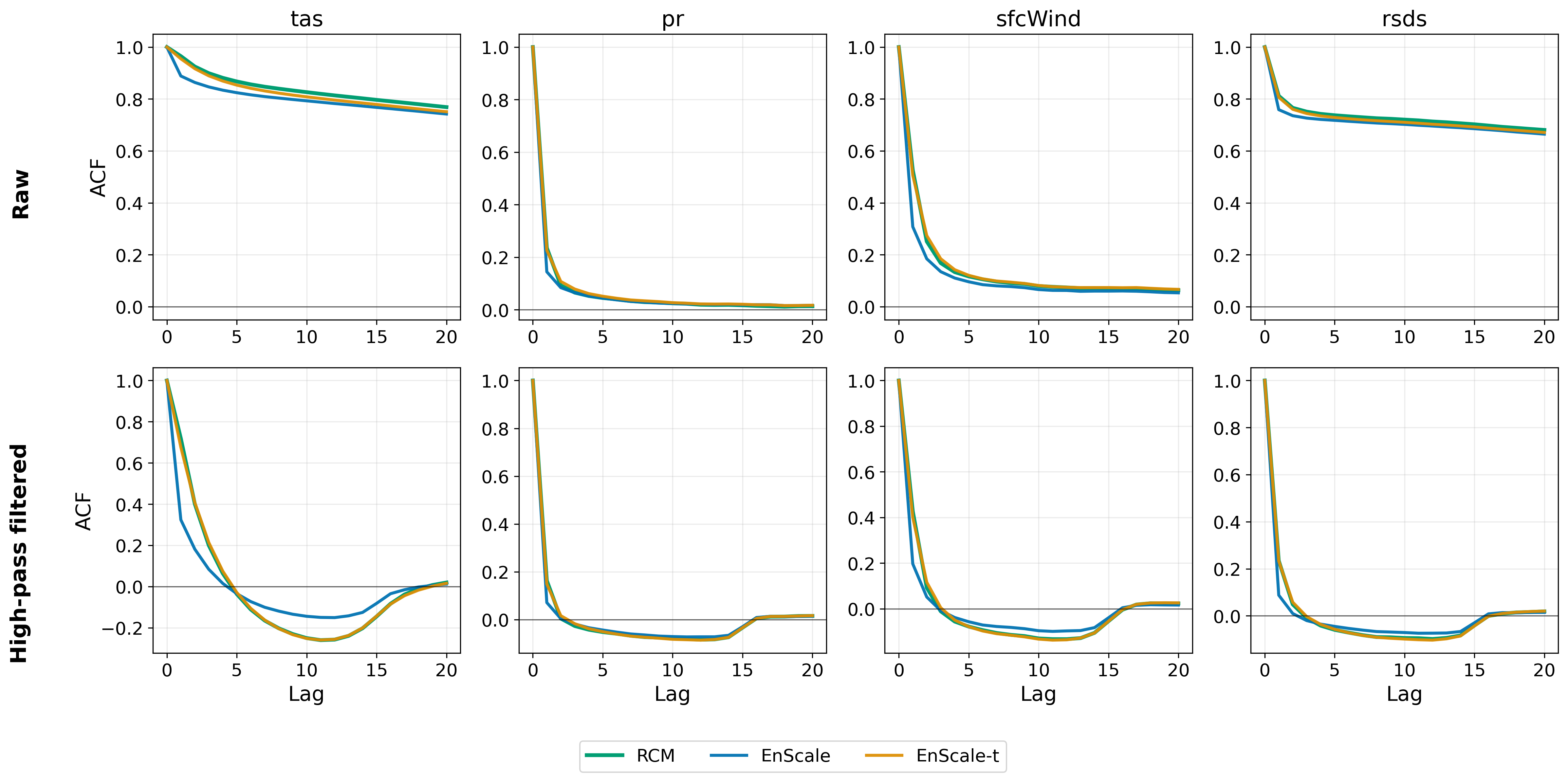}
    \caption{Autocorrelation functions (ACFs). We calculate ACFs for lags up to 20 days for the raw time series (top row) and for a time series after high-pass filtering (bottom row), i.e. removing a 30-day-running mean from the time series before ACF calculation. ACFs are averaged across all locations and all GCM-RCM pairs.}
    \label{fig:results:acfs}
\end{figure}

Autocorrelation functions (ACFs) are shown in Fig.~\ref{fig:results:acfs}. \ourmethod underestimates local autocorrelation for all variables, with the effect being strongest for temperature. \ourmethodtemp is able to reduce the error substantially and matches true ACFs very closely, both for the raw time series as well as after high-pass filtering.

To compare \ourmethod and \ourmethodtemp to the benchmarks, Tab.~\ref{tab:appendix:temporal-acf-both-errors} shows both signed as well as absolute errors in lag-1 (i.e.,~a single time step) ACFs. We find that \ourmethodtemp is superior to the others and achieves the lowest deviation from the true ACFs across all variables. All methods except NN-det and CorrDiff have negative differences, i.e.,~they underestimate the local autocorrelation. CorrDiff accurately reproduces autocorrelations for temperature, but overestimates them for all other variables.

Capturing autocorrelations is particularly relevant when analyzing statistics aggregated over multiple days, such as in 5-day heatwaves. Underestimating autocorrelation can lead to an underrepresentation of variability in multi-day averages, therefore potentially underestimating risk. These findings point to limitations of existing ML-based downscaling approaches that fail to include temporal correlation on a local scale.

\subsection{Evaluation on extrapolation test period}

\begin{figure}
    \centering
    \includegraphics[width=0.9\linewidth]{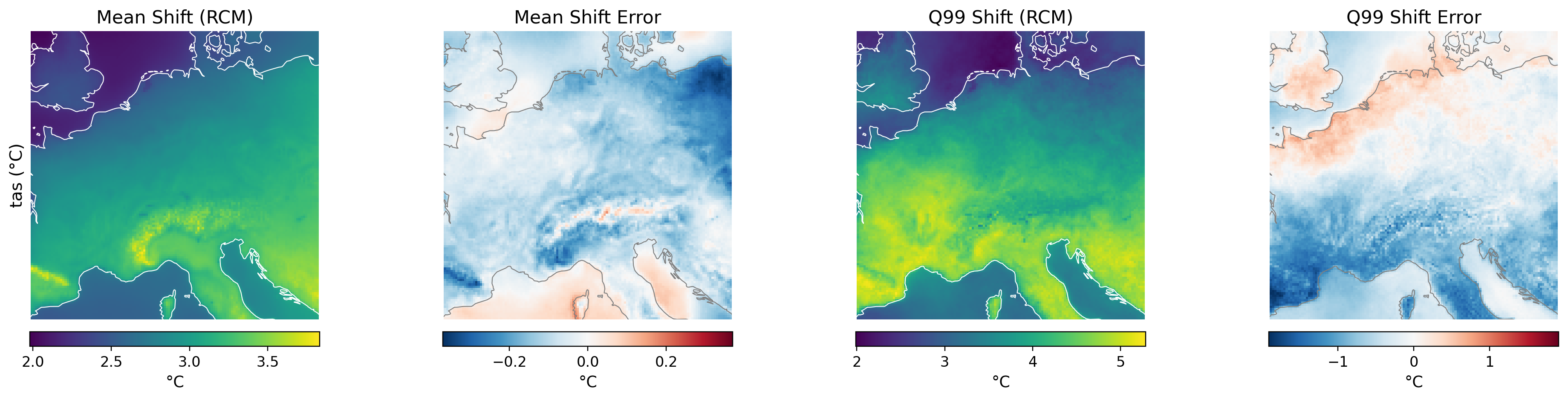}
    \caption{Climate change trend for the average temperature between 2030-2039 and 2090-2099. The trend is computed per GCM-RCM pair first and then averaged across pairs.}
    \label{fig:results:climate-shift}
\end{figure}

So far, the evaluation has focused on the interpolation test period (2030-2039). We also provide scores on the extrapolation test set (2090-2099) (see Sec.~\ref{sec:appendix:extrapolation-results}). We find that scores are very similar and the conclusions are consistent. This shows that our method is able to extrapolate in a mild extrapolation case. Note that training data extends until 2089, so 2090-99 represents only a small extrapolation. For example, in GCM-RCM pairs with relatively strong climate change signal, the extrapolation test data is, on average, only \SI{0.77}{K} warmer than the last decade of the train data, with warming of around \SI{1.2}{K} in a few locations.

We also show climate change patterns in Fig.~\ref{fig:results:climate-shift}. We present only \ourmethod; \ourmethodtemp yields similar results. For the mean shift, errors of \ourmethod are small. We slightly underestimate the climate change signal over land, by around 5 or at most 10 \% of the true climate change signal. Considering climate change trends in the extremes, we find that \ourmethod broadly captures the trend in the RCM. However, the spatial pattern of the temperature trend is not correct for the samples: Over land, the trend is underestimated.

We checked climate change trends also for other variables, but we found that there are large inter-model differences, yielding very small absolute values for the average climate change signals. This lack of clear signal in the data also leads to larger relative errors for \ourmethod, and we found it difficult to reach robust conclusions.

Our current data transformation (Sec.~\ref{sec:methods:data-transformation}) relies on fitted ECDFs and samples data within the support of the training data. Thus, it is not designed for strong out-of-support-extrapolation setups. This might be one reason for the slight underestimation of climate change signals for temperature. Further improving the extrapolation capabilities of \ourmethod would require an adjustment of the data transformation, which we leave to future work.

\subsection{Computational cost}
\label{sec:results:computational-cost}

\begin{table}[]
    \caption{Training and inference times for \ourmethod, the GAN, and CorrDiff/CorrDiff-s$^a$.}
    \centering
    \begin{tabular}{l|c|c}
        \toprule
         Method & Training & Inference\\
         \midrule
         EnScale & 26 h & 1.75 h \\
         EnScale-t & 31.4 h & 1.8 h \\
         CorrDiff(-s) & 216 h & 36 h \\ 
         GAN & 125 h & 20 h \\ \bottomrule
    \multicolumn{3}{@{}p{0.5\textwidth}}{$^a$ All runtimes were checked on a single NVIDIA A100 GPU (80GB) and exclude data loading. Inference times refer to producing a high-resolution ensemble of 100 years for a single GCM-RCM pair with 10 samples per day.}
    \end{tabular}
    \label{tab:results:compute-times}
\end{table}

\ourmethod aims to be a computationally lightweight alternative to state-of-the-art generative models. As an example, we compare train and inference times of \ourmethod to the best-performing and most recent ML-baseline, CorrDiff (Tab.~\ref{tab:results:compute-times}). CorrDiff-s only adjusts the sampling scheme compared to CorrDiff, so train and inference times are identical. We find that training \ourmethod only takes about 1 day, whereas the diffusion model requires almost 10 times as long. The difference for inference is even larger, with \ourmethod being around 20 times faster. However, it should be noted that inference times can likely be improved for both models. For \ourmethod, the sampling can be done in a few minutes, but the current bottleneck is the inversion of the data-transformation. \ourmethodtemp is only slightly more computationally intensive than \ourmethod. Note that, in the current implementation, we also use \ourmethod to generate a starting value for \ourmethodtemp, hence, sampling from \ourmethodtemp requires training both \ourmethod and \ourmethodtemp.

Train times depend on the batch size and stopping criterion. We train all models until convergence, i.e. until the test loss flattens. For CorrDiff, the deterministic UNet and the stochastic residual model both are trained with a batch size of 32. Training the residual model is more expensive and takes around 0.8s/batch or, equivalently, 38 samples/second. During training of the GAN, we use a batch size of 16, and the model takes around 3s/batch or 5 samples/second. For \ourmethod, the last super-resolution step is the most computationally intensive. It is trained with a batch size of 128, taking around 0.2s/batch or 581 samples/second.

\ourmethod is parameter-efficient due to its sparse layers (Sec.~\ref{sec:methods:architecture}). Despite modeling effects specific to each location, the network only has 63 million parameters. \ourmethodtemp needs an additional 3 million parameters, yielding 66 million in total. The UNet in our CorrDiff implementation has around 80 million parameters (as in \citeA{Mardani2025ResidualDownscaling}) and is used in both the regression as well as the residual modeling step, yielding a total of around 160 million parameters. The GAN has around 69 million parameters.

Computational efficiency gains for \ourmethod stem from both the overall setup and the model architecture. The sparse local layers were designed for progressive, stochastic resolution refinement, are effective when combined with the energy score loss and our multi-step setup (Sec. \ref{sec:setup:2step-marginal}), and are key to achieving the short training time. For example, training a fully convolutional architecture for the last steps of the \textit{super-resolution model} took several days.

\section{Variability in \ourmethod's samples}
\label{sec:results:variability}

The conditional variance for the full GCM-RCM path is large, as can be seen from diversity in the emulator's sampled fields (Fig.~\ref{fig:results:samples-all}). This variability arises from internal variability in the RCM given the GCM's boundary conditions (see Sec.~\ref{sec:results:example-samples}). As each RCM downscales each GCM input only once, we cannot quantify this variability in the RCM data directly. Instead, we use \ourmethod to analyze this spread. Firstly, we split the variance of the full path from GCM to RCM into variance on coarse scales (path from GCM to coarsened RCM) and on local scales (path from coarsened RCM to RCM). Secondly, we compare internal variability with structural differences between RCMs.

\subsection{Coarse-scale vs. local-scale variability}

\label{sec:results:variability:coarse-local}

\begin{table}[]
\caption{Variance decomposition coarse scale vs high-resolution scale$^a$}
    \centering
    \begin{tabular}{lrrr}
\toprule
Variable & Coarse-scale variance & Local-scale variance & Ratio (coarse/local) \\
\midrule
tas & 3.72 & 0.24 & 15.61 \\
pr & 15.63 & 8.63 & 1.86 \\
sfcWind & 1.82 & 0.30 & 6.06 \\
rsds & 1307.13 & 287.18 & 4.55 \\
\bottomrule
\multicolumn{4}{@{}p{0.8\textwidth}}{$^a$We split the full variance in \ourmethod's samples into a contribution stemming from the variance of the \textit{coarse model} (first column), and one of the \textit{super-resolution model} (second column). The decomposition is formalized in Eq. \eqref{eq:variability:coarse-local} in the methods. The last column shows the ratio of coarse / local, indicating the level of variance that is present on the coarse compared to the high-resolution scale. Variances are given in the square of the native units ($[\mathrm{K}^2]$ for temperature, $[\mathrm{(mm/day)}^2]$ for precipitation, $[\mathrm{(m/s)}^2]$ for surface wind, $[\mathrm{(W/m^2)}^2]$ for solar radiation), and the ratio is dimensionless.}
\end{tabular}

    \label{tab:variability:coarse-super}
\end{table}

\ourmethod naturally allows us to separate sources of variability on different scales, as it splits the full GCM-RCM map into two steps. First, the \textit{coarse model} maps GCM predictors to average-pooled RCM data and second, the \textit{super-resolution model} learns the distribution of the high-resolution target given the average-pooled predictors (Fig.~\ref{fig:methods:2step}). This means we can split the full variability into contributions from the two parts: (i) coarse-scale variability, quantifying uncertainties in the pooled RCM data given GCM inputs, and (ii) local-scale variability, capturing variance in high-resolution RCM fields when conditioned on their coarse averages (see Sec.~\ref{sec:methods:variability:coarse-super} for the formal variance decomposition).

For a first qualitative insight, we present samples of the pure \ourmethod's \textit{super-resolution model}, where average-pooled versions of the target RCM data are used as inputs and the coarse correction step is bypassed. In this setup, the samples agree very closely with the target (see Fig.~\ref{fig:results:variability:pure-super}), and show much smaller variability than the samples from \ourmethod's full GCM-RCM map (as shown in Fig.~\ref{fig:results:samples-all}). The different samples from the \textit{super-resolution model} show only small discrepancies on a local scale and variability is greatly reduced in this case compared to the full GCM-RCM model. This highlights that the full RCM emulation is a very different task from a pure super-resolution setup. In addition, it indicates that variability between GCM and RCM on coarse scales is a main source of the mismatches that we observed in samples by \ourmethod (see Fig.~\ref{fig:results:samples-all}) and benchmarks (see Fig.~\ref{fig:appendix:analogues} -- \ref{fig:appendix:gan}).

The quantitative result of the decomposition is presented in Tab.~\ref{tab:variability:coarse-super}, for each variable separately, averaged over locations. For all variables, the ratio of coarse-scale to local-scale variance greatly exceeds 1, indicating that most variability stems from the transfer of GCMs to average-pooled RCMs. In other words, the GCM boundary conditions contain comparatively little information about the coarsened RCM data, whereas the coarsened RCM already well determines the high-resolution output. We suspect that this is because the GCM only drives the RCM at the boundaries, leaving a wide range of plausible climate conditions inside the RCM's domain.

The ratio is between 2.0 and 16.8 for the different variables. It is largest for temperature and wind and smallest for precipitation. For wind, the high ratio is primarily due to large-scale variability over sea areas, where local variance is small. In contrast, precipitation exhibits greater local-scale variability, particularly in the Alpine regions. Since these small-scale effects cannot be fully captured by the average-pooled data $Z$, the contribution of local-scale variance remains relatively large, leading to a lower overall ratio.

Individual variances can be interpreted. We consider temperature, which is approximately normally distributed. Calculating $\pm 2 \sqrt{\var}$ gives the range around the mean in which approx. 95 \% of data points lie. For example, for the local-scale variance $\pm 2 \sqrt{\var} = \pm 0.98$. That means, after fixing the average-pooled RCM data, local temperatures fluctuate by around $\pm$ \SI{1}{K} in each grid point. Conditioning on the GCM input, however, temperatures on coarse scales can still vary by about $\pm$ \SI{3.7}{K}.

\subsection{Internal variability vs. model differences}

\begin{table}[]
\caption{Variance decomposition for different RCMs with the same driving GCM$^a$}
    \centering
    \begin{tabular}{lrrr}
\toprule
Variable & Variance of RCM & Model difference & Ratio (Variance / model diff.) \\
\midrule
tas & 4.10 & 0.44 & 9.41 \\
pr & 27.03 & 1.27 & 21.29 \\
sfcWind & 2.05 & 0.15 & 13.60 \\
rsds & 1745.99 & 303.14 & 5.76 \\
\bottomrule
\multicolumn{4}{@{}p{0.8\textwidth}}{$^a$Using \ourmethod, we compare variance within samples for individual RCMs to the structural difference between RCMs, following the decomposition derived in Eq. \eqref{eq:variability:internal-model-diff} in the methods. Variances are expressed conditional on GCM inputs $X$, and averaged over all inputs in the test set. We use \ourmethod's samples for ALADIN63 as a baseline run and compare to samples for CCLM4-8-17 (both driven by CNRM-CM5). The last column shows the ratio the variance of RCMs vs. the model differences. The units for the variance and model differences are again squares of the native units, as in Tab.~\ref{tab:variability:coarse-super}, and the ratio is dimensionless.}
\end{tabular}

    \label{tab:variability:groups}
\end{table}

We use \ourmethod to analyze internal variability relative to model disagreements. Recall that \ourmethod emulates the distribution of each RCM specifically ( see Sec.~\ref{sec:setup:2step-marginal}). We compare variability within \ourmethod's samples for individual RCMs to differences between samples for two RCMs, as described in Sec.~\ref{sec:methods:variability:groups}, in particular Eq.~\eqref{eq:variability:internal-model-diff}. Tab.~\ref{tab:variability:groups} presents the results. 

We find that variances within samples for an individual RCM largely dominate over differences between samples for two distinct RCMs (i.e.,~ratio larger than 1). The ratio is largest for wind, where the variance is high for \ourmethod's samples for each RCM alone already. We show results for samples for two RCMs only, both with the same GCM input. Overall findings are similar also for other pairs.

The variances here can be interpreted as well, similar to the discussion in Sec.~\ref{sec:results:variability:coarse-local}. For example, for temperature, the variance of one RCM is \SI{4.1}{K^2}. That means, for a fixed GCM input, temperatures simulated by the RCM in each location can fluctuate by $\pm \SI{4.0}{K}$ in an approximate 95 \% confidence interval (based on an approximation with a normal distribution, and calculated as above as $2 \sqrt{\var}$). However, the disagreements between RCMs are much smaller than that. In each location, the means of the RCMs differ by about \SI{1.3}{K}, when conditioning on a GCM input (calculated as $2 \sqrt{\text{model diff.}}$, compare with the formula \eqref{eq:variability:internal-model-diff} in Sec.~\ref{sec:methods:variability:groups}).

This finding points to the fact that internal variability seems to dominate over structural model differences for regional climate model ensembles. It is similar to \citeA{Rampal2025DownscalingExtremes}, who use an RCM-emulator to present an uncertainty quantification into internal variability, scenario uncertainty and model uncertainty and also find a prominent role of internal variability. In addition, our variance decomposition underlines the benefits of jointly training one ML model for downscaling based on multiple GCM-RCM combinations. Even though inter-model differences can be rather large, intrinsic conditional variance of $Y|X$ dominates. Hence, there is sufficient similarity between GCM-RCM pairs such that the ML training can benefit from a joint training.

\section[Discussion]{Discussion}
\label{sec:discussion}
In this study, we present \ourmethod, a method for emulating RCMs. We compare \ourmethod to various benchmarks, both physically-motivated (analogues) and machine learning methods (GAN and a diffusion model, CorrDiff(-s)). We find that \ourmethod achieves competitive performance compared to the state-of-the-art diffusion model w.r.t. a wide range of metrics, both when evaluated per grid point and per variable separately, as well as for spatial structure and dependencies between variables. To our knowledge, \ourmethodtemp is the first downscaling approach to model temporal dependencies in the high-resolution output, approximating temporal consistency of dynamical downscaling. This improves the temporal consistency of the output substantially compared to all other benchmarks.

Compared with physics-based climate models as well as state-of-the-art machine learning models, our emulator is highly computationally efficient. Training \ourmethod completes in approximately one day. Afterwards, generating a downscaled ensemble with ten members for 100 years of GCM data is completed in under two hours. CorrDiff instead needs about 10 days for training and 36 hours for inference. Running the RCM for the same period would typically take weeks and still yield only one ensemble member. The sparse architecture in \ourmethod is efficient and scalable also for larger domains and more climate variables.

Our work presented a comprehensive evaluation framework of generative downscaling approaches for calibration, spatial and temporal structure, extremes, and multivariate dependencies. Notably, evaluating such methods is not a straightforward task. The output comprises many locations, variables and time steps, and capturing all these aspects is challenging \cite{Maraun2015VALUEStudies}. Model evaluation and statements about skill are necessarily specific for each question and application. So far, our evaluation methods are primarily statistically motivated and inspired by the field of forecast evaluation. There remain open questions, in particular for evaluating the dependencies for many variables at once and for the temporal structure. Evaluating emulators in application-relevant case studies is an interesting avenue of further research \cite{Isphording2024ASimulations}.

We found large conditional variability in RCM fields given GCM inputs, arising from internal variability within RCMs. This highlights the value of stochastic emulators: by generating an ensemble of many RCM trajectories for one given driving GCM run, they can capture this internal variability and characterize its spread. This can enable better uncertainty quantification in high-resolution predictions, analogous to the benefits of large ensembles from global climate models \cite{Deser2020InsightsProspects, Rampal2025DownscalingExtremes}.

Our downscaling via coarse correction is similar to some existing works: \citeA{Price2022IncreasingModels} also split downscaling of precipitation with a GAN into two steps: a coarse correction step and a pure super-resolution step. The target is radar measurements. Their setup is simpler than ours because they only use data from one numerical model, but still they find that isolating the correction step in a low-dimensional task improves the results. \citeA{Wan2024StatisticalModels} downscale data from a single GCM stochastically, considering historical GCM data and using ERA5 reanalysis as a target. Hence, their first step corrects large-scale mismatches between GCMs and ERA5. In a similar fashion, CorrDiff's mean-residual split (Sec.~\ref{sec:methods:corrdiff}) corrects some mismatches between input and target data in a simpler task (the mean prediction, first step), before learning the correct variability (residual prediction, second step). 

We trained our method on a combined dataset, comprising multiple GCMs and RCMs, while conditioning on the RCM in question. This allows the model to benefit from a large dataset and learn shared properties. However, a potential limitation is that the model might regress toward the multi-model mean, underestimating differences among RCMs and failing to capture all model-specific details.

We choose to learn the full path from GCM to RCM, which differs from RCM-emulation in \citeA{Doury2024OnPrecipitation, Doury2022}. Previous work argues that emulators which learn a map from GCM to RCM data might pick up patterns specific to each GCM-RCM pair, and are thus less transferable to other GCMs \cite{Hernanz2023OnIssues, Rampal2024EnhancingLearning, Bano-Medina2023TransferabilityApplications}. However, our setup is different because we (i) train with multiple GCM-RCM pairs, possibly enabling the model to learn shared properties across different climate models, and (ii) train a generative model rather than a deterministic one.

\ourmethod can readily be applied to emulate the three RCMs that were used during training for any GCM input. It can be used to downscale data from CMIP6 for the three GCMs in the train data \cite{Effenberger2025BridgingEurope}. As a preliminary result, we also tested extrapolation to two unseen GCMs, HadGEM2-ES and CanESM2, and present the results in Sec.~\ref{sec:appendix:gcm-extrapolation}. For most variables, scores on the unseen GCMs are within the range of scores on the training pairs. Temperature is the only exception, where scores on new GCMs are higher than scores on all GCM-RCM pairs in the training data. We find that, even within the training dataset, scores can differ substantially both between different driving GCMs as well as between different RCMs for the same driving GCM. Hence, studying these differences is an interesting direction of future research to assess the extrapolation scores more rigorously. In summary, however, these results show promising capabilities of \ourmethod and suggest that the learned map can be generalized across GCMs.

\ourmethod leverages both past and future data, enabling it to learn future relationships between low-resolution predictors and high-resolution targets directly from physics-based models. In contrast, model training in perfect prognosis downscaling relies on historical data and aims to extrapolate these learned relationships to future conditions. RCM emulation like in \ourmethod can readily be combined with a state-of-the art bias-correction method to obtain ``future observations''. A promising direction for further development is to explore more sophisticated bias correction algorithms that preserve multivariate structure across time, space, and several variables or integrate observational data also in model training \cite{Hess2025FastLearning, Schmidt2025ASimulation}.

\ourmethod is computationally efficient and can be re-trained also on other datasets from other GCMs and RCMs or for other regions. Our sparse architecture would allow \ourmethod to scale to larger domains while still requiring only a reasonable amount of memory. In addition, inference times can be decreased even further with an adjusted data transformation. Also, our emulator could be used to downscale to a higher resolution, for example approximating convection-permitting models as a target. These benefits establish \ourmethod as a powerful tool for data-driven future climate projections on a local scale and open new opportunities for climate impact studies, where reliable high-resolution projections are essential.

\paragraph{Acknowledgments}
Maybritt Schillinger, Maxim Samarin, Nicolai Meinshausen, and Reto Knutti are part of SPEED2ZERO, a Joint Initiative co-financed by the ETH Board. In some instances, ChatGPT was used to rephrase sentences for clarity. We thank Ruth Lorenz and Urs Beyerle for their help with the data-preprocessing and Sebastian Sippel for valuable inputs during the early stages of this project. Further, we benefitted from many helpful and inspiring discussions with numerous people, among them were Antoine Doury, Jose González-Abad, Shahine Bouabid, Sam Allen, Anna Merrifield, Dominik Schumacher, Daniele Nerini, Niklas Boers and his group. We thank Nina Effenberger and Anya Fries for their constructive feedback on the draft which greatly improved the readability of this work. Last but not least, we thank three anonymous reviewers for their careful and constructive feedback on this manuscript.

\paragraph{Open Research Section}
All datasets are publicly available from the Earth System Grid Federation (ESGF) infrastructure, via \url{https://esgf-metagrid.cloud.dkrz.de/search}. Code is available at \url{https://github.com/m-schillinger/enscale}. The repository will be updated to include the final version of the code upon publication.

\appendix
\counterwithin{figure}{section} % reset figure numbering per section
\counterwithin{table}{section}
\clearpage

\clearpage
\section{Additional results}

\subsection{Further samples}
\begin{figure}[H]
	\centering
	\includegraphics[width=0.65\linewidth]{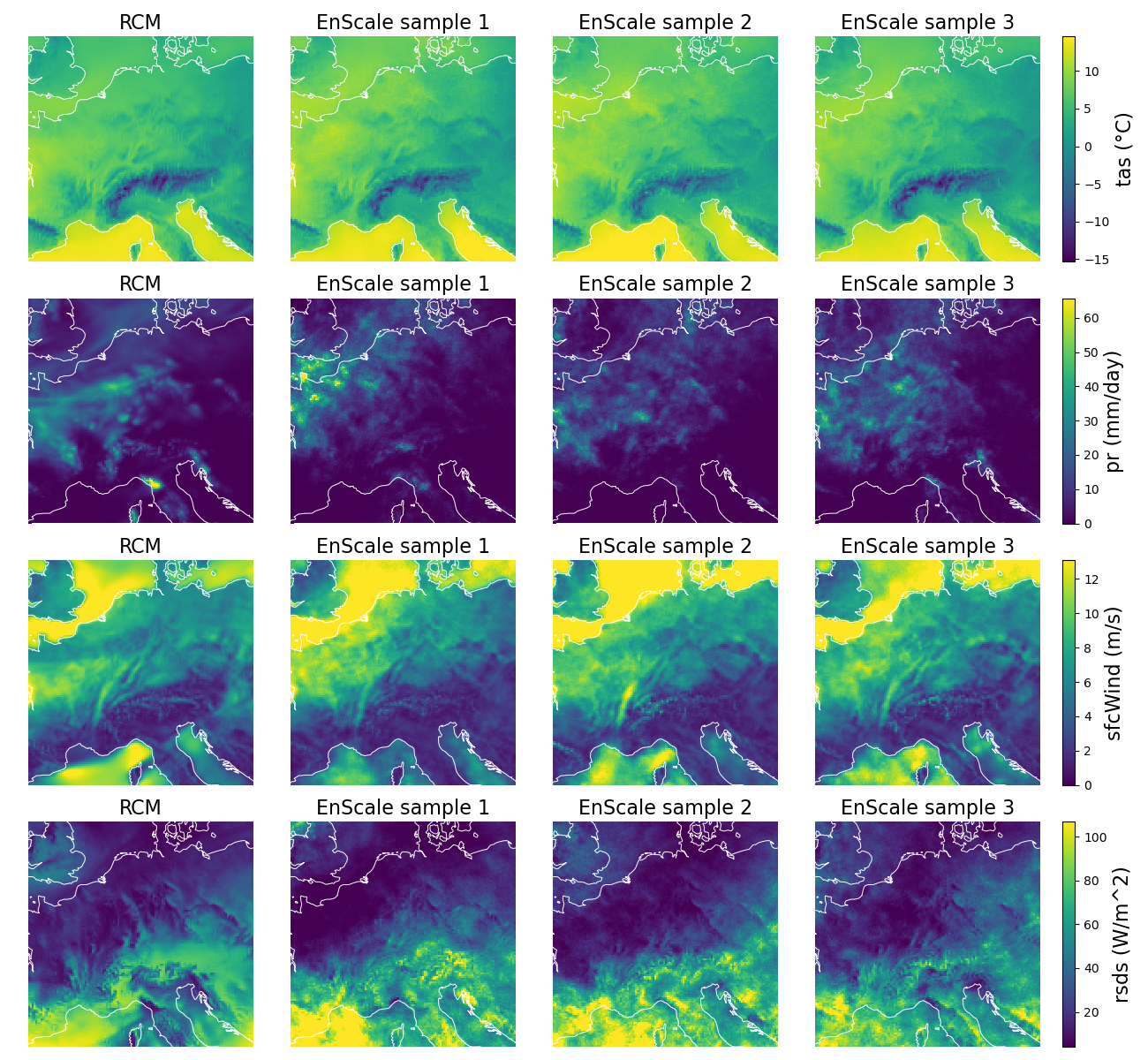}
	\caption{Examples for \ourmethod: Like Fig.~\ref{fig:results:samples-all}, but showing a day with particularly good scores (about 10-th percentile).}
	\label{fig:appendix:enscale-good}
\end{figure}

\begin{figure}[H]
	\centering
	\includegraphics[width=0.65\linewidth]{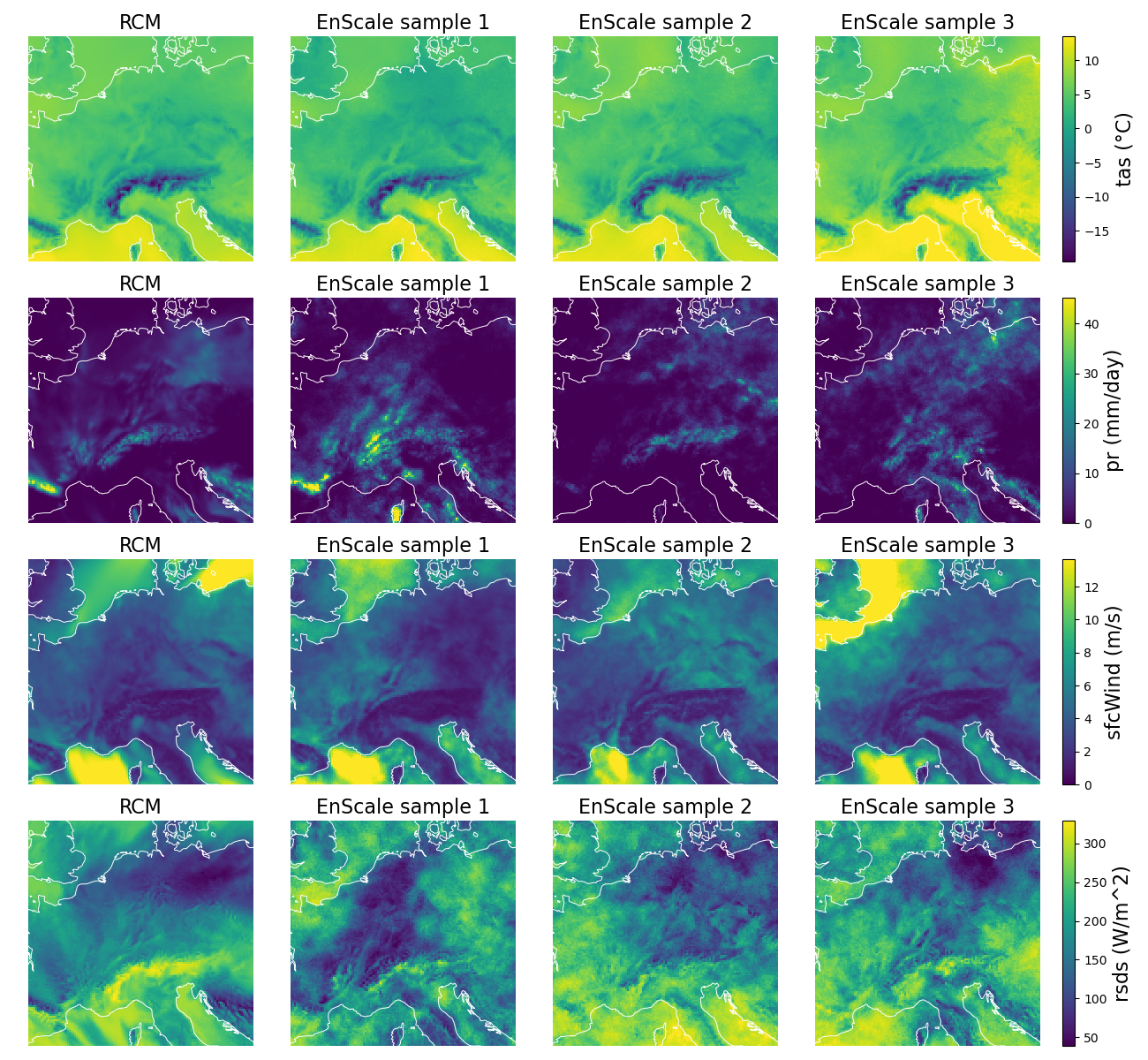}
	\caption{Examples for \ourmethod: Like Fig.~\ref{fig:results:samples-all}, but showing days with particularly bad scores (about 90-th percentile).}
	\label{fig:appendix:enscale-bad}
\end{figure}

\begin{figure}[H]
	\centering
	\includegraphics[width=0.65\linewidth]{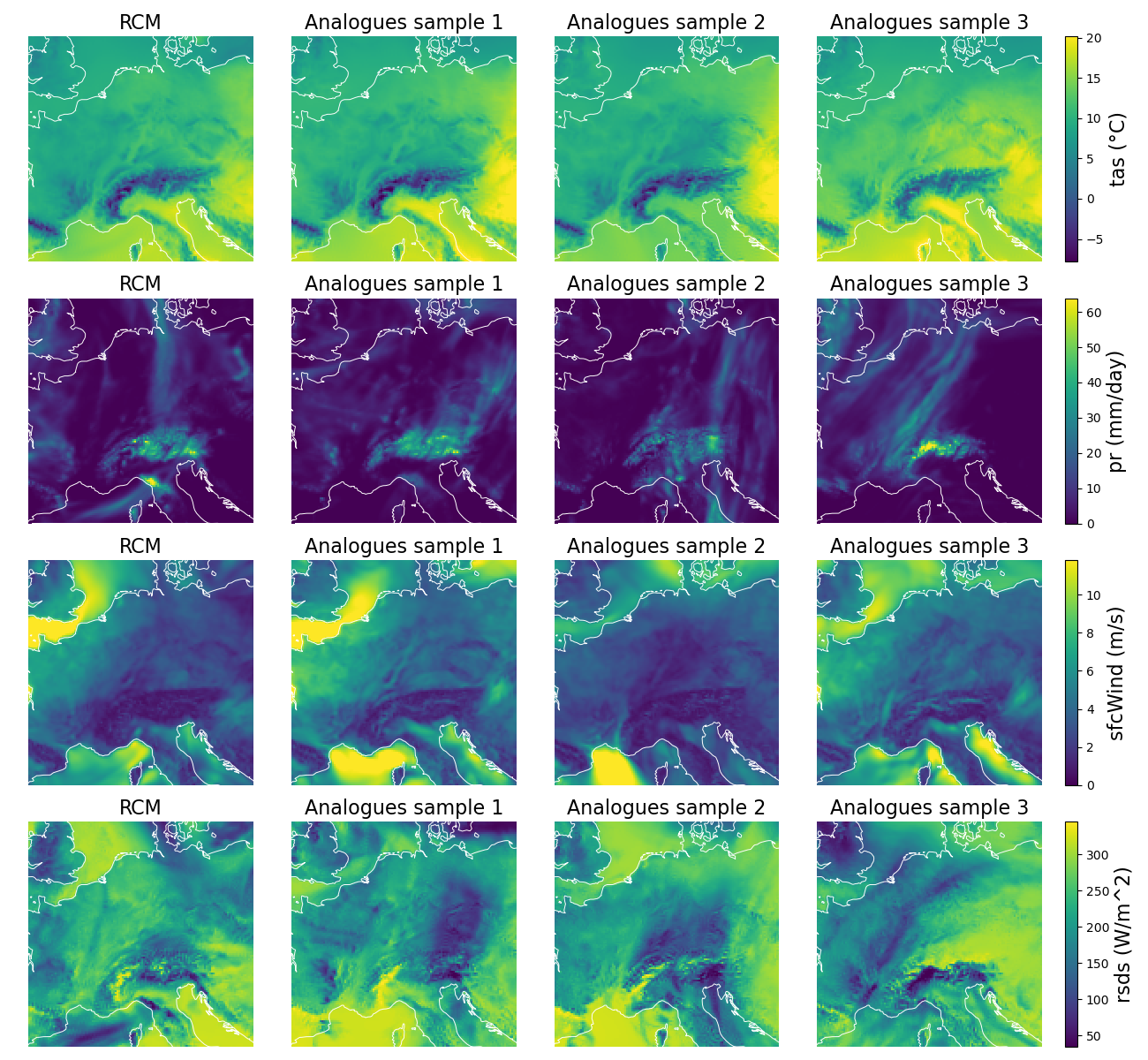}
	\caption{Examples for the analogues: Like Fig.~\ref{fig:results:samples-all}, but showing examples for the analogue benchmark.}
	\label{fig:appendix:analogues}
\end{figure}

\begin{figure}[H]
	\centering
	\includegraphics[width=0.65\linewidth]{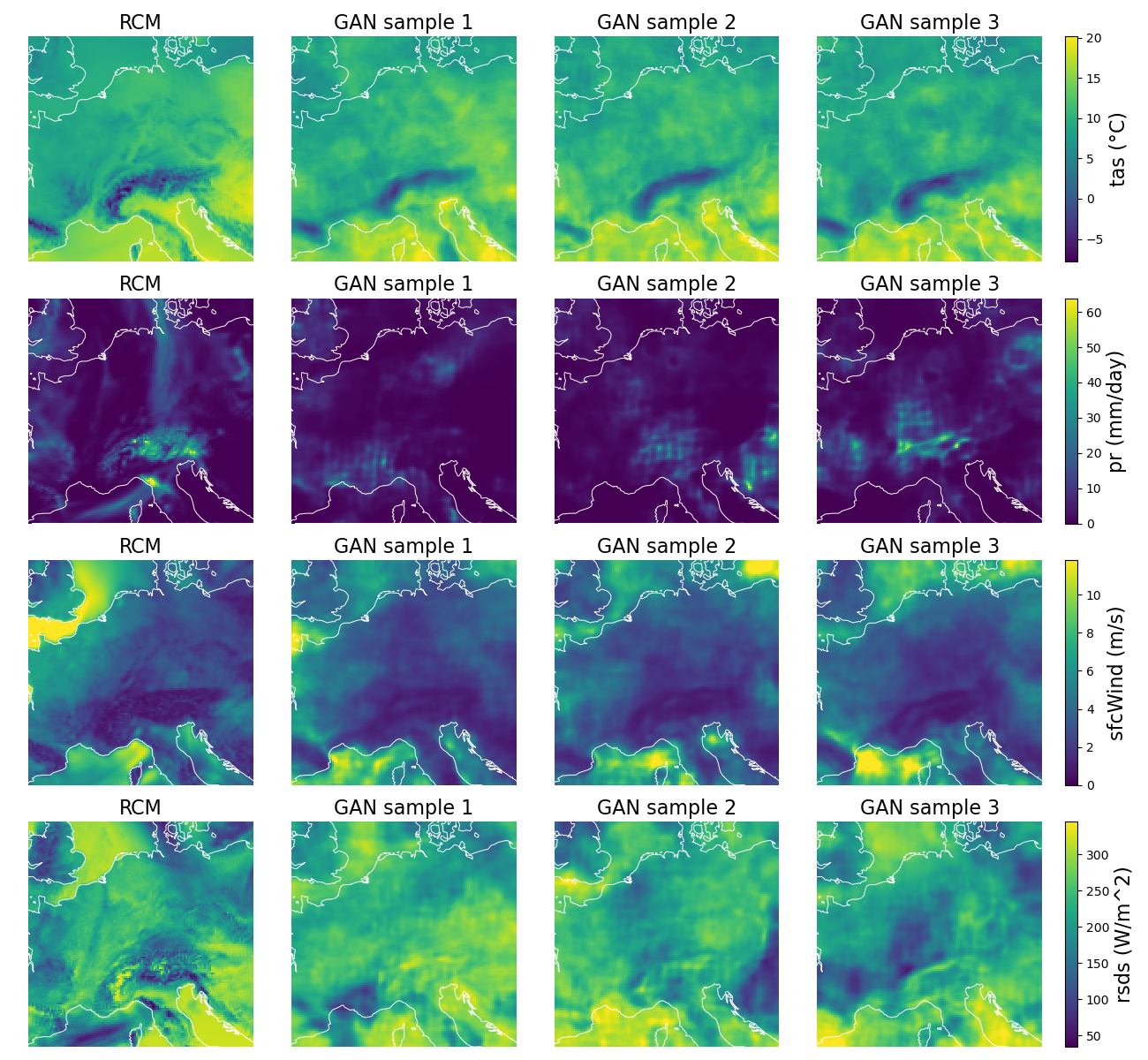}
	\caption{Examples for the GAN: Like Fig.~\ref{fig:results:samples-all}, but showing examples for the GAN benchmark.}
	\label{fig:appendix:gan}
\end{figure}

\begin{figure}[H]
	\centering
	\includegraphics[width=0.65\linewidth]{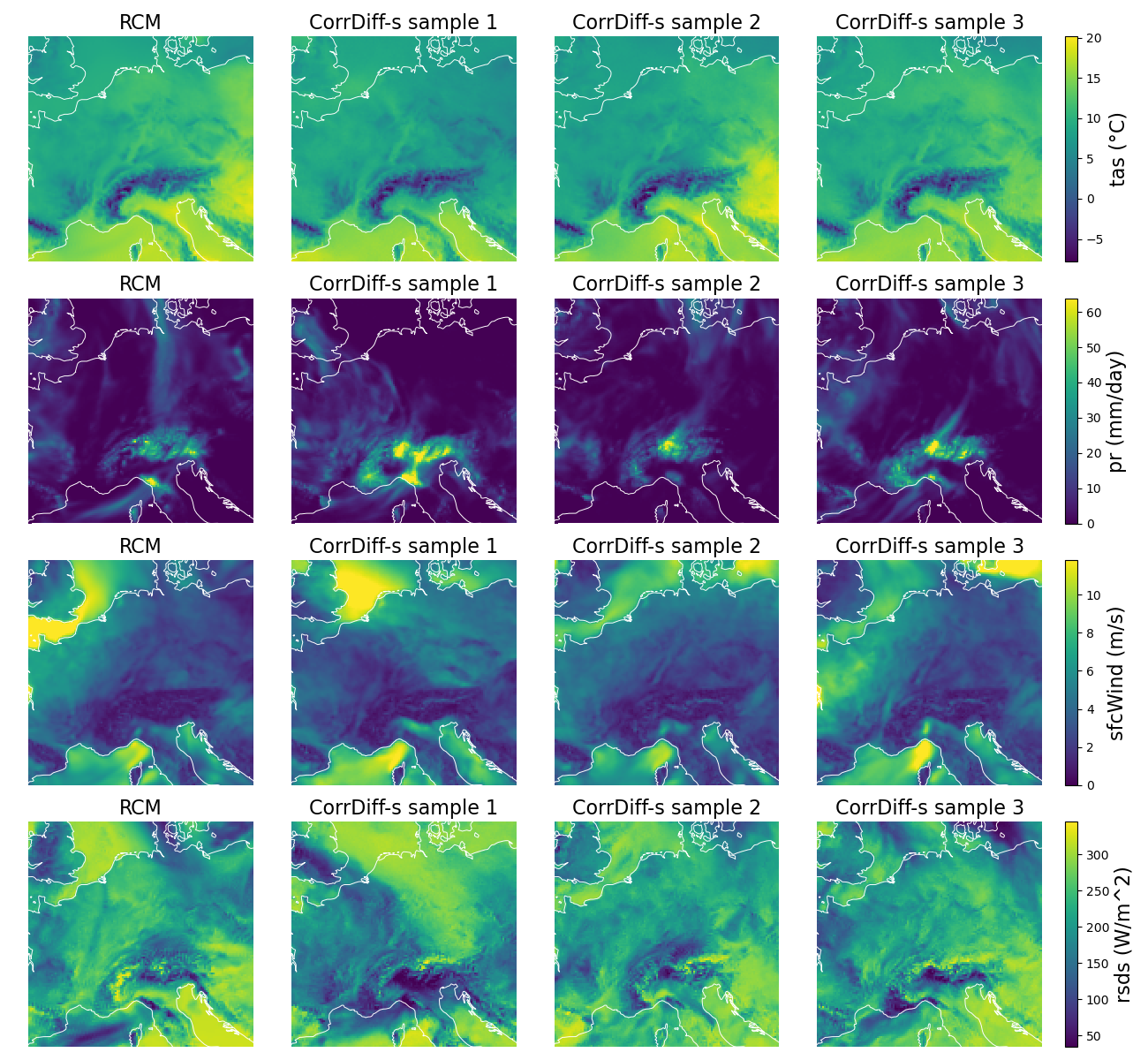}
	\caption{Examples for CorrDiff: Like Fig.~\ref{fig:results:samples-all}, but showing examples for the CorrDiff benchmark. These examples are representative for both CorrDiff and CorrDiff-s.}
	\label{fig:appendix:corrdiff}
\end{figure}

\begin{figure}[H]
	\centering
	\includegraphics[width=0.65\linewidth]{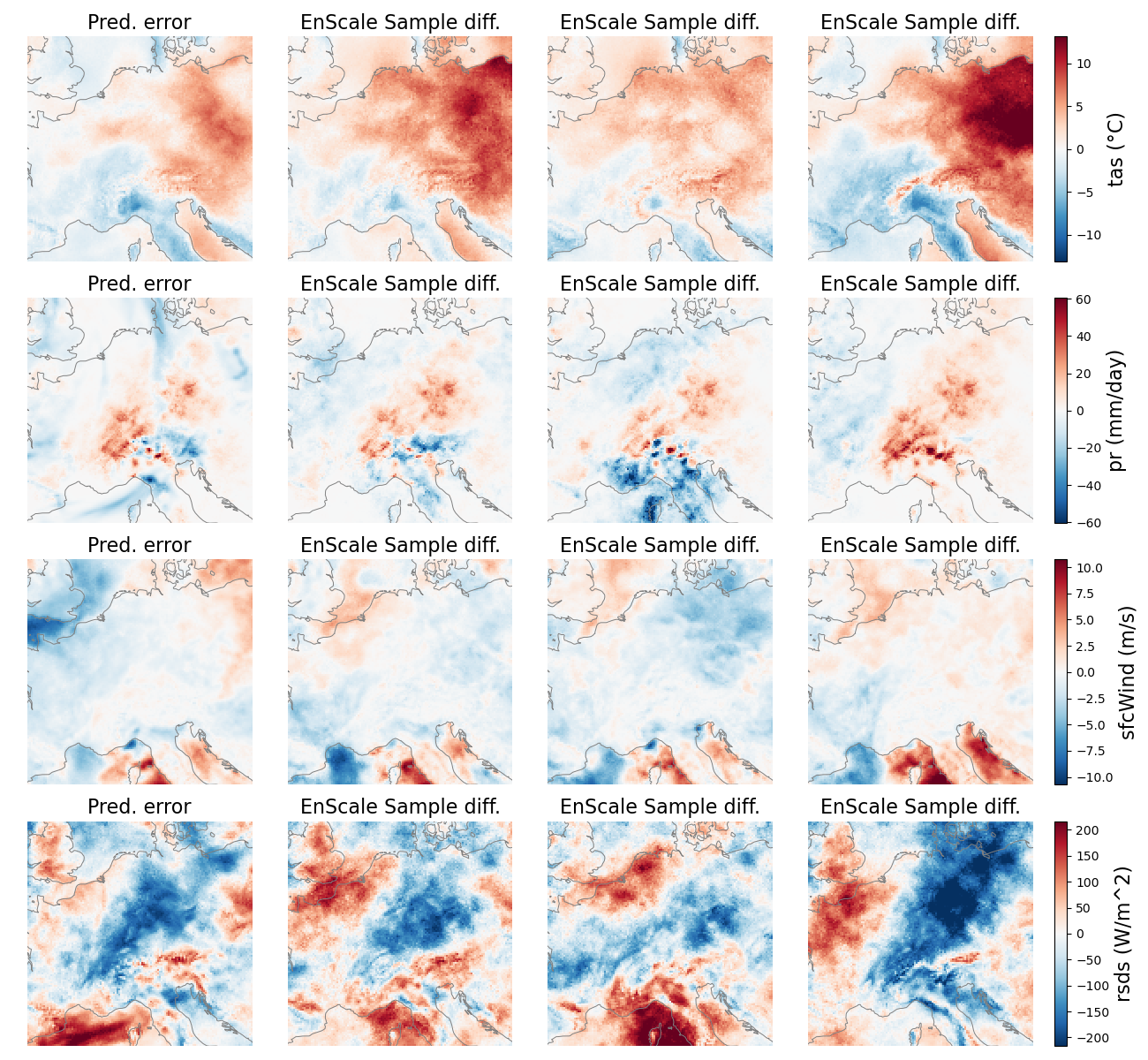}
	\caption{Balance of prediction error vs. variability for \ourmethod: The first column shows the prediction error for one of the samples from \ourmethod, i.e. the difference of the generated sample to the RCM truth. Columns 2-4 show the variability, i.e. difference between three pairs of \ourmethod-samples.}
	\label{fig:appendix:enscale-pred-diff}
\end{figure}

\begin{figure}[H]
	\centering
	\includegraphics[width=0.65\linewidth]{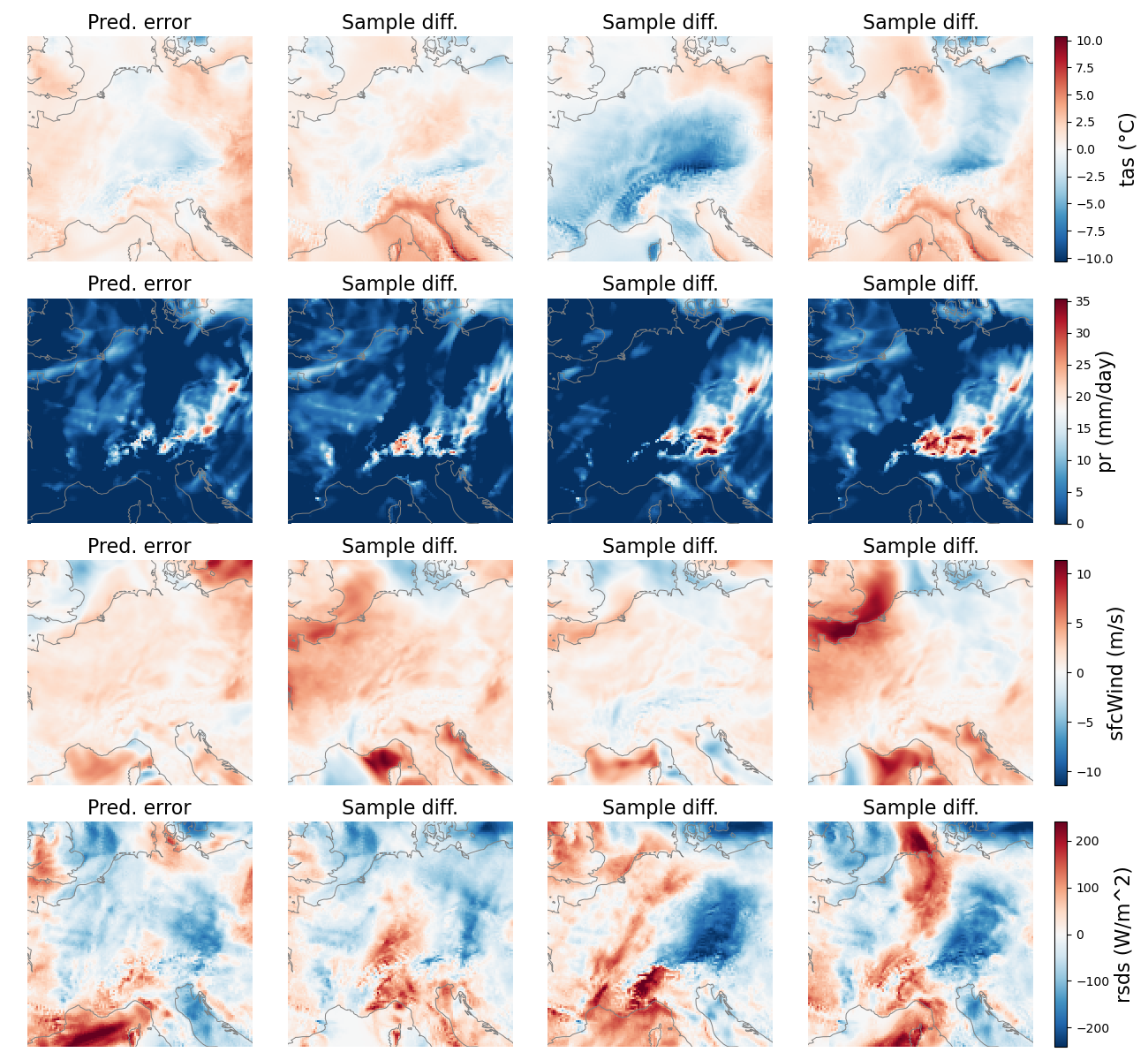}
	\caption{Like Fig.~\ref{fig:appendix:analogues-pred-diff}, but for the analogue benchmark.}
	\label{fig:appendix:analogues-pred-diff}
\end{figure}

\begin{figure}[H]
	\centering
	\includegraphics[width=0.65\linewidth]{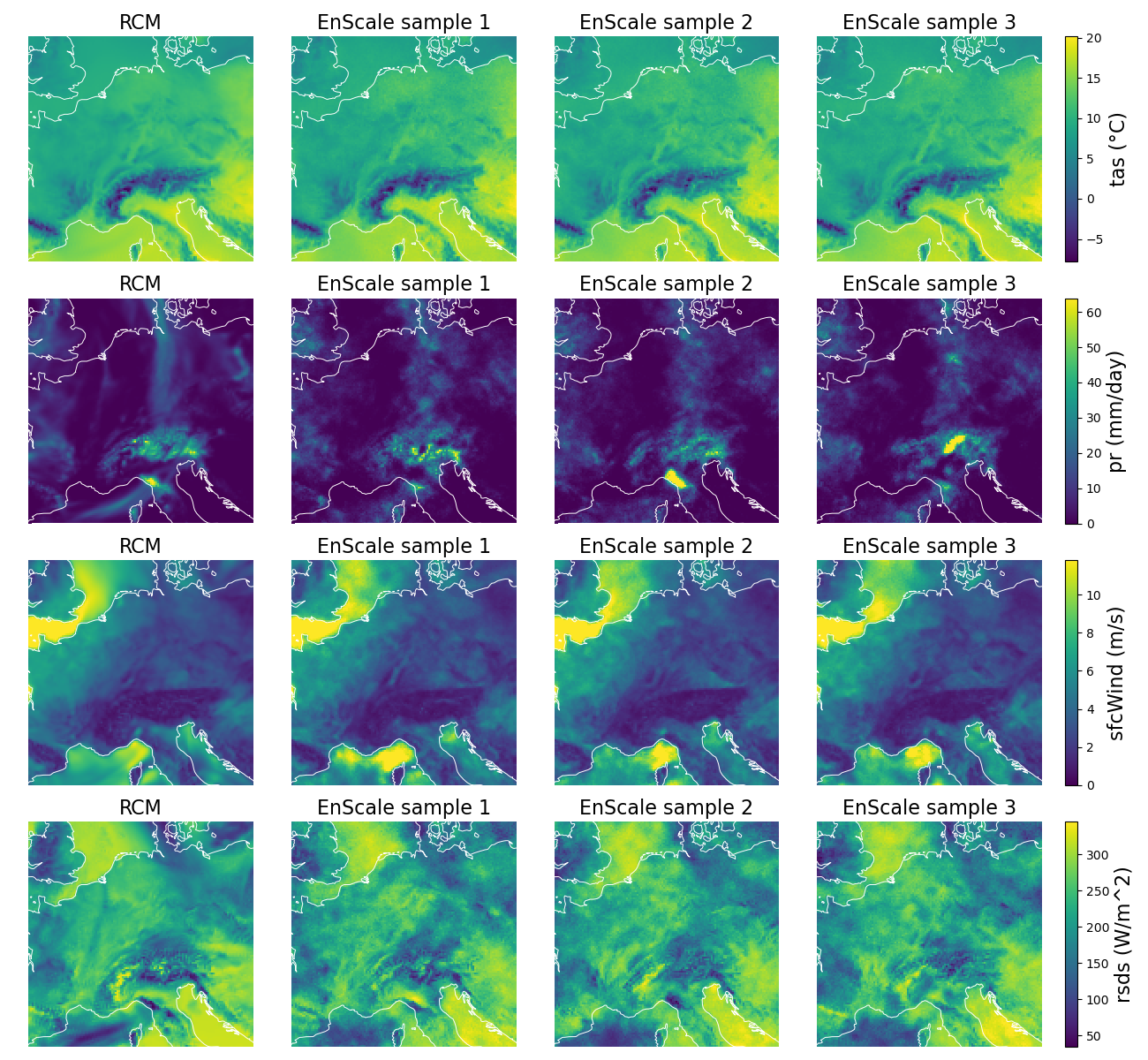}
	\caption{Examples for pure super-resolution map: Like Fig.~\ref{fig:results:samples-all}, but instead of presenting the full map from GCM data $X$ to RCM data $Y$, we show the results of only the super-resolution map from coarsened RCM data $Z$ to RCM $Y$.}
	\label{fig:results:variability:pure-super}
\end{figure}

We present further examples for \ourmethod in Fig.~\ref{fig:appendix:enscale-good} and Fig.~\ref{fig:appendix:enscale-bad}, for the benchmarks in Fig.~\ref{fig:appendix:analogues} -- Fig.~\ref{fig:appendix:corrdiff}. In addition, we present the balance of prediction error and variability in Fig.~\ref{fig:appendix:enscale-pred-diff} and  Fig.~\ref{fig:appendix:analogues-pred-diff}. Finally, Fig.~\ref{fig:results:variability:pure-super} presents results for \ourmethod's \textit{super-resolution model}.

\begin{figure}[H]
	\centering
	\includegraphics[width=0.65\linewidth]{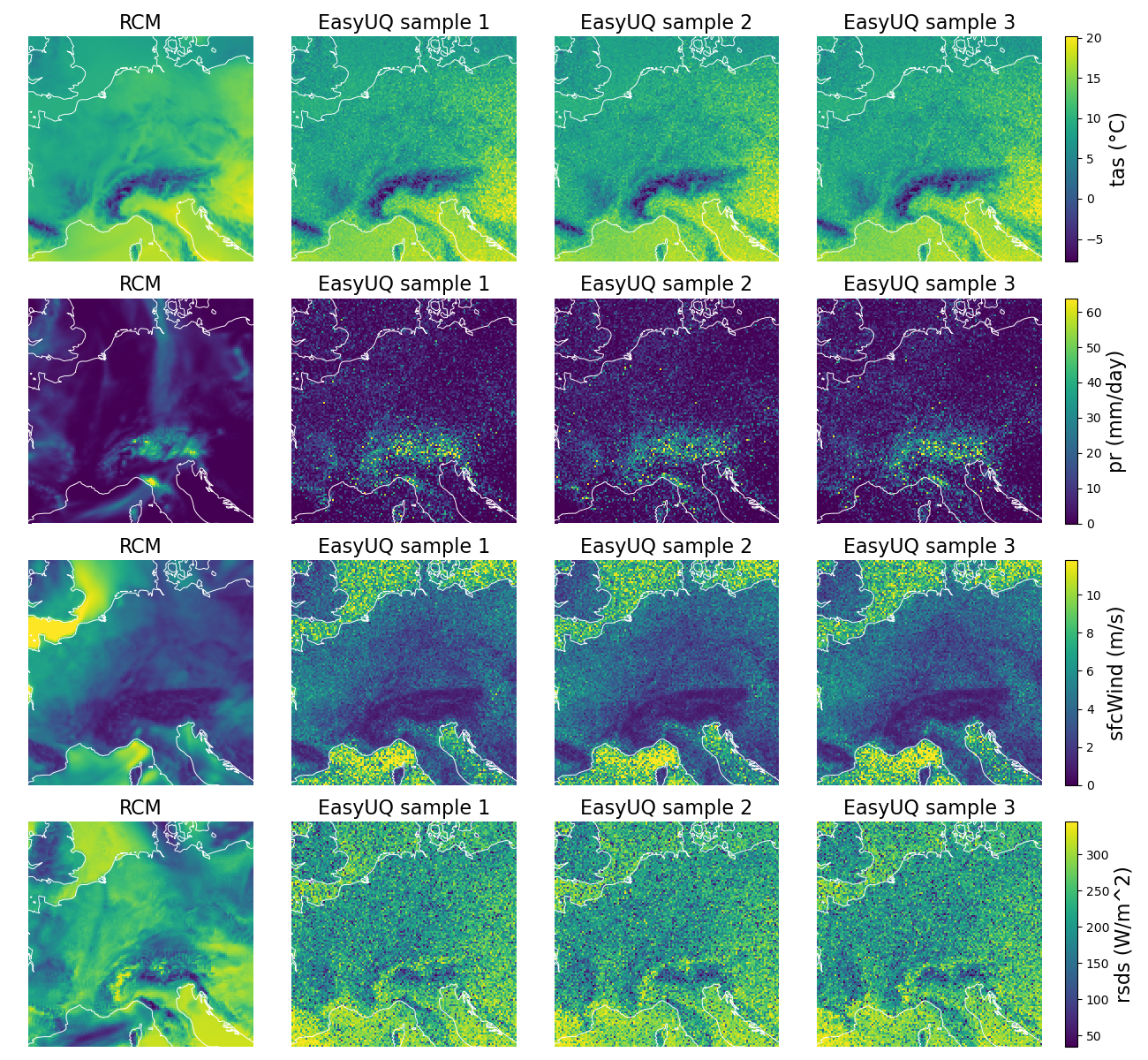}
	\caption{Examples for EasyUQ: Like Fig.~\ref{fig:results:samples-all}, but showing examples for the EasyUQ benchmark.}
	\label{fig:appendix:easyuq}
\end{figure}

\begin{figure}[H]
	\centering
	\includegraphics[width=0.35\linewidth]{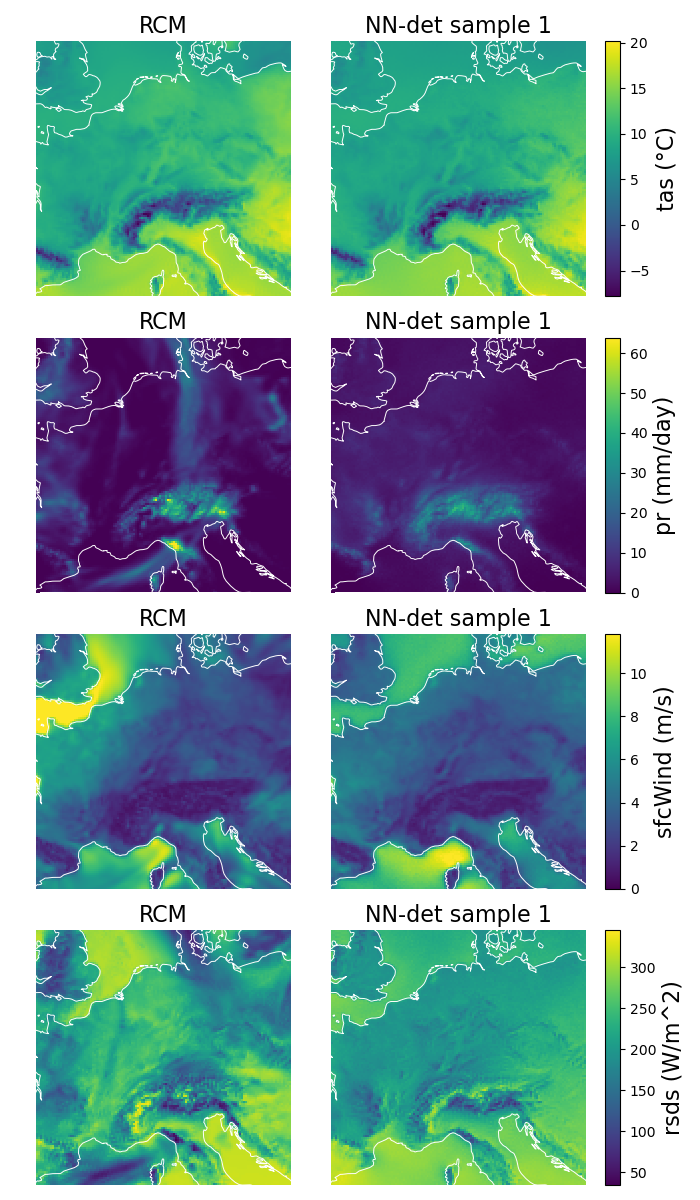}
	\caption{Examples for NN-det: Like Fig.~\ref{fig:results:samples-all}, but showing examples for the NN-det benchmark.}
	\label{fig:appendix:nndet}
\end{figure}

%\begin{figure}
%	\centering
%	\includegraphics[width=0.7\linewidth]{figures/enscale_examples_medium_2297_viridis_remo.png}
	%\caption{Examples for \ourmethod: Like Fig.~\ref{fig:results:samples-all}, but showing a different day from REMO2015, driven by MIROC5.}
	%\label{fig:appendix:enscale-remo}
%\end{figure}

\Needspace{5\baselineskip}

\clearpage
\subsection{Further evaluation of spatial structure}

% \begin{figure}[H]
%     \centering
%     \includegraphics[width=0.9\linewidth]{figures/psd_joint_twopanelvreview1.png}
%     \caption{Power spectral densities  \red{new}}
%     \label{fig:appendix:psd}
% \end{figure}

\begin{figure}[H]
    \centering
    \includegraphics[width=0.9\linewidth]{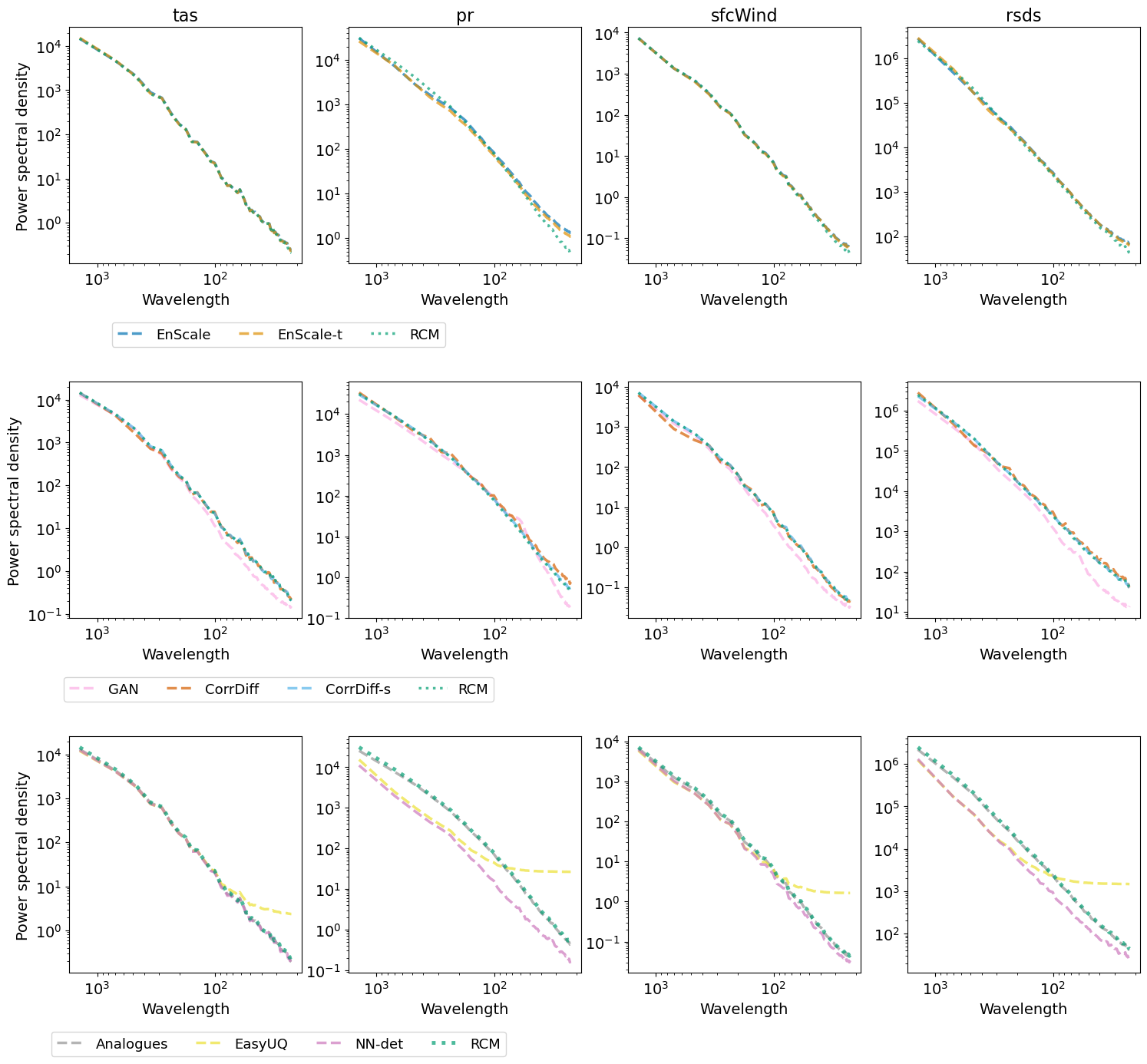}
    \caption{Power spectral densities}
    \label{fig:appendix:psd}
\end{figure}

% \begin{table}[H]
%     \caption{RALSD-avg \red{new}}
%     \centering
%     \input{tables/table_psd_vreview1}
%     \label{tab:appendix:lsd-avg}
% \end{table}

\begin{table}[H]
    \caption{RALSD-avg}
    \centering
    \begin{tabular}{lllll}
\toprule
Method & tas & pr & sfcWind & rsds \\
\midrule
NN-det & 0.465 & 8.600 & 1.390 & 4.048 \\
EasyUQ & 5.959 & 7.614 & 8.503 & 9.969 \\
Analogues & \underline{0.150} & \textbf{0.342} & \textbf{0.152} & \textbf{0.169} \\
GAN & 1.999 & 6.894 & 3.185 & 4.066 \\
CorrDiff & 0.393 & 0.948 & 0.432 & 0.751 \\
CorrDiff-s & \textbf{0.142} & \underline{0.526} & \underline{0.309} & \underline{0.342} \\
EnScale & 0.517 & 1.299 & 0.659 & 1.251 \\
EnScale-t & 0.405 & 1.443 & 0.573 & 1.112 \\
\bottomrule
\end{tabular}

    \label{tab:appendix:lsd-avg}
\end{table}

\Needspace{5\baselineskip}

\subsection{Histograms for benchmark methods}

\begin{figure}
    \centering
    \includegraphics[width=0.99\linewidth]{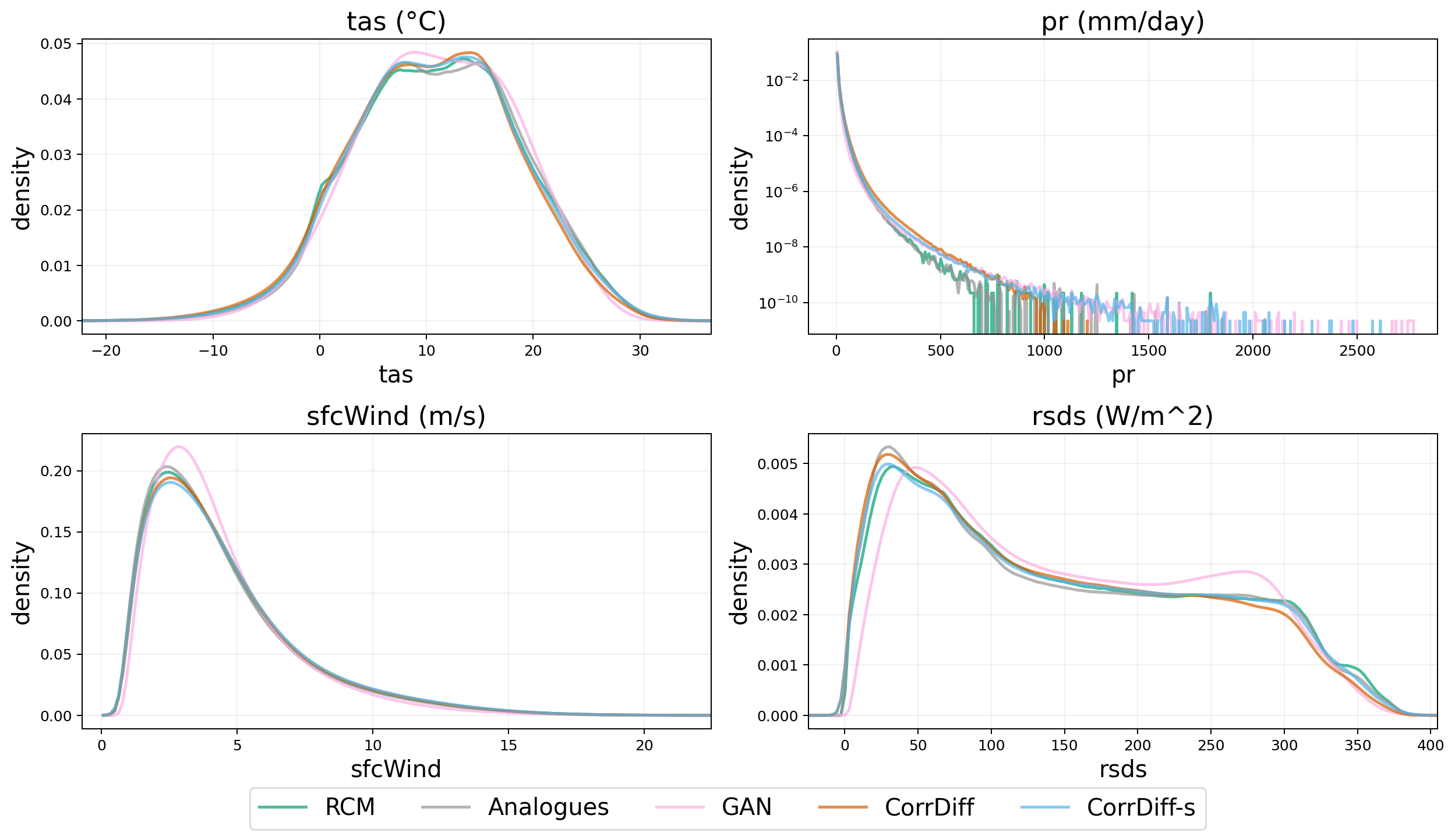}
    \caption{Histograms for the benchmark methods. As Fig.~\ref{fig:results:histograms}, but showing a selection of benchmarks.}
    \label{fig:results:histograms-benchmarks}
\end{figure}

\clearpage
\subsection{Further evaluation of extremes}

% \begin{table*}
%     \caption{Extreme evaluation regarding marginal quantiles as well as miscalibration$^a$ \red{new}
%     }
%      \centering
% 	\input{tables/table_quantiles_with_mcb_vreview1_99}
% 	\label{tab:results:quantiles}
% \end{table*}

\begin{table*}
    \caption{Extreme evaluation regarding marginal quantiles as well as miscalibration$^a$
    }
     \centering
	\begin{tabular}{lllll|llll}
\midrule
Method&\multicolumn{4}{c}{tas}&\multicolumn{3}{|c}{pr}\\
\midrule
 & AE $Q^{0.01}_{\text{Wi}}\downarrow$ & AE $Q^{0.99}_{\text{Su}}\downarrow$ & MCB $Q^{0.1}$ $\downarrow$ & MCB $Q^{0.9}$ $\downarrow$ & AE $Q^{0.01}_{\text{Wi}}\downarrow$ & AE $Q^{0.99}_{\text{Su}}\downarrow$ & MCB $Q^{0.1}$ $\downarrow$ & MCB $Q^{0.9}$ $\downarrow$ \\
\midrule
NN-det & 2.005 & 1.366 & N/A & N/A & 0.001 & 16.963 & N/A & N/A \\
EasyUQ & 1.094 & \textbf{0.664} & 0.060 & 0.064 & \underline{0.000} & 6.362 & \textbf{0.028} & 0.039 \\
Analogues & 1.131 & \underline{0.708} & 0.103 & 0.054 & \textbf{0.000} & \underline{3.852} & 0.068 & 0.085 \\
GAN & 2.484 & 1.816 & 0.097 & 0.063 & 0.003 & 10.481 & 0.274 & 0.038 \\
CorrDiff & 1.478 & 1.228 & 0.056 & 0.050 & 0.071 & 10.345 & 0.088 & 0.047 \\
CorrDiff-s & \textbf{0.825} & 0.771 & \textbf{0.017} & \textbf{0.019} & 0.090 & \textbf{3.574} & \underline{0.030} & \textbf{0.013} \\
EnScale & 1.032 & 0.727 & 0.018 & 0.026 & 0.000 & 7.972 & 0.047 & \underline{0.014} \\
EnScale-t & \underline{0.982} & 1.036 & \underline{0.017} & \underline{0.022} & 0.000 & 7.472 & 0.048 & 0.017 \\
\midrule
Method&\multicolumn{4}{c}{sfcWind}&\multicolumn{3}{|c}{rsds}\\
\midrule
 & AE $Q^{0.01}_{\text{Wi}}\downarrow$ & AE $Q^{0.99}_{\text{Su}}\downarrow$ & MCB $Q^{0.1}$ $\downarrow$ & MCB $Q^{0.9}$ $\downarrow$ & AE $Q^{0.01}_{\text{Wi}}\downarrow$ & AE $Q^{0.99}_{\text{Su}}\downarrow$ & MCB $Q^{0.1}$ $\downarrow$ & MCB $Q^{0.9}$ $\downarrow$ \\
\midrule
NN-det & 0.588 & 1.204 & N/A & N/A & 6.153 & 12.345 & N/A & N/A \\
EasyUQ & \underline{0.111} & \textbf{0.333} & 0.041 & 0.049 & \underline{1.155} & \underline{2.473} & 0.023 & 0.038 \\
Analogues & \textbf{0.106} & \underline{0.358} & 0.076 & 0.095 & \textbf{0.907} & \textbf{2.345} & 0.064 & 0.119 \\
GAN & 0.266 & 0.988 & 0.064 & 0.068 & 5.465 & 9.512 & 0.091 & 0.038 \\
CorrDiff & 0.264 & 0.591 & 0.059 & 0.043 & 2.836 & 7.049 & 0.051 & 0.097 \\
CorrDiff-s & 0.120 & 0.649 & 0.022 & \underline{0.014} & 2.138 & 3.861 & \textbf{0.013} & 0.033 \\
EnScale & 0.136 & 0.699 & \textbf{0.016} & \textbf{0.012} & 3.184 & 10.222 & 0.015 & \textbf{0.022} \\
EnScale-t & 0.167 & 0.488 & \underline{0.016} & 0.016 & 2.744 & 10.302 & \underline{0.014} & \underline{0.024} \\
\bottomrule
\multicolumn{9}{@{}p{0.99\textwidth}}{$^a$ For each method, we compute extreme quantiles marginally across all days for two seasons (winter / summer). We report absolute errors (AE) compared to RCM data (``AE Q'' columns). The miscalibration metric for quantiles (``MCB Q'' columns) evaluates reliability of quantiles in the conditional distribution of the samples given GCM input.}
\end{tabular}

	\label{tab:results:quantiles}
\end{table*}

\clearpage
\subsection{Evaluation of variable dependencies}
\label{sec:appendix:dependencies}

Firstly, we consider errors in pairwise correlations between variables for three selected variable pairs. For all pairs, the analogue benchmark reaches best or second-best scores. This is because the analogues only reproduce existing RCM data. By construction, the dependencies between variables are correct for the existing data. Scores by \ourmethod and \ourmethodtemp are slightly higher. Also CorrDiff-s yields low errors, while CorrDiff performs substantially worse. In contrast, EasyUQ which models each variable independently of others, shows much higher errors. This is another example for the benefits of our multivariate approach.

The next metrics evaluates dependencies of the variables conditional on GCM data $X$: we consider calibration of the pairwise difference of variables (last three columns of Tab.~\ref{tab:results:multivariable}, see Sec.~\ref{sec:methods:validation:variable-dep} for details). Now, performance of the analogues is worse. Additional evaluation shows that they are again underdispersed (not shown). Instead, \ourmethod or \ourmethodtemp achieve the best results, followed by CorrDiff-s. This finding again highlights the difference when analyzing marginals vs. conditional distributions, as in Sec.~\ref{sec:results:extremes}.

% \begin{table}[]
%     \caption{Multi-variable evaluation$^a$ \red{new}}
%     \centering
%     \input{tables/table_multivariate_updated_vreview1}
%     \label{tab:results:multivariable}
% \end{table}
\begin{table}[]
    \caption{Multi-variable evaluation$^a$}
    \centering
    \begin{tabular}{lllllll}
\toprule
Method & \multicolumn{3}{r}{Correlations} & \multicolumn{3}{r}{Miscalibration} \\
 & pr-sfcWind & sfcWind-rsds & tas-pr & pr-sfcWind & sfcWind-rsds & tas-pr \\
\midrule
NN-det & 0.125 & 0.062 & 0.060 & N/A & N/A & N/A \\
EasyUQ & 0.166 & 0.217 & 0.055 & 0.562 & 0.790 & 0.388 \\
Analogues & \textbf{0.026} & \underline{0.029} & \underline{0.025} & 0.355 & 0.332 & 0.325 \\
GAN & 0.063 & 0.056 & 0.042 & 0.097 & 0.093 & 0.123 \\
CorrDiff & 0.126 & 0.094 & 0.117 & 0.195 & 0.194 & 0.214 \\
CorrDiff-s & \underline{0.026} & 0.046 & \textbf{0.024} & 0.061 & \textbf{0.052} & 0.084 \\
EnScale & 0.043 & 0.033 & 0.035 & \textbf{0.055} & \underline{0.056} & \underline{0.069} \\
EnScale-t & 0.038 & \textbf{0.027} & 0.043 & \underline{0.056} & 0.057 & \textbf{0.067} \\
\bottomrule
\multicolumn{7}{@{}p{0.8\textwidth}}{$^a$We calculate pairwise correlations between two variables in both the samples and the RCM data, and report absolute errors averaged over 1000 randomly chosen locations (first three columns). The last three columns show miscalibration of the pairwise differences of variables. Bold and underlined values highlight the best and second-best results, respectively.}
\end{tabular}

    \label{tab:results:multivariable}
\end{table}

\clearpage
\Needspace{5\baselineskip}
\subsection{Further time series evaluation}

\begin{figure}
	\centering
	\includegraphics[width=0.9\linewidth]{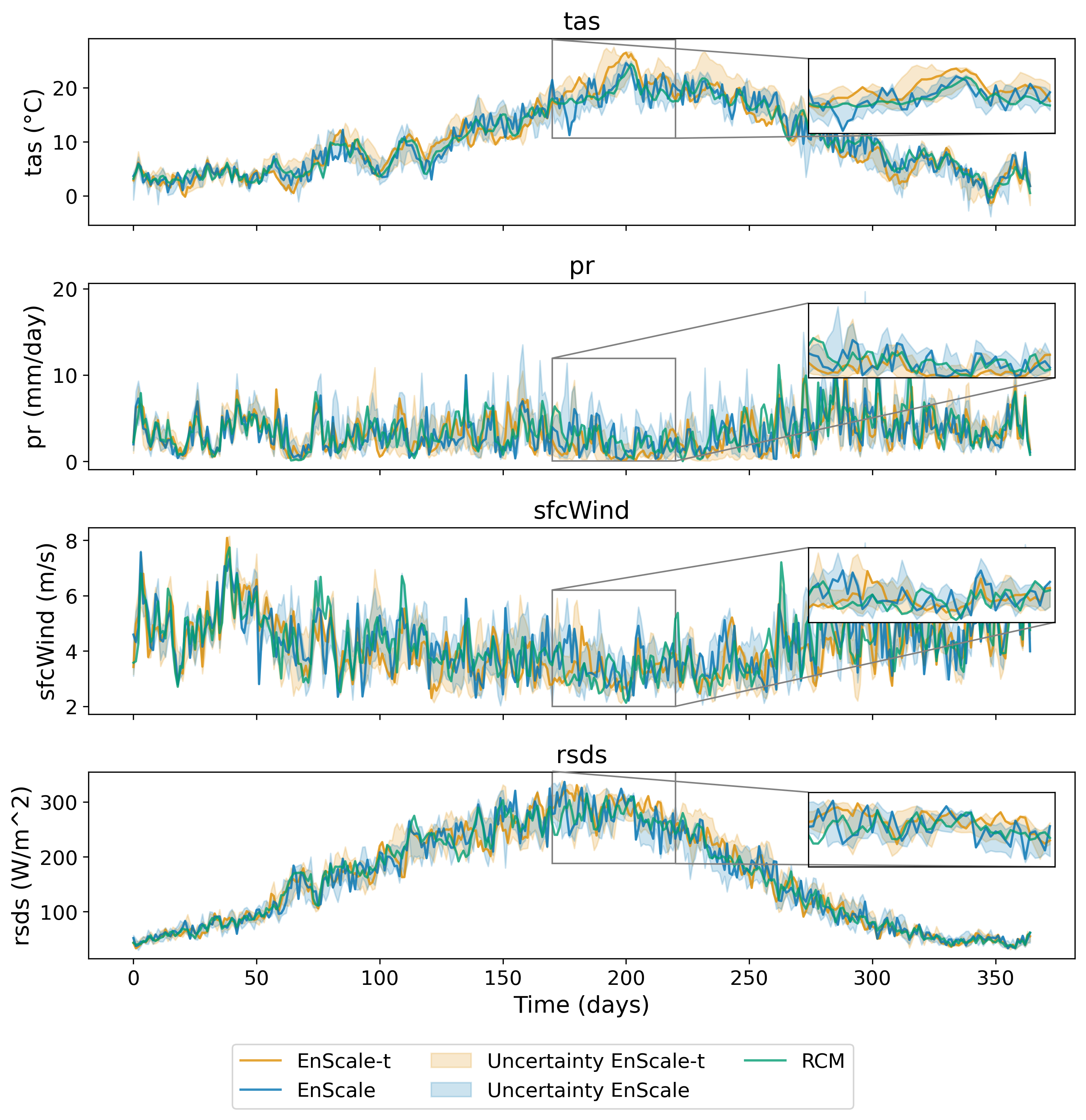}
	\caption{As Fig.~\ref{fig:results:overview-time-series}, omitting CorrDiff, but presenting all four target variables instead.}
	\label{fig:results:time-series}
\end{figure}

% \begin{table}[H]
% \caption{Errors in lag-1 autocorrelations (ACF) for each variable}
%     \centering
%     \input{tables/table_acfs_signed_and_abs_error_vreview1}
%     \label{tab:appendix:temporal-acf-both-errors}
% \end{table}

\begin{table}[H]
\caption{Signed and absolute errors in lag-1 autocorrelations (ACF) for each variable$^a$}
    \centering
    \begin{tabular}{lllllllll}
\toprule
Method & \multicolumn{2}{c}{tas} & \multicolumn{2}{c}{pr} & \multicolumn{2}{c}{sfcWind} & \multicolumn{2}{c}{rsds} \\
\cmidrule(lr){2-3} \cmidrule(lr){4-5} \cmidrule(lr){6-7} \cmidrule(lr){8-9}
& Error & Abs. Error & Error & Abs. Error & Error & Abs. Error & Error & Abs. Error \\
\midrule
NN-det & 0.01 & 0.01 & 0.17 & 0.17 & 0.08 & 0.08 & 0.11 & 0.11 \\
EasyUQ & -0.03 & 0.03 & -0.07 & 0.07 & -0.15 & 0.15 & -0.03 & 0.03 \\
Analogues & -0.1 & 0.1 & -0.12 & 0.12 & -0.29 & 0.29 & -0.05 & 0.05 \\
GAN & -0.06 & 0.06 & -0.1 & 0.1 & -0.21 & 0.21 & 0.02 & 0.02 \\
CorrDiff & 0.01 & 0.01 & 0.22 & 0.22 & 0.1 & 0.1 & 0.09 & 0.09 \\
CorrDiff-s & -0.06 & 0.06 & -0.09 & 0.09 & -0.21 & 0.21 & -0.04 & 0.04 \\
EnScale & -0.08 & 0.08 & -0.09 & 0.09 & -0.22 & 0.22 & -0.05 & 0.05 \\
EnScale-t & -0.01 & 0.01 & -0.01 & 0.04 & -0.02 & 0.03 & -0.01 & 0.02 \\
\bottomrule
\multicolumn{9}{@{}p{0.85\textwidth}}{$^a$ We calculate ACFs in each location for both the samples and the RCM data. We average both the signed difference of samples' ACF minus target ACF (columns ``Error'') as well as the absolute errors in ACFs for samples minus target (columns ``Abs. Error.''). Both metrics are averaged across locations.}
\end{tabular}

    \label{tab:appendix:temporal-acf-both-errors}
\end{table}

\clearpage
\subsection{Results on extrapolation test period}

\label{sec:appendix:extrapolation-results}

\begin{figure}[H]
    \centering
    \includegraphics[width=0.99\linewidth]{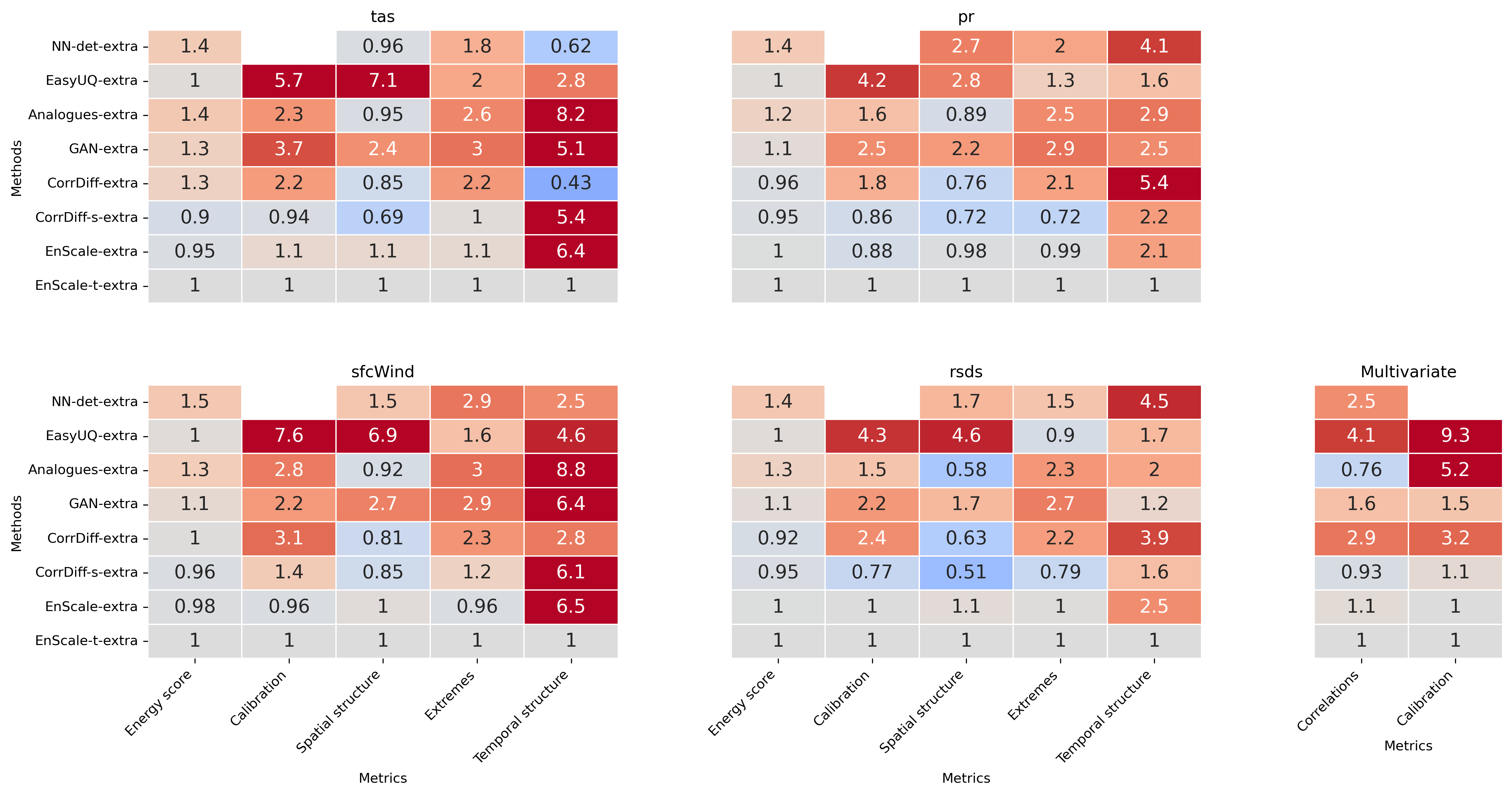}
    \caption{As Fig. \ref{fig:results:overview-summary}, but showing results for the extrapolation test period.}
    \label{fig:appendix:overview-summary-extra}
\end{figure}

\begin{figure}[H]
    \centering
    \includegraphics[width=0.99\linewidth]{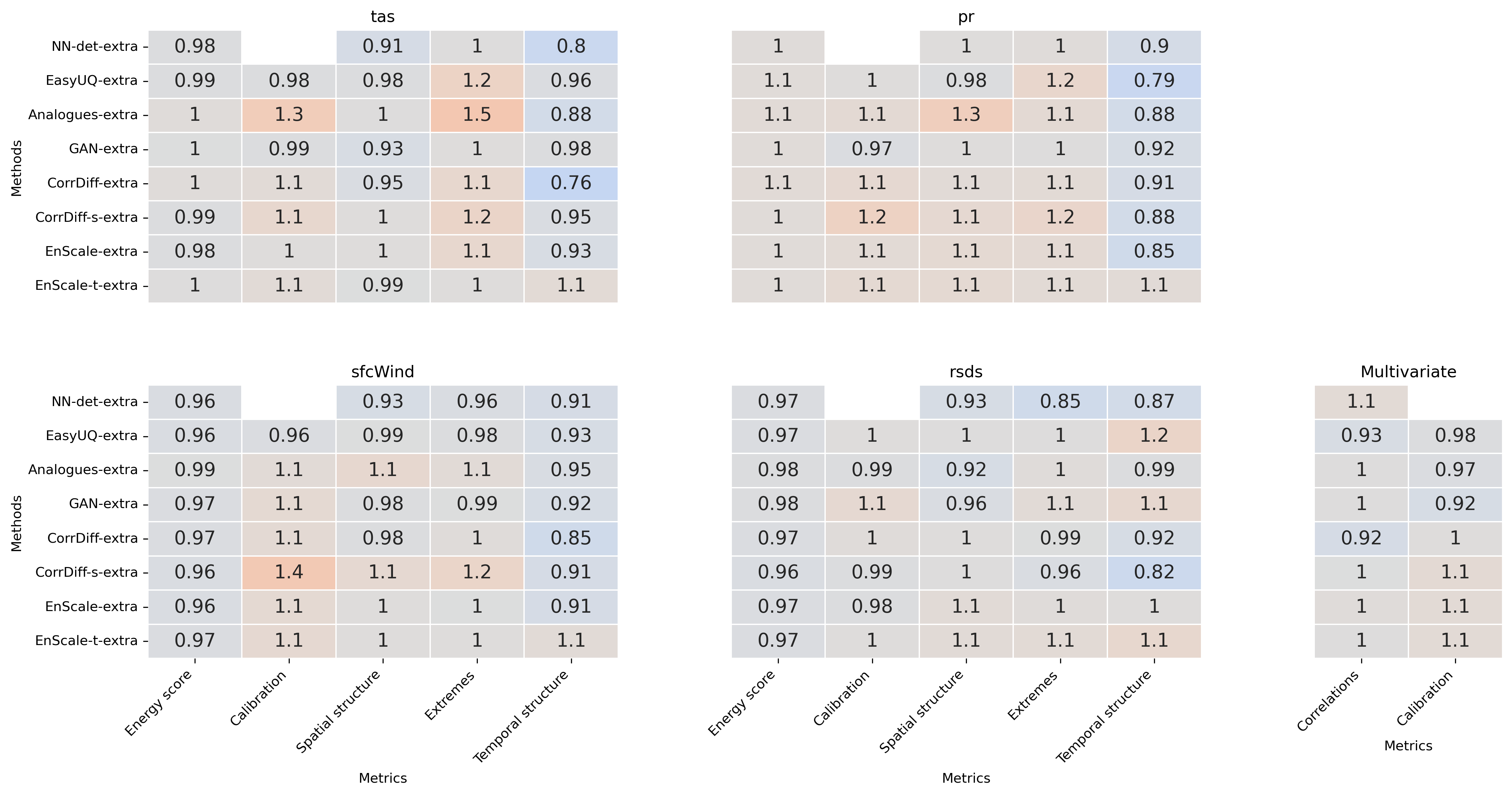}
    \caption{As Fig. \ref{fig:results:overview-summary}, but comparing performance on extrapolation vs. interpolation test set: For each method and each metric, we calculate the ratio of the metric on the extrapolation test set vs. that on the interpolation test set.}
    \label{fig:appendix:overview-summary-extra-vs-inter-ratio}
\end{figure}

We present a selection of results also on the extrapolation test period (2090-2099). Fig. \ref{fig:appendix:overview-summary-extra} summarizes the evaluation and Tab. \ref{tab:app:results:eval1-all-extra}  (as Tab. \ref{tab:results:eval1-all}  in the main text) presents performance w.r.t. MSE, energy score, CRPS per location and after max-pooling as well as RALSD. The ordering of the methods remains consistent with the interpolation period, and absolute scores are very similar. To directly compare scores on interpolation and extrapolation test set, we compute the ratios between the two test periods (Fig. \ref{fig:appendix:overview-summary-extra-vs-inter-ratio}). We find that most ratios are close to one, indicating at most small differences in performance between the two periods. Also, different methods seem to extrapolate similarly well. This is expected here since this is only a mild extrapolation case: the training data extend until 2089, and the test period (2090–2099) differs only slightly in climate conditions, with long-term climate change signals being small compared to daily variability.

\begin{table}[H]
\caption{As Tab. \ref{tab:results:eval1-all}, but for the extrapolation test set.}
\begin{tabular}{llllllll}
\midrule
\multicolumn{7}{c}{tas} \\
\midrule
Method & MSE $\downarrow$ & ES $\downarrow$ & ES$_\text{pred}$ & ES$_\text{var}$ & CRPS $\downarrow$ & CRPS-mp $\downarrow$ & RALSD $\downarrow$ \\
\midrule
NN-det-extra & \textbf{4.4} ± \footnotesize{0.08} & 240.1 ± \footnotesize{1.83} & 239.9 & 0.0 & 1.51 ± \footnotesize{0.01} & 1.51 ± \footnotesize{0.01} & \textbf{0.8} ± \footnotesize{0.01} \\
EasyUQ-extra & 6.3 ± \footnotesize{0.08} & 179.6 ± \footnotesize{1.49} & 305.3 & 251.3 & 1.20 ± \footnotesize{0.01} & 2.14 ± \footnotesize{0.02} & 6.1 ± \footnotesize{0.02} \\
Analogues-extra & 8.0 ± \footnotesize{0.12} & 249.2 ± \footnotesize{2.91} & 415.6 & 332.7 & 1.74 ± \footnotesize{0.01} & 1.72 ± \footnotesize{0.01} & 1.2 ± \footnotesize{0.01} \\
GAN-extra & 9.3 ± \footnotesize{0.09} & 223.3 ± \footnotesize{1.43} & 377.4 & 308.1 & 1.48 ± \footnotesize{0.01} & 1.45 ± \footnotesize{0.01} & 2.2 ± \footnotesize{0.01} \\
CorrDiff-extra & 6.8 ± \footnotesize{0.11} & 219.7 ± \footnotesize{2.36} & 394.1 & 348.6 & \underline{1.09} ± \footnotesize{0.01} & \underline{1.08} ± \footnotesize{0.01} & 0.9 ± \footnotesize{0.01} \\
CorrDiff-s-extra & 7.6 ± \footnotesize{0.13} & \textbf{157.2} ± \footnotesize{1.69} & 313.9 & 313.4 & \textbf{1.08} ± \footnotesize{0.01} & \textbf{1.07} ± \footnotesize{0.01} & \underline{0.9} ± \footnotesize{0.01} \\
EnScale-extra & \underline{4.5} ± \footnotesize{0.08} & \underline{166.8} ± \footnotesize{1.5} & 343.3 & 352.7 & 1.18 ± \footnotesize{0.01} & 1.19 ± \footnotesize{0.01} & 1.3 ± \footnotesize{0.01} \\
EnScale-t-extra & 4.8 ± \footnotesize{0.08} & 174.8 ± \footnotesize{1.78} & 348.2 & 346.7 & 1.22 ± \footnotesize{0.01} & 1.22 ± \footnotesize{0.01} & 1.2 ± \footnotesize{0.01} \\
\midrule
\multicolumn{7}{c}{pr} \\
\midrule
Method & MSE $\downarrow$ & ES $\downarrow$ & ES$_\text{pred}$ & ES$_\text{var}$ & CRPS $\downarrow$ & CRPS-mp $\downarrow$ & RALSD $\downarrow$ \\
\midrule
NN-det-extra & \textbf{35.1} ± \footnotesize{0.8} & 632.3 ± \footnotesize{6.3} & 632.6 & 0.0 & 2.38 ± \footnotesize{0.03} & 6.41 ± \footnotesize{0.07} & 10.5 ± \footnotesize{0.13} \\
EasyUQ-extra & 54.2 ± \footnotesize{0.93} & 469.8 ± \footnotesize{5.29} & 838.8 & 737.5 & 1.98 ± \footnotesize{0.02} & 11.38 ± \footnotesize{0.07} & 9.6 ± \footnotesize{0.05} \\
Analogues-extra & 45.7 ± \footnotesize{0.88} & 567.2 ± \footnotesize{6.36} & 935.8 & 737.9 & 2.42 ± \footnotesize{0.02} & 6.30 ± \footnotesize{0.06} & 5.7 ± \footnotesize{0.07} \\
GAN-extra & 53.0 ± \footnotesize{1.24} & 481.2 ± \footnotesize{5.27} & 799.1 & 635.0 & 1.98 ± \footnotesize{0.02} & 5.38 ± \footnotesize{0.06} & 8.8 ± \footnotesize{0.07} \\
CorrDiff-extra & 55.1 ± \footnotesize{1.0} & \underline{438.8} ± \footnotesize{4.93} & 832.6 & 788.1 & \underline{1.90} ± \footnotesize{0.02} & \underline{4.92} ± \footnotesize{0.05} & \textbf{4.0} ± \footnotesize{0.06} \\
CorrDiff-s-extra & 61.7 ± \footnotesize{1.31} & \textbf{434.2} ± \footnotesize{5.11} & 854.8 & 842.4 & \textbf{1.85} ± \footnotesize{0.02} & \textbf{4.75} ± \footnotesize{0.05} & \underline{4.5} ± \footnotesize{0.06} \\
EnScale-extra & 36.2 ± \footnotesize{0.81} & 454.4 ± \footnotesize{5.19} & 849.1 & 788.6 & 1.93 ± \footnotesize{0.02} & 5.08 ± \footnotesize{0.05} & 5.4 ± \footnotesize{0.06} \\
EnScale-t-extra & \underline{35.5} ± \footnotesize{0.75} & 455.8 ± \footnotesize{5.38} & 831.3 & 750.7 & 1.92 ± \footnotesize{0.02} & 5.02 ± \footnotesize{0.05} & 5.3 ± \footnotesize{0.06} \\
\midrule
\multicolumn{7}{c}{sfcWind} \\
\midrule
Method & MSE $\downarrow$ & ES $\downarrow$ & ES$_\text{pred}$ & ES$_\text{var}$ & CRPS $\downarrow$ & CRPS-mp $\downarrow$ & RALSD $\downarrow$ \\
\midrule
NN-det-extra & \underline{2.6} ± \footnotesize{0.03} & 196.5 ± \footnotesize{1.03} & 196.6 & 0.0 & 1.13 ± \footnotesize{0.01} & 1.44 ± \footnotesize{0.01} & 2.0 ± \footnotesize{0.02} \\
EasyUQ-extra & 4.2 ± \footnotesize{0.03} & 141.2 ± \footnotesize{0.78} & 255.6 & 229.0 & 0.88 ± \footnotesize{0.01} & 2.22 ± \footnotesize{0.01} & 8.9 ± \footnotesize{0.02} \\
Analogues-extra & 3.9 ± \footnotesize{0.04} & 180.9 ± \footnotesize{1.56} & 304.1 & 246.5 & 1.13 ± \footnotesize{0.01} & 1.39 ± \footnotesize{0.01} & 2.0 ± \footnotesize{0.02} \\
GAN-extra & 4.7 ± \footnotesize{0.04} & 151.4 ± \footnotesize{0.84} & 269.9 & 236.9 & 0.95 ± \footnotesize{0.0} & 1.16 ± \footnotesize{0.01} & 3.6 ± \footnotesize{0.03} \\
CorrDiff-extra & 3.8 ± \footnotesize{0.04} & 137.3 ± \footnotesize{0.9} & 269.4 & 264.2 & \underline{0.82} ± \footnotesize{0.0} & \textbf{1.00} ± \footnotesize{0.01} & \textbf{1.4} ± \footnotesize{0.01} \\
CorrDiff-s-extra & 4.4 ± \footnotesize{0.05} & \textbf{129.2} ± \footnotesize{0.87} & 255.4 & 252.4 & \textbf{0.81} ± \footnotesize{0.0} & \underline{1.00} ± \footnotesize{0.01} & \underline{1.6} ± \footnotesize{0.02} \\
EnScale-extra & \textbf{2.6} ± \footnotesize{0.03} & \underline{133.1} ± \footnotesize{0.84} & 263.3 & 260.5 & 0.85 ± \footnotesize{0.0} & 1.06 ± \footnotesize{0.01} & 1.8 ± \footnotesize{0.02} \\
EnScale-t-extra & 2.7 ± \footnotesize{0.03} & 135.2 ± \footnotesize{0.9} & 264.1 & 257.8 & 0.86 ± \footnotesize{0.0} & 1.07 ± \footnotesize{0.01} & 1.8 ± \footnotesize{0.01} \\
\midrule
\multicolumn{7}{c}{rsds} \\
\midrule
Method & MSE $\downarrow$ & ES $\downarrow$ & ES$_\text{pred}$ & ES$_\text{var}$ & CRPS $\downarrow$ & CRPS-mp $\downarrow$ & RALSD $\downarrow$ \\
\midrule
NN-det-extra & \textbf{1917.0} ± \footnotesize{32.37} & 4886.2 ± \footnotesize{45.21} & 4889.5 & 0.0 & 29.93 ± \footnotesize{0.27} & 25.74 ± \footnotesize{0.23} & 3.5 ± \footnotesize{0.03} \\
EasyUQ-extra & 3216.3 ± \footnotesize{42.23} & 3518.3 ± \footnotesize{31.72} & 6613.3 & 6186.4 & 23.59 ± \footnotesize{0.21} & 46.95 ± \footnotesize{0.29} & 10.3 ± \footnotesize{0.03} \\
Analogues-extra & 2557.9 ± \footnotesize{38.29} & 4275.5 ± \footnotesize{47.61} & 7228.4 & 5889.6 & 28.22 ± \footnotesize{0.25} & 21.04 ± \footnotesize{0.2} & 1.8 ± \footnotesize{0.02} \\
GAN-extra & 2903.0 ± \footnotesize{43.52} & 3563.3 ± \footnotesize{31.19} & 6196.1 & 5271.6 & 23.46 ± \footnotesize{0.19} & 17.73 ± \footnotesize{0.14} & 3.7 ± \footnotesize{0.03} \\
CorrDiff-extra & 3457.1 ± \footnotesize{52.99} & \textbf{3100.1} ± \footnotesize{26.18} & 6240.6 & 6292.5 & \underline{21.85} ± \footnotesize{0.21} & \underline{15.58} ± \footnotesize{0.14} & \textbf{1.5} ± \footnotesize{0.01} \\
CorrDiff-s-extra & 3143.7 ± \footnotesize{52.35} & \underline{3207.3} ± \footnotesize{31.0} & 6299.2 & 6166.7 & \textbf{20.95} ± \footnotesize{0.2} & \textbf{15.00} ± \footnotesize{0.14} & \underline{1.6} ± \footnotesize{0.01} \\
EnScale-extra & \underline{1932.3} ± \footnotesize{29.54} & 3363.9 ± \footnotesize{29.12} & 6844.5 & 6945.6 & 23.24 ± \footnotesize{0.19} & 20.98 ± \footnotesize{0.14} & 2.9 ± \footnotesize{0.02} \\
EnScale-t-extra & 1947.6 ± \footnotesize{31.29} & 3373.0 ± \footnotesize{32.02} & 6885.7 & 7028.8 & 23.01 ± \footnotesize{0.2} & 21.40 ± \footnotesize{0.16} & 2.8 ± \footnotesize{0.02} \\
\end{tabular}
\label{tab:app:results:eval1-all-extra}
\end{table}

\section{Extrapolation to unseen GCMs}
\label{sec:appendix:gcm-extrapolation}
\begin{table}[H]
\caption{Scores per GCM-RCM pair. GCM-RCM combinations that were \textbf{not} included in the train data are marked in bold.}
    \label{tab:appendix:gcm-extrapolation}
\begin{tabular}{lllll}
\midrule
\multicolumn{5}{c}{tas} \\
\midrule
RCM & GCM & MSE & ES & CRPS \\
\midrule
ALADIN63 & CNRM-CM5 & 5.46 & 184.86 & 1.29 \\
\textbf{ALADIN63}$^\star$ & \textbf{HadGEM2-ES}$^\star$ & 7.31 & 214.38 & 1.53 \\
ALADIN63 & MPI-ESM-LR & 4.89 & 175.68 & 1.22 \\
CCLM4-8-17 & CNRM-CM5 & 3.68 & 154.43 & 1.09 \\
\textbf{CCLM4-8-17}$^\star$ & \textbf{CanESM2}$^\star$ & 6.70 & 216.06 & 1.52 \\
CCLM4-8-17 & MIROC5 & 4.09 & 162.19 & 1.16 \\
CCLM4-8-17 & MPI-ESM-LR & 2.53 & 130.81 & 0.93 \\
REMO2015 & MIROC5 & 6.25 & 195.85 & 1.39 \\
RegCM4-6 & CNRM-CM5 & 5.46 & 186.48 & 1.31 \\
RegCM4-6 & MPI-ESM-LR & 4.82 & 173.66 & 1.24 \\
\midrule
\multicolumn{5}{c}{pr} \\
\midrule
RCM & GCM & MSE & ES & CRPS \\
\midrule
ALADIN63 & CNRM-CM5 & 40.04 & 502.29 & 2.33 \\
\textbf{ALADIN63}$^\star$ & \textbf{HadGEM2-ES}$^\star$ & 40.74 & 517.32 & 2.26 \\
ALADIN63 & MPI-ESM-LR & 29.03 & 413.21 & 1.86 \\
CCLM4-8-17 & CNRM-CM5 & 31.01 & 432.09 & 1.85 \\
\textbf{CCLM4-8-17}$^\star$ & \textbf{CanESM2}$^\star$ & 20.21 & 326.74 & 1.37 \\
CCLM4-8-17 & MIROC5 & 32.92 & 441.67 & 1.73 \\
CCLM4-8-17 & MPI-ESM-LR & 26.21 & 394.10 & 1.69 \\
REMO2015 & MIROC5 & 52.25 & 557.67 & 2.20 \\
RegCM4-6 & CNRM-CM5 & 21.93 & 366.25 & 1.79 \\
RegCM4-6 & MPI-ESM-LR & 21.49 & 356.76 & 1.73 \\
\midrule
\multicolumn{5}{c}{sfcWind} \\
\midrule
RCM & GCM & MSE & ES & CRPS \\
\midrule
ALADIN63 & CNRM-CM5 & 2.82 & 140.44 & 0.87 \\
\textbf{ALADIN63}$^\star$ & \textbf{HadGEM2-ES}$^\star$ & 3.55 & 159.83 & 1.00 \\
ALADIN63 & MPI-ESM-LR & 2.25 & 124.98 & 0.78 \\
CCLM4-8-17 & CNRM-CM5 & 2.55 & 133.71 & 0.84 \\
\textbf{CCLM4-8-17}$^\star$ & \textbf{CanESM2}$^\star$ & 3.06 & 147.64 & 0.94 \\
CCLM4-8-17 & MIROC5 & 2.57 & 133.34 & 0.84 \\
CCLM4-8-17 & MPI-ESM-LR & 1.99 & 116.87 & 0.75 \\
REMO2015 & MIROC5 & 3.24 & 149.61 & 0.97 \\
RegCM4-6 & CNRM-CM5 & 3.59 & 158.37 & 1.04 \\
RegCM4-6 & MPI-ESM-LR & 3.39 & 152.52 & 1.01 \\
\midrule
\multicolumn{5}{c}{rsds} \\
\midrule
RCM & GCM & MSE & ES & CRPS \\
\midrule
ALADIN63 & CNRM-CM5 & 2141.77 & 3659.38 & 25.04 \\
\textbf{ALADIN63}$^\star$ & \textbf{HadGEM2-ES}$^\star$ & 1990.96 & 3567.66 & 24.57 \\
ALADIN63 & MPI-ESM-LR & 1860.73 & 3395.63 & 23.05 \\
CCLM4-8-17 & CNRM-CM5 & 2121.95 & 3478.42 & 24.30 \\
\textbf{CCLM4-8-17}$^\star$ & \textbf{CanESM2}$^\star$ & 2136.50 & 3537.87 & 24.29 \\
CCLM4-8-17 & MIROC5 & 2135.70 & 3496.39 & 24.28 \\
CCLM4-8-17 & MPI-ESM-LR & 1895.41 & 3327.91 & 22.88 \\
REMO2015 & MIROC5 & 3177.68 & 4340.34 & 30.12 \\
RegCM4-6 & CNRM-CM5 & 1589.19 & 3113.20 & 21.66 \\
RegCM4-6 & MPI-ESM-LR & 1482.80 & 2988.44 & 20.82 \\
\midrule
\multicolumn{5}{l}{\footnotesize $^\star$ GCM unseen during training (extrapolation case)} \\
\end{tabular}
\end{table}

\clearpage
\section{Details on methods}

\subsection{Conditional independence assumptions for multi-step approach}
\label{sec:appendix:marginal-cond-ind}
We explain downscaling via coarse correction in Sec. \ref{sec:setup:2step-marginal}. Here, we approximate the full distribution by assuming that $Y$ is independent of $X$ given $Z$. Formally, this means $Y \ind X | Z$. One can show that the concatenation of the coarse model predicting $p_{Z|X}$ with the model for $p_{Y|Z}$ yields $p_{Y|X}$ under this assumption (see proof below).
To verify validity of this approximation, we compare the loss when predicting $Y$ from $Z$ only versus from $Z, X$. If $p_{Y | X, Z} \neq p_{Y | Z},$ the loss should decrease in the second case. For simplicity, we checked these losses in a univariate setup, predicting only one variable at a time, and using one dense model as the super-resolution model (without $\tZ$ as an intermediate step). We checked precipitation and temperature. For precipitation, we could not observe any difference in loss. For temperature, the loss improves by about 2\% when adding $X$ as an additional predictor. This indicates that the conditional independence is a close approximation. 

The final setup is more complicated. In Sec. \ref{sec:methods:architecture}, we additionally describe the splitting of the super-resolution model into several intermediate steps. This requires additional approximations, which we have not verified explicitly.

Under the assumption $Y \ind X | Z$, the concatenation of \textit{coarse model} and \textit{super-resolution model} yields $p_{Y|X}$. We prove it in the following. We use that $A \ind B | C$ if $p_{A | B, C} = p_{A | C}$ to simplify the full conditional distribution. Here, we assume all distributions to have a density and denote those by small $p$. To model the conditional density $p_{Y|X}$, we integrate the density of the full distribution $p_{Y, Z}$ over all possible $Z$. In the following, we drop the subscripts of the densities for simplicity and use the corresponding lower case letters as arguments, i.e. $p_{Y|X}(y|x)$ will be written as $p(y|x)$.
\begin{align*}
	p(y|x) &= \int p(y, z | x) dz \\
	&= \int p(y|x, z) p(z | x) dz \\
	&\stackrel{(1)}{=} \int p(y|z) p(z | x) dz
\end{align*}

where (1) holds because $Y \ind X | Z$.
The integrand in final line is exactly the joint distribution $(Z, Y) | X$ that the concatenation of the two models yields. Then, integrating out $Z$ corresponds to selecting the marginal $Y|X$, equivalent to dropping the components $Z$. Thus, our three-step approach learns $p_{Y|X}$ under the given conditional independence assumptions.

\subsection{Conditional energy score loss}
\label{sec:methods:gen-model-energy-score:loss-cond}

%\quest{Xinwei correctly adjusted my notation in some cases from $p_{Y|X}$ to $p_{Y|X=x}$, but now it's inconsistent because I also used $p_{Y|X}(.|x)$ for the same thing. I guess I find Xinwei's suggestion easier to read and will adjust to that. Let me know if I missed something.}

Our generative model should learn to sample from the RCM's conditional distribution $p^\iota_{Y \vert X=x}$ for all GCM inputs $x$ and for each GCM-RCM pair $\iota$. Thus, we need to adjust our loss function to this conditional case. For simplicity, we drop the superscript $\iota$ in the following description. We define the loss for inputs $X$ and output $Y$, but apply the equivalent loss function to the individual steps in our pipeline (see Sec.~\ref{sec:methods:architecture}). The generative model, conditioned on $X$, samples from the conditional distribution $q_{Y|X}$. 

The objective is to ensure that the learned conditionals match with those from the RCM, i.e.,~$q_{Y|X=x} = p_{Y|X=x}$ for all $x$ and for each RCM. However, for each GCM input $x$ and for each RCM, we only have one corresponding RCM realization $Y \sim p_{Y|X=x}$. Thus, we draw multiple GCM inputs $X$ from the GCM distribution $p_X$ and average over those in the loss function:
\begin{align}
\label{eq:loss-cond-exp}
\text{Loss}_{\text{cond}} =  
\underbrace{E_{X \sim p_X} \Bigg( E_{Y \sim p_{Y|X}, \hat{Y} \sim q_{Y|X}}(\lVert Y - \hat{Y} \rVert) \Bigg)}_{\text{ES}_\text{pred}}  
- \frac{1}{2} \underbrace{E_{X \sim p_X} \Bigg( E_{\hat{Y}, \hat{Y}' \sim q_{Y|X}}(\lVert \hat{Y} - \hat{Y}' \rVert)}_{\text{ES}_\text{var}}  
\Bigg)
\end{align}
We call this the \condenscoreloss. As before, we call the prediction error (first term) $\text{ES}_\text{pred}$ and the variability (second term) $\text{ES}_\text{var}$. This loss is minimized if and only if, for each predictor $x$, the predicted conditional distributions $q_{Y|X=x}$ indeed match the target conditionals $p_{Y|X=x}$. More formally, the score is minimized if and only if $q_{Y|X=x} = p_{Y \vert X=x}$ $p_X$-almost everywhere \cite{Pacchiardi2024ProbabilisticMinimization}. As for the unconditional loss in Eq. \eqref{eq:exp-loss-uncond}, if $p_{Y|X=x}=q_{Y|X=x}$ for each $x$, it holds that $\text{ES}_\text{pred} = \text{ES}_\text{var}$. During training, we use minibatching. We describe the standard batch-wise approximation of the loss in App.~\ref{sec:appendix:batchwise-approx}, where we use two samples by the generative model for each input $x$. Each batch contains data from multiple GCM-RCM pairs, such that the loss is averaged across pairs.

\subsection{Batch-wise approximation}
\label{sec:appendix:batchwise-approx}
The theoretical expected values (e.g. Eq. \eqref{eq:loss-cond-exp}) are impossible to calculate. However, during training, they can be replaced by averages over mini-batches of data, yielding an unbiased estimate of the expectation. Let $(x_i, y_i) \, i \in \{1,...,n\}$ be a mini-batch of data, and $\hat{y_i}, \hat{y_i}' \sim Q_{Y|X}(.|x_i)$ two samples from the conditional distribution of the RCM data given the GCM input $x_i$. Then the finite-sample approximation of the loss is
\begin{align*}
	\hat{\text{Loss}}_{\text{cond}} = \frac{1}{n} \sum_{i=1}^n (\lVert y_i - \hat{y_i} \rVert - \frac{1}{2} \lVert \hat{y_i} - \hat{y_i}' \rVert).
\end{align*}
Note that, per input $x_i$, we only use two samples from the generative model, as we found that increasing the batch size has a similar effect as increasing the number of samples. This loss can be minimised via stochastic gradient descent, as usual in machine learning.

\subsection{Model architecture}
\label{sec:appendix:model-architecture}

\subsubsection{Coarse model}
The first coarse model is a dense multilayer perceptron (MLP) and its architecture is as in \citeA{Shen2024Engression:Regression}. To enable the model to capture difference between GCM-RCM pairs, we integrate the index for the RCM considered: We one-hot-encode the RCM index and append it to the input. In addition, the MLP is modified to include random noise that allows for stochastic output. %\red{Need a note on $\tZ$, reference Sec. \ref{sec:appendix:training2step}.} 
As explained above, the generative model needs to take random initialization as input. Instead of just adding this noise to the input at the beginning, we add the randomness at each layer. We concatenate the hidden units with standard Gaussian noise, where the noise is scaled by a learnable parameter. This concatenation is passed into the next layer. Adding noise in multiple hidden layers allows to learn more complex representations of the Gaussian noise. For example, we find that this yields spatially correlated variability in the output.

Each layer consists of a linear map plus a ReLU activation function. We include skip connections that allow the input to bypass two layers, facilitating gradient flow and improving training stability.

Both the coarse model as well as its temporally-consistent extension consist of 5 hidden layers with sizes 200, 200, 200, 256, 256. The concatenated Gaussian noise has dimension 10 in each layer. We also experimented with larger noise dimension, but found no improvement in performance. We found better performance when adding a first layer to pre-process the inputs: Each input variable is compressed separately with a linear layer and a ReLU-activation, and only afterwards all variables are concatenated. In this way, the model is less sensitive to differences in units or scales in different input variables. 

\subsubsection{Super-resolution model}

To clarify, we explain how to calculate the number of learnable parameters in our sparse local layers. The first linear upsampling step uses $\text{number of high-res pixels} \times \text{low-res neighborhood size} \times \text{number of variables}$ parameters. The second linear aggregation uses $\text{number of high-res pixels} \times (\text{number of variables} + \text{number of noise channels}) \times \text{high-res neighborhood size}$ parameters. In addition, there are the learnable weights for the MLP (once, as this is the same for all locations) and it is applied to the full latent space (Fig.~\ref{fig:methods:sparselayers}), yielding the full multivariate output.

We choose 9 nearest neighbors of the low-resolution input in the first upsampling step. Afterwards, 5 noise channels of Gaussian i.i.d. noise are concatenated. For the stochastic refinement, 25 nearest neighbors in the high-resolution grid are used and linearly aggregated to a latent space of dimension 12 for each pixel. The shared MLP has two hidden layers of hidden dimension 12 each. We add a small weight decay penalty (scaling 1e-3), as we find that this improves image quality.

As for the coarse model, we enable learning RCM-specific details by integrating the index for the RCM into the sparse local layers (see Sec.~\ref{sec:methods:architecture}). In the first linear upsampling step, the learnable weights are RCM-specific, i.e. the model can learn a different set of weights for each RCM identifier. The stochastic refinement does not depend on the RCM and, thus, learned stochasticity is RCM-agnostic. 
In earlier stages of this work, we also tested learning a separate set of weights per GCM-RCM pair in both the first deterministic as well as the stochastic part, but found that using a joint model for the stochastic part reaches the same loss, while being more efficient computationally.

For \ourmethodtemp, conditioning on the previous timestep is again integrated only in the first linear upsampling step: For each variable, we interpolate the low-resolution input to the next resolution, i.e., we increase resolution by a factor of 2 in each dimension (with nearest neighbor interpolation). We then concatenate the high-resolution prediction from the previous day as a separate channel and use both channels as an input for the linear upsampling.

\subsection{Training}

We train all models using the Adam optimizer with learning rate 0.0001. We check convergence of the test loss. The coarse and temporal coarse model are trained for 200 epochs. The sparse layers in the super-resolution model are also trained for the 200 epochs (stages 1--3) and for 100 epochs (last stage).

\subsection{Details on benchmarks}

\label{sec:appendix:details-benchmarks}
\subsubsection{Deterministic neural network}

We use a similar model architecture to the dense models in our generative approach (see Sec. \ref{sec:appendix:model-architecture}), using a fully-connected neural network with alternating linear maps and ReLU activation. However, unlike for the generative models, there is no noise input needed for the deterministic benchmark. The layers have sizes 3612, 500, 500, 500, 1024, 4096, 16384.

The data transformations are the same as for \ourmethod: The predictors are scaled with a simple standardization, and the targets are transformed to uniform. The deterministic network converges very quickly. We use the model checkpoints after 25 epochs. Afterwards, the test loss increases again.

\subsubsection{EasyUQ}
\label{sec:appendix:details-easyuq}

EasyUQ's predictions are optimal under a monotonicity assumption, in the sense that they minimize the CRPS. The monotonicity assumption describes that the targets $Y$ tend to increase as the regression model's predictions $\hat{\mu}$ also increase. This is a natural requirement for a useful statistical model. However, if the regression model is deficient, this assumption might be violated. EasyUQ will still yield predictive distributions, but optimality is not guaranteed in this case. For more details on EasyUQ and its mathematical guarantees, see \citeA{Walz2024EasyOutput} and \citeA{Henzi2021IsotonicRegression}.

We implement EasyUQ based on the Python package \textit{isodisreg} \cite{WalzIsodisreg}. We train EasyUQ on the training data that was also used to train the NN-det, separately for each location. After training, we retrieve the predicted distributions on the test data and sample from them, independently in each location, time point and for each variable. For this, we slightly adjust the default prediction method from the Python package in order to be able to take in a different set of probabilities for each point in time. These samples can then be compared to the samples from \ourmethod.

\subsection{Analogues}
\label{sec:appendix:details-analogues}

We provide more details on our implementation of the analogues in the following: before generating samples, we calculate a distance matrix capturing pairwise similarities between days in a GCM dataset, including train and test periods. To reduce computational complexity, we restrict comparisons to days within a 50-calendar-day window. For each pair of days, we compute the Euclidean distance of the GCM's spatial fields separately per variable, considering all four target variables as well as sea level pressure.
To aggregate across variables, we follow a rank-based approach: for each variable and day, we sort the distances to the other days, convert them to ranks. Finally, we average the ranks over all variables. This approach standardizes the distances, avoiding scale issues from different units and dependence on data transformations. In summary, two days are considered to be ``close'' if they are similar w.r.t. all the five GCM variables.

Using this distance matrix, we can approximate $p^\iota_{Y|X}$ independently for each GCM-RCM pair. For each day in the test data, we identify the $k = 5$ days in the train data with the smallest distance, so-called \textit{nearest neighbors} (again within a 50-day calendar window). We approximate the conditional $p^\iota_{Y|X}$ by the discrete distribution with these $k$ data points. To generate an ensemble, we sample with replacement from these $k$ nearest neighbors.

The choice of $k$ balances the prediction error with the variability: smaller values of $k$ imply closer neighbors and smaller predictions errors but low variability. Larger $k$ increase diversity at the cost of larger prediction error. We found $k=5$ to be a good compromise. 

We are aware that a single GCM dataset with about 30k train data points is rather small for an analogue-based approach. However, combining data from multiple GCMs or RCMs gives further issues due to distribution shifts, which is why we include the simple version as a conceptually easy benchmark.

\subsection{Proof of variance decomposition}
\label{sec:appendix:var-coarse-super}
We prove the decomposition from the main text using law of total variance $\var(Y) = E[\var(Y | Z)] + \var(E[Y|Z])$. However, we apply it to $\var^\iota(Y|X)$, s.t. we don't only condition on $Z$ but also on $X$:
\begin{equation*}
	\var^\iota(Y|X) = E[\var^\iota(Y|Z, X) | X] + \var^\iota(E^\iota[Y|Z, X] | X),
\end{equation*}
We explained above (see App.~\ref{sec:appendix:marginal-cond-ind}) that our approach to downscaling via coarse correction assumes $Y \ind X |Z$. Hence, $\var^\iota(Y|Z, X) = \var^\iota(Y|Z)$ and $E^\iota[Y|Z, X] = E^\iota[Y|Z]$, which concludes the proof.

\printbibliography

\end{document}